\documentclass[twoside]{article}
\usepackage{amsfonts,amssymb,amsmath,amsthm,qic}

\renewcommand{\thefootnote}{\fnsymbol{footnote}}  

\usepackage{algpseudocode}
\usepackage[section]{algorithm}
\usepackage{xspace}
\usepackage{multirow}
\usepackage{rotating}
\usepackage[usenames, dvipsnames, table]{xcolor}
\usepackage{tikz}
\usepackage{tikz-3dplot}

\input{Qcircuit}
\renewcommand{\Qcircuit}[1][0em]{\xymatrix @*=<#1>}

\newcommand{\braket}[2]{\langle #1 | #2 \rangle}

\newcommand{\cC}{\mathcal{C}}
\newcommand{\cM}{\mathcal{M}}
\newcommand{\cA}{\mathcal{A}}

\newcommand{\cH}{\mathcal{H}}

\newcommand{\normform}{basis form}

\makeatletter
\renewcommand*\env@matrix[1][*\c@MaxMatrixCols c]{%
  \hskip -\arraycolsep
  \let\@ifnextchar\new@ifnextchar
  \array{#1}}
\makeatother

\newtheorem{proposition}[theorem]{Proposition}
\newtheorem{observation}[theorem]{Observation}
\newtheorem{example}[theorem]{Example}

\DeclareMathOperator*{\argmax}{arg\,max}

\newcommand{\myfcaption}[1]{
        \refstepcounter{figure}
\centering
        \parbox{7.5cm}{
\footnotesize\smalllineskip Fig.~\thefigure. #1
}
}

\begin{document}
\setlength{\textheight}{8.0truein}    

\runninghead{On the Geometry of Stabilizer States}
            {H. J. Garc\'{i}a, I. L. Markov, A. W. Cross}

\normalsize\textlineskip
\thispagestyle{empty}
\setcounter{page}{1}

\copyrightheading{0}{0}{2003}{000--000}

\vspace*{0.88truein}

\alphfootnote

\fpage{1}

\centerline{\bf
ON THE GEOMETRY OF STABILIZER STATES}
\vspace*{0.25truein}
\centerline{\footnotesize H\'{E}CTOR J. GARC\'{I}A,$^1$ IGOR L. MARKOV,$^1$ ANDREW W. CROSS$^2$}
\centerline{\footnotesize\it $^1$ University of Michigan, EECS, Ann Arbor, MI 48109-2121}
\centerline{\footnotesize\it \{hjgarcia, imarkov\}@eecs.umich.edu}
\centerline{\footnotesize\it $^2$ IBM T. J. Watson Research Center}
\centerline{\footnotesize\it Yorktown Heights, NY 10598}
\vspace*{0.225TRUEIN}


\abstracts{
Large-scale quantum computation is likely to require massive quantum error correction (QEC). 
QEC codes and circuits are described via the stabilizer formalism, which represents
{\em stabilizer states} by keeping track of the operators that preserve them. Such states 
are obtained by {\em stabilizer circuits} (consisting of CNOT, Hadamard and Phase gates)
and can be represented compactly on conventional computers 
using $\Theta (n^2)$ bits, where $n$ is the number of qubits \cite{Gottes98}. 
As an additional application, the work in~\cite{Aaron, AaronGottes} suggests the use 
of superpositions of stabilizer states to represent arbitrary quantum states.
To aid in such applications and improve our understanding of stabilizer states,
we characterize and count nearest-neighbor stabilizer states, quantify the distribution of angles between pairs of stabilizer states, study succinct stabilizer superpositions and stabilizer bivectors, explore the approximation of non-stabilizer states by single stabilizer states and short linear combinations of stabilizer states, develop an improved inner-product computation for stabilizer states via synthesis of compact canonical stabilizer circuits, propose an orthogonalization procedure for stabilizer states, and evaluate several of these algorithms empirically.
}{}{}

\vspace*{10pt}

\vspace*{3pt}
\communicate{to be filled by the Editorial}

\vspace*{1pt}\textlineskip    

\setcounter{footnote}{0}
\renewcommand{\thefootnote}{\alph{footnote}}

\section{Introduction} \label{sec:intro}

\noindent
Gottesman \cite{Gottes} and Knill showed that for certain types of 
quantum circuits known as {\em stabilizer circuits}, efficient simulation on 
classical computers is possible. Stabilizer circuits are exclusively composed 
of {\em stabilizer gates}~---~Controlled-NOT, Hadamard and Phase gates 
(Figure~\ref{fig:chp_pauli}a)~---~followed by measurements in the
computational basis. Such circuits are applied to a computational basis state
(usually $\ket{00...0}$) and produce output states called {\em stabilizer states}.
The case of unitary stabilizer circuits
(without measurement gates) is considered often, e.g., by consolidating
measurements at the end \cite{Jozsa}. Stabilizer circuits can be simulated in
poly-time by keeping track of a set Pauli operators that stabilize\footnote{An 
operator $U$ is said to stabilize a state iff $U\ket{\psi}=\ket{\psi}$.} 
\ the quantum state. Such {\em stabilizer operators}
uniquely represent a stabilizer state up to an unobservable global phase. 
Equation~\ref{eq:stab_count} shows that the number of $n$-qubit stabilizer states
grows as $2^{n^2/2}$, therefore, describing a generic stabilizer state
requires at least $n^2/2$ bits. Despite their compact representation, 
stabilizer states can exhibit multi-qubit entanglement and are
encountered in many quantum information applications
such as Bell states, GHZ states, error-correcting codes and 
one-way quantum computation. To better understand the role stabilizer states 
play in such applications, researchers have designed techniques to quantify 
the amount of entanglement \cite{Fattal, Hein, Wunder} 
in such states and studied related topics such as purification schemes \cite{Dur}, 
Bell inequalities \cite{Guehne} and equivalence classes \cite{Vanden04}. 
In particular, 
the authors of~\cite{DiVince, Klapp, Montanaro} 
show that the uniform distribution over the $n$-qubit stabilizer 
states is an exact $2$-design\footnote{
A state $k$-design is an ensemble of states such that, when one state is 
chosen from the ensemble and copied $k$ times, it is indistinguishable 
from a uniformly random state \cite{Harrow, DiVince}.}. Therefore, 
similar to general random quantum states, stabilizer states exhibit
a distribution that is close to uniform.
Furthermore, the results from \cite{Dahlsten, Smith}
show that the entanglement of stabilizer states is nearly maximal
and similar to that of general random states. This suggests the 
possibility of using stabilizer states as proxies for generic
quantum states, e.g., represent arbitrary states by superpositions
of stabilizer states.  
To this end, the work in~\cite{Aaron} proposes a hierarchy for quantum 
states based on their complexity~---~the number of classical
bits required to describe the state. Such a hierarchy can
be used to describe which classes of states admit polynomial-size 
classical descriptions of various kinds. To explore 
the boundaries between these classes, one can design
new stabilizer-based representations (e.g., superpositions
of stabilizer states and tensor products of such superpositions) 
that ($i$) admit polynomial-size descriptions
of practical non-stabilizer states, and ($ii$) facilitate efficient 
simulation of generic quantum circuits.
Our work advances the understanding of the geometry 
of stabilizer states and develops efficient computation of distances, 
angles and volumes between them. This line of research
can help identify efficient techniques for representing and manipulating
new classes of quantum states, rule out inefficient techniques, and 
quantify entanglement of relevant states.
Our work contributes the following:

	\begin{figure}[!b]
		\centering
		\begin{tabular}{ll}
		\scalebox{.8}[.8]{\tdplotsetmaincoords{70}{110}
\begin{tikzpicture}[scale=3, tdplot_main_coords]
	\coordinate (O) at (0,0,0);
	\draw[thick,->] (O) -- (1,0,0) node[anchor=north east]{$\ket{01}$};
	\draw[thick,->] (O) -- (0,1,0) node[anchor=north west]{$\ket{10}$};
	\draw[thick,->] (O) -- (0,0,1) node[anchor=south]{$\ket{00}$};
	
	\filldraw [draw=black,fill=black,opacity=0.3] 
         (0,0,0)--(0,1,0)--(1,1,0)--(1,0,0)--(0,0,0)--cycle ;

	\pgfmathsetmacro{\ax}{1/sqrt(2)}
	\pgfmathsetmacro{\ay}{0}
	\pgfmathsetmacro{\az}{1/sqrt(2)}
	\draw[thick,->,blue] (O) -- (\ax,\ay,\az) node[anchor=south east, color=black]
	{$\ket{s_2}=\frac{\ket{00}+\ket{01}}{\sqrt{2}}$};
	
	\draw[thick] (0,.15,0) -- (.15,.15,0);
	\draw[thick] (.15,.15,0) -- (.15,0,0);
	
	\pgfmathsetmacro{\ax}{1/sqrt(2)}
	\pgfmathsetmacro{\ay}{1/sqrt(2)}
	\pgfmathsetmacro{\az}{0}
	\draw[thick,->,blue] (0,0,0) -- (\ax,\ay,\az) node[anchor=south, color=black]
	{ };
	
	\pgfmathsetmacro{\ax}{0}
	\pgfmathsetmacro{\ay}{1/sqrt(2)}
	\pgfmathsetmacro{\az}{1/sqrt(2)}
	\draw[thick,->,blue] (0,0,0) -- (\ax,\ay,\az) node[anchor=south, color=black]
	{$\ket{s_1}=\frac{\ket{00}+\ket{10}}{\sqrt{2}}$};
	
	\tdplotdefinepoints(0,0,0)(1/sqrt(2),0,1/sqrt(2))(0,1/sqrt(2),1/sqrt(2));
	\tdplotdrawpolytopearc[thick, color=blue]{.5}{anchor=south}{$60^\circ$}
	
	\tdplotdefinepoints(0,0,0)(0,1/sqrt(2),0)(0,1/sqrt(2),1/sqrt(2));
	\tdplotdrawpolytopearc[thick, color=black]{.2}{anchor=west}{$45^\circ$}
	
	\tdplotdefinepoints(0,0,0)(1/sqrt(2),0,0)(1/sqrt(2),0,1/sqrt(2));
	\tdplotdrawpolytopearc[thick, color=black]{.2}{anchor=east}{$45^\circ$}
	

\end{tikzpicture}}
		&
		\raisebox{55pt}{
		\myfcaption{\label{fig:basis_angle} The angle between any stabilizer
		state and its nearest neighbors is $45^\circ$ and the distance
		is $\sqrt{2-\sqrt{2}}$. Here, $\ket{s_1}$ is a nearest
		neighbor of both $\ket{00}$ and $\ket{10}$. Similarly,
		$\ket{s_2}$ is a nearest neighbor of $\ket{00}$ and $\ket{01}$.
		The angle between these two nearest neighbors of $\ket{00}$ 
		is $60^\circ$. Consider the linearly-dependent triplets 
		$\{\ket{00}, \ket{10}, \ket{s_1}\}$ and $\{\ket{00}, \ket{01}, \ket{s_2}\}$.
		Each set contains two pairs of nearest neighbors and one pair
		of orthogonal states. 
		}
		}
		\end{tabular}
	\end{figure}

\begin{itemize}
	\item[({\bf 1})] We quantify the distribution 
	of angles between pairs of stabilizer states
	and characterize {\em nearest-neighbor
	stabilizer states}.
	\item[({\bf 2})] We study linearly-dependent sets 
	of stabilizer states and show that every 
	linearly-dependent triplet of such states
	that are non-parallel to each other includes 
	two pairs of nearest neighbors and one pair of orthogonal states.
	Such triplets are illustrated in Figure~\ref{fig:basis_angle}. 
	We also describe an orthogonalization procedure for stabilizer states
	that exploits this rather uniform nearest-neighbor structure.
	\item[({\bf 3})] We show that: ($i$)~for any $n$-qubit stabilizer state 
	$\ket{\psi}$, there are at least $5(2^n-1)$ states $\ket{\varphi}$ 
	such that $\ket{\psi\wedge\varphi}$ is a {\em stabilizer bivector},
	and ($ii$)~the norm of the wedge product between any two stabilizer
	states is $\sqrt{1-2^{-k}}$, where $0\leq k\leq n$.
	\item[({\bf 4})] We explore the approximation of 
	non-stabilizer states by single stabilizer states 
	and short superpositions of stabilizer states. 
	\item[({\bf 5})] We exploit our analysis of the 
	geometry of stabilizer states to design 
	algorithms for the following fundamental tasks:
	($i$)~synthesizing new {\em canonical stabilizer circuits} 
	that reduce the number of gates required for QECC encoding 
	procedures, ($ii$)~computing the inner product between 
	stabilizer states, and ($iii$)~computing stabilizer bivectors.
\end{itemize}

\noindent
{\bf Geometric properties of stabilizer states}. 
We define {\em nearest neighbor} of an $n$-qubit stabilizer state $\ket{\psi}$
as a state $\ket{\phi}$ such that $|\braket{\psi}{\phi}| = 1/\sqrt{2}$, 
the largest possible value $\neq 1$ (Corollary~\ref{cor:minmaxip}). 
In Theorem~\ref{th:num_near}, we prove that every stabilizer state has
exactly $4(2^n - 1)$ nearest neighbors. Theorem~\ref{th:near_count}
generalizes this result to the case of $k$-neighbors, where $0 < k \leq n$. 
We use these results to quantify the distribution of angles 
between any one stabilizer state and all other stabilizer states.
We show that, for sufficiently large $n$, $1/3$ of all stabilizer states 
are orthogonal to $\ket{\psi}$ (Corollary~\ref{cor:stab_ortho})
and the fraction of $k$-neighbors tends to zero for $0 < k < n-4$ (Theorem~\ref{th:stab_knear}).
Therefore, $2/3$ of all stabilizer states are oblique
to $\ket{\psi}$ and this fraction is dominated by the $k$-neighbors, where 
$n-4 \leq k \leq n$. These findings suggest a rather uniform geometric
structure for stabilizer states. This is further evidenced by two additional
facts. First, for any $n$-qubit stabilizer state 
$\ket{\psi}$, there exists a set of $2^n-1$ 
nearest neighbors to $\ket{\psi}$ that lie at $60^\circ$ angles to 
each other (Corollary~\ref{cor:basis_angle}). Second, every 
linearly-dependent triplet of stabilizer states that are non-parallel 
to each other includes two pairs of nearest neighbors and one pair of 
orthogonal states (Corollary~\ref{cor:stabtriplets}). Additionally, 
Theorem~\ref{th:num_wedge} shows that there are $5(2^n - 1)$ states 
$\ket{\varphi}$ (including all nearest-neighbor states) such that 
the wedge product $\ket{\phi}=\ket{\psi\wedge\varphi}$ can also 
be represented compactly (up to a phase) using the stabilizer 
formalism. We call such a state a {\em stabilizer bivector}. In Proposition~\ref{prop:bivec_norm},
we prove that the norm of any stabilizer bivector and thus the area of the parallelogram 
formed by any two stabilizer states is $\sqrt{1-2^{-k}}$ for $0\leq k\leq n$.

\  \\
\noindent
{\bf Embedding of stabilizer geometry in the Hilbert space}. In Section~\ref{sec:embedding},
we describe how the discrete embedding of stabilizer geometry in Hilbert space
complicates several natural geometric tasks. Our results on the geometric properties
of stabilizer states imply that there are 
significantly more stabilizer states than the dimension of 
the Hilbert space in which they are embedded (Theorem~\ref{th:ipexpect}), and that 
they are arranged in a fairly uniform pattern (Corollaries~\ref{cor:basis_angle} and \ref{cor:stabtriplets}).
These factors suggest that, if one seeks a stabilizer state closest to a given arbitrary 
quantum state, local search appears a promising candidate. To the contrary, we show that 
local search {\em does not} guarantee finding such stabilizer states (Section~\ref{sec:locsrch}).
The second natural task we consider is representing or approximating a given arbitrary quantum state
by a short linear combination of stabilizer states. Again, having considerably more
stabilizer states than the dimension of the Hilbert space 
may suggest a positive result at first.
However, we demonstrate a family of quantum states
that exhibits {\em asymptotic orthogonality} to all stabilizer states and can thus be neither
represented nor approximated by short superpositions of stabilizer states (Theorem~\ref{th:stab_evade}).
Furthermore, we show in Proposition~\ref{prop:stabgap} that the maximal radius 
of any $2^n$-dimensional ball centered at a point on the unit sphere that does 
not contain any $n$-qubit stabilizer states cannot exceed $\sqrt{2}$,
but approaches $\sqrt{2}$ as $n\rightarrow\infty$.

\noindent
{\bf Computational geometry of stabilizer states}. Angles between stabilizer states
were discussed in \cite{AaronGottes}, where the authors describe possible values for 
such angles and outline an inner-product computation that involves the synthesis 
of a {\em basis-normalization stabilizer circuit} that maps a stabilizer state 
to a computational basis state. We observe that this circuit-synthesis procedure is the 
computational bottleneck of the algorithm and thus any 
improvements to this synthesis process translate into increased 
performance of the overall algorithm. It was also shown 
in \cite{AaronGottes} that, for any unitary
stabilizer circuit, there exists an equivalent block-structured 
{\em canonical circuit} that applies a block of Hadamard ($H$) gates 
only, followed by a block of CNOT ($C$) only, then a block of 
Phase ($P$) gates only, and so on in the $7$-block sequence 
$H$-$C$-$P$-$C$-$P$-$C$-$H$. The number of gates in any $H$ and $P$ 
block is at most $2n$ and $4n$, respectively, so the total number 
of gates for such canonical circuits is dominated by the $C$ 
blocks, which contain $O(n^2/\log n)$ gates \cite{PatelMarkov}. Therefore, reducing the 
number of $C$ blocks (or those involving other controlled operations)
leads to more compact circuits that minimize
the build-up of decoherence when implementing QECC encoding procedures.
Using an alternative representation
for stabilizer states, the work in \cite{Vanden} proves the existence of 
canonical circuits with the shorter sequence $H$-$C$-$X$-$P$-$CZ$, where the $X$
and $CZ$ blocks consist of NOT and Controlled-$Z$ (CPHASE) gates, respectively. 
Such circuits contain only two controlled-op blocks (as compared to three $C$ 
blocks in the $7$-block sequence from \cite{AaronGottes}) but no algorithms are available 
in literature to synthesize these circuits for an arbitrary stabilizer state. 
We describe an algorithm for synthesizing canonical circuits with 
the block sequence $H$-$C$-$CZ$-$P$-$H$. Our circuits are
therefore close to the smallest known circuits proposed in \cite{Vanden}.
Our canonical circuits improve the computation of inner products, helping 
us outperform the inner-product algorithm 
based on non-canonical circuits proposed in \cite{Audenaert}. Furthermore, 
we leverage our circuit-synthesis technique and inner-product algorithm 
to compute stabilizer bivectors efficiently. 

For comparison, we implemented the circuit-synthesis portion 
of the algorithm from \cite{Audenaert}. Our algorithm produces circuits with 
less than half as many gates on average and runs roughly $2\times$ faster.
Furthermore, our synthesis approach produces canonical circuits given any 
input stabilizer state by first obtaining a {\em canonical generator set} 
for the state. We describe a separate algorithm for reducing an arbitrary
stabilizer generator set to its canonical form. 

This paper is structured as follows. Section \ref{sec:background} reviews
the stabilizer formalism and relevant algorithms for 
manipulating stabilizer-based representations of quantum states.
Our findings related to geometric structure of stabilizer states
are described in Sections~\ref{sec:stabneighbors} and \ref{sec:embedding}.
Section \ref{sec:inprod_stab} describes our algorithms for circuit synthesis,
inner product computation and construction of stabilizer bivectors.
In Section \ref{sec:empirical},
we evaluate the performance of our inner-product algorithm, and
Section \ref{sec:conclude} closes with concluding remarks.

	\begin{figure*}[!t]\footnotesize\centering
	\begin{tabular}{cc}
	\hspace{-4cm}
    $
        H = \frac{1}{\sqrt{2}}\begin{pmatrix}
            1 & 1 \\
            1 & -1 \end{pmatrix} \quad
        P = \begin{pmatrix}
            1 & 0 \\
            0 & i \end{pmatrix}
    $ 
    & \hspace{-2cm}
    $ X = \begin{pmatrix}
                0 & 1 \\
                1 & 0 \end{pmatrix} \quad
            Y = \begin{pmatrix}
                0 & -i \\
                i &  0 \end{pmatrix}
    $ \\ \\
    \hspace{-4cm}
    $     CNOT = \begin{pmatrix}
            1 & 0 & 0 & 0 \\
            0 & 1 & 0 & 0 \\
            0 & 0 & 0 & 1 \\
            0 & 0 & 1 & 0 \end{pmatrix}
    $ 
    &
    \hspace{-2cm}
    $   Z = \begin{pmatrix}
                1 & 0 \\
                0 & -1 \end{pmatrix}
    $ \\
    & \\
    
    \hspace{-4cm} \fcaption{\centering\label{fig:chp_pauli} 
    (a) Stabilizer gates Hadamard (H), Phase (P)}
    & \hspace{-2cm} Fig. 1. (b) The Pauli matrices. \\
    \hspace{-4cm} and Controlled-NOT (CNOT). &
    \end{tabular}
    \vspace{-.3cm}
	\end{figure*}
	
\section{Background and Previous Results}  
\label{sec:background}

Gottesman~\cite{Gottes98} developed a description for 
quantum states involving the {\em Heisenberg representation} often 
used by physicists to describe atomic phenomena. In this model,
quantum states are described by keeping track of their symmetries 
rather than explicitly maintaining the amplitudes of exponentially-large vectors. 
The symmetries are operators for which these states are $1$-eigenvectors.
Algebraically, symmetries form {\em group} structures,
which can be specified compactly by group generators.
This approach, also known 
as the {\em stabilizer formalism}, facilitates particularly 
efficient manipulation of an important class of quantum states.

\subsection{The stabilizer formalism} 
\label{sec:stab}

A unitary operator $U$ {\em stabilizes} a state $\ket{\psi}$ if
$\ket{\psi}$ is a $1$--eigenvector of $U$, i.e., $U\ket{\psi} 
= \ket{\psi}$ \cite{Gottes, NielChu}. We are interested in operators $U$
derived from the Pauli matrices shown in Figure~\ref{fig:chp_pauli}b.
The following table lists the one-qubit states stabilized
by the Pauli matrices.

\begin{center}
	\begin{tabular}{lccclc}
		$X$ : & $(\ket{0}+\ \ket{1})/\sqrt{2}$ &&& $-X$ : & $(\ket{0}-\ \ket{1})/\sqrt{2}$ \\
		$Y$ : & $(\ket{0}+i\ket{1})/\sqrt{2}$ &&&$-Y$ : & $(\ket{0}-i\ket{1})/\sqrt{2}$ \\
		$Z$ : & $\ket{0}$ &&& $-Z$ : & $\ket{1}$ \\
	\end{tabular}
\end{center}

Observe that $I$ stabilizes all states and $-I$ does not stabilize any state. 
As an example, the entangled state $(\ket{00} + \ket{11})/\sqrt{2}$ is stabilized by 
the Pauli operators $X\otimes X$, $-Y\otimes Y$, $Z\otimes Z$ and $I\otimes I$.  
As shown in Table~\ref{tab:pauli_mult}, it turns out that the Pauli matrices 
along with $I$ and the multiplicative factors $\pm1$, $\pm i$, form a 
{\em closed group} under matrix multiplication \cite{NielChu}. Formally, the {\em Pauli group} 
$\mathcal{G}_n$ on $n$ qubits consists of the $n$-fold tensor product 
of Pauli matrices, $P = i^kP_1\otimes\cdot\cdot\cdot\otimes P_n$ 
such that $P_j\in\{I, X, Y, Z\}$ and $k\in\{0,1,2,3\}$. For brevity, the tensor-product symbol 
is often omitted so that $P$ is denoted by a string of $I$, $X$, $Y$ 
and $Z$ characters or {\em Pauli literals} and a separate integer value
$k$ for the phase $i^k$. This string-integer pair representation allows us to compute 
the product of Pauli operators without explicitly 
computing the tensor products,\footnote{\scriptsize This holds true due to the 
identity: $(A\otimes B)(C \otimes D)=(AC\otimes BD)$.} ~e.g., 
$(-IIXI)(iIYII) = -iIYXI$. Since $\mid \mathcal{G}_n\mid= 4^{n+1}$, 
$\mathcal{G}_n$ can have at most $\log_2 \mid \mathcal{G}_n \mid = 
\log_2 4^{n+1} = 2(n + 1)$ irredundant generators \cite{NielChu}.
The key idea behind the stabilizer formalism is to represent an 
$n$-qubit quantum state $\ket{\psi}$ by its {\em stabilizer group} 
$S(\ket{\psi})$~---~the subgroup of $\mathcal{G}_n$ that stabilizes $\ket{\psi}$.
As the following theorem shows, if $|S(\ket{\psi})|=2^n$, the group uniquely specifies 
$\ket{\psi}$. 

	\begin{table}[b]
        \centering
    		\tcaption{\label{tab:pauli_mult} \centering Multiplication table for Pauli matrices. \\
    		Shaded cells indicate anticommuting products.}
        \begin{tabular}{|c||c|c|c|c|}
            \hline
                &   $I$ & $X$                         & $Y$   & $Z$ \\ \hline\hline
            $I$ &   $I$ & $X$                         & $Y$   &  $Z$ \\ \hline
            $X$ &   $X$ & $I$   & \cellcolor[gray]{0.85} $iZ$  & \cellcolor[gray]{0.85} $-iY$ \\ \hline
            $Y$ &   $Y$ & \cellcolor[gray]{0.85} $-iZ$ & $I$   & \cellcolor[gray]{0.85} $iX$ \\ \hline
            $Z$ &   $Z$ & \cellcolor[gray]{0.85} $iY$  & \cellcolor[gray]{0.85} $-iX$ & $I$ \\
            \hline
        \end{tabular}
    \end{table}

    \begin{theorem} \label{th:gen_commute}
        For an $n$-qubit pure state $\ket{\psi}$, there exists a $k\leq n$
        such that $S(\ket{\psi}) \cong {\mathbb Z}_2^k$. 
        If $k=n$, $\ket{\psi}$ is specified uniquely by $S(\ket{\psi})$ and is called a stabilizer state.
	\end{theorem}
	\noindent
	{\bf Proof.} 
     		($i$) To prove that $S(\ket{\psi})$ is commutative,
            let $P, Q \in S(\ket{\psi})$ such that $PQ\ket{\psi} = \ket{\psi}$. If $P$ and $Q$ anticommute,
            $-QP\ket{\psi} = -Q(P\ket{\psi}) = -Q\ket{\psi} = -\ket{\psi} \neq \ket{\psi}$.
            Thus, $P$ and $Q$ cannot both be elements of $S(\ket{\psi})$.
			
			\noindent
            ($ii$) To prove that every element of $S(\ket{\psi})$ is of
            degree $2$, let $P \in S(\ket{\psi})$ such that $P\ket{\psi} = \ket{\psi}$.
            Observe that $P^2 = i^lI$ for $l\in\{0,1,2,3\}$.
            Since $P^2\ket{\psi} = P(P\ket{\psi}) = P\ket{\psi} = \ket{\psi}$, we obtain
            $i^l = 1$ and $P^2 = I$.
			
			\noindent
            ($iii$) From group theory, a finite Abelian group with
            $a^2 = a$ for every element must be $\cong {\mathbb Z}_2^k$.
			
			\noindent
            ($iv$) We now prove that $k \leq n$.
            First note that each independent generator $P \in S(\ket{\psi})$
            imposes the linear constraint $P\ket{\psi}=\ket{\psi}$
            on the $2^n$-dimensional vector space.
            The subspace of vectors that satisfy such a constraint
            has dimension $2^{n-1}$, or half the space. Let $gen(\ket{\psi})$
            be the set of generators for $S(\ket{\psi})$.
            We add independent generators to $gen(\ket{\psi})$ one by one and impose
  			their linear constraints, to limit $\ket{\psi}$ to the shared
			$1$-eigenvector. Thus the size of $gen(\ket{\psi})$ is at most $n$.
			In the case $|gen(\ket{\psi})| = n$, the $n$ independent 
			generators reduce the subspace of possible states to dimension
        		one. Thus, $\ket{\psi}$ is uniquely specified.
	\square

The proof of Theorem \ref{th:gen_commute} shows that $S(\ket{\psi})$ 
is specified by only $\log_2 2^{n} = n$ {\em irredundant stabilizer generators}. 
Therefore, an arbitrary $n$-qubit stabilizer state can be represented by
a {\em stabilizer matrix} $\cM$ whose rows represent a set of 
generators $g_1,\ldots,g_n$ for $S(\ket{\psi})$. (Hence we use the terms 
{\em generator set} and {\em stabilizer matrix} interchangeably.) Since each $g_i$ is a string 
of $n$ Pauli literals, the size of the matrix is $n\times n$. The 
phases of each $g_i$ are stored separately using a vector of $n$ integers.
Therefore, the storage cost for $\cM$ is 
$\Theta(n^2)$, which is an {\em exponential 
improvement} over the $O(2^n)$ cost often encountered in 
vector-based representations.

Theorem \ref{th:gen_commute} suggests that Pauli literals can be 
represented using only two bits, e.g., $00 = I$, $01 = Z$, $10 = X$ and $11 = Y$. 
Therefore, a stabilizer matrix can be encoded using an $n\times2n$ binary matrix or {\em tableau}.
The advantage of this approach is that this literal-to-bits mapping induces an isomorphism
${\mathbb Z}_2^{2n} \rightarrow \mathcal{G}_n$ because vector addition
in ${\mathbb Z}_2^{2}$ is equivalent to multiplication of Pauli operators
up to a global phase. The tableau implementation of the stabilizer formalism
is covered in \cite{AaronGottes, NielChu}. 

    \begin{proposition} \label{prop:stab_count}
        The number of $n$-qubit pure stabilizer states is given by
        \begin{equation} \label{eq:stab_count}
    	   \mathcal{N}(n) = 2^n\prod_{k=0}^{n-1}(2^{n-k} + 1) = 2^{(.5 + o(1))n^2}
    	   \vspace{-2pt}
	    \end{equation}
    \end{proposition}
    
The proof of Proposition~\ref{prop:stab_count} can be found in~\cite{AaronGottes}. 
An alternate interpretation of Equation \ref{eq:stab_count} is
given by the simple recurrence relation $\mathcal{N}(n)=2(2^n+1) \mathcal{N}(n-1)$ with base case
$\mathcal{N}(1) = 6$. For example, for $n=2$ the number
of stabilizer states is $60$, and for $n=3$ it is $1080$. This recurrence
relation stems from the fact that there are $2^n+1$ ways of combining the
generators of the $\mathcal{N}(n-1)$ states with additional 
Pauli matrices to form valid
$n$-qubit generators. The factor of $2$ accounts for the increase in the number
of possible sign configurations. Table~\ref{tab:two_qbssts} and 
Appendix~A list all two-qubit and three-qubit 
stabilizer states, respectively.

	\begin{table}[!t]
		\centering
        \tcaption{\label{tab:two_qbssts} 
     Sixty two-qubit stabilizer states and their corresponding Pauli generators. 
     Shorthand notation represents a stabilizer state as $\alpha_0, \alpha_1, \alpha_2, \alpha_3$ 
     where $\alpha_i$ are the normalized amplitudes of the basis states. The basis 
     states are emphasized in bold. The first column lists states whose generators 
     do not include an upfront minus sign, and other columns introduce the signs. 
     A sign change creates an orthogonal vector. Therefore, each row of the table 
     gives an orthogonal basis. The cells in dark grey indicate stabilizer states 
     with four non-zero basis amplitudes, i.e., $\alpha_i \neq 0\ \forall\ i$. 
     The $\angle$ column indicates the angle between that state and $\ket{00}$, 
     which has twelve nearest-neighbor states (light gray) and $15$ orthogonal 
     states ($\perp$).\vspace{5mm}
     } 
     \scalebox{.73}[.73]{\begin{tabular}{|r|c|c|c||c|c|c||c|c|c||c|c|c|}
            \hline
            & \sc State & \hspace{-2.5mm} \sc Gen'tors \hspace{-2.5mm} & $\angle$
            & \sc State & \hspace{-2.5mm} \sc Gen'tors \hspace{-2.5mm} & $\angle$
            & \sc State & \hspace{-2.5mm} \sc Gen'tors \hspace{-2.5mm} & $\angle$
            & \sc State & \hspace{-2.5mm} \sc Gen'tors\hspace{-2.5mm}  & $\angle$ \\ \hline\hline
            \multirow{9}{2mm}{\rotatebox{90}{\sc Separable}}

            & \cellcolor[gray]{0.7} $1,1,1,1$  & IX, XI & \hspace{-3mm} $\pi/3$ \hspace{-3mm} & \cellcolor[gray]{0.7}  \hspace{-3mm} $1,-1,1,-1$  \hspace{-3mm} & -IX, XI  & \hspace{-3mm} $\pi/3$ \hspace{-3mm} & \cellcolor[gray]{0.7}  \hspace{-3mm} $1,1,-1,-1$  \hspace{-3mm} &  IX, -XI  & \hspace{-3mm} $\pi/3$ \hspace{-3mm} &  \cellcolor[gray]{0.7} $1,-1,-1,1$ &  -IX, -XI & \hspace{-3mm} $\pi/3$ \hspace{-3mm} \\ 
            & \cellcolor[gray]{0.7} $1,1,i,i$  &  IX, YI & \hspace{-3mm} $\pi/3$ \hspace{-3mm} & \cellcolor[gray]{0.7} $1,-1,i,-i$ & -IX, YI & \hspace{-3mm} $\pi/3$ \hspace{-3mm} & \cellcolor[gray]{0.7} $1,1,-i,-i$  &  IX, -YI & \hspace{-3mm} $\pi/3$ \hspace{-3mm} & \cellcolor[gray]{0.7} $1,-1,-i,i$  & -IX, -YI & \hspace{-3mm} $\pi/3$ \hspace{-3mm} \\ 
            & $1,1,0,0$ & IX, ZI & \cellcolor[gray]{0.85}\hspace{-3mm} $\pi/4$ \hspace{-3mm} & $1,-1,0,0$  & -IX, ZI & \cellcolor[gray]{0.85}\hspace{-3mm} $\pi/4$ \hspace{-3mm} & $0,0,1,1$  &  IX, -ZI & \hspace{-3mm} $\perp$ \hspace{-3mm} & $0,0,1,-1$ & -IX, -ZI & \hspace{-3mm} $\perp$ \hspace{-3mm} \\ 

            & \cellcolor[gray]{0.7} $1,i,1,i$  &  IY, XI & \hspace{-3mm} $\pi/3$ \hspace{-3mm} & \cellcolor[gray]{0.7} $1,-i,1,-i$ & -IY, XI & \hspace{-3mm} $\pi/3$ \hspace{-3mm} & \cellcolor[gray]{0.7} $1,i,-1,-i$ &  IY, -XI & \hspace{-3mm} $\pi/3$ \hspace{-3mm} & \cellcolor[gray]{0.7} $1,-i,-1,i$ & -IY, -XI & \hspace{-3mm} $\pi/3$ \hspace{-3mm} \\ 
            & \cellcolor[gray]{0.7} $1,i,i,-1$ &  IY, YI & \hspace{-3mm} $\pi/3$ \hspace{-3mm} & \cellcolor[gray]{0.7} $1,-i,i,1$  & -IY, YI & \hspace{-3mm} $\pi/3$ \hspace{-3mm} & \cellcolor[gray]{0.7} $1,i,-i,1$  &  IY, -YI & \hspace{-3mm} $\pi/3$ \hspace{-3mm} & \cellcolor[gray]{0.7} \hspace{-3mm} $1,-i,-i,-1$  \hspace{-3mm} & -IY, -YI & \hspace{-3mm} $\pi/3$ \hspace{-3mm}  \\ 
            & $1,i,0,0$ & IY, ZI & \cellcolor[gray]{0.85}\hspace{-3mm} $\pi/4$ \hspace{-3mm} & $1,-i,0,0$  & -IY, ZI & \cellcolor[gray]{0.85}\hspace{-3mm} $\pi/4$ \hspace{-3mm} & $0,0,1,i$ &  IY, -ZI & \hspace{-3mm} $\perp$ \hspace{-3mm} & $0,0,1,-i$ & -IY, -ZI & \hspace{-3mm} $\perp$ \hspace{-3mm} \\ 

            & $1,0,1,0$ &  IZ, XI & \cellcolor[gray]{0.85}\hspace{-3mm} $\pi/4$ \hspace{-3mm} & $0,1,0,1$  &  -IZ, XI & \hspace{-3mm} $\perp$ \hspace{-3mm} & $1,0,-1,0$  &  IZ, -XI & \cellcolor[gray]{0.85}\hspace{-3mm} $\pi/4$ \hspace{-3mm} & $0,1,0,-1$  & -IZ, -XI & \hspace{-3mm} $\perp$ \hspace{-3mm} \\ 
            & $1,0,i,0$ &  IZ, YI & \cellcolor[gray]{0.85}\hspace{-3mm} $\pi/4$ \hspace{-3mm} & $0,1,0,i$  &  -IZ, YI & \hspace{-3mm} $\perp$ \hspace{-3mm} & $1,0,-i,0$  &  IZ, -YI & \cellcolor[gray]{0.85}\hspace{-3mm} $\pi/4$ \hspace{-3mm} & $0,1,0,-i$  &  -IZ, -YI & \hspace{-3mm} $\perp$ \hspace{-3mm} \\ 
            & $\mathbf{1,0,0,0}$ & \bf IZ, ZI & $0$ & $\mathbf{0,1,0,0}$ & \bf -IZ, ZI  & \hspace{-3mm} $\perp$ \hspace{-3mm} & $\mathbf{0,0,1,0}$ & \bf IZ, -ZI  & \hspace{-3mm} $\perp$ \hspace{-3mm} &  $\mathbf{0,0,0,1}$  & \bf -IZ, -ZI & \hspace{-3mm} $\perp$ \hspace{-3mm} \\ \hline\hline

            \multirow{6}{2mm}{\rotatebox{90}{\sc Entangled}}

            & $0,1,1,0$ &  XX, YY  & \hspace{-3mm} $\perp$ \hspace{-3mm} &  $1,0,0,-1$ & -XX, YY & \cellcolor[gray]{0.85}\hspace{-3mm} $\pi/4$ \hspace{-3mm} & $1,0,0,1$  &  XX, -YY & \cellcolor[gray]{0.85}\hspace{-3mm} $\pi/4$ \hspace{-3mm} & $0,1,-1,0$ &  -XX, -YY & \hspace{-3mm} $\perp$ \hspace{-3mm} \\ 
            & $1,0,0,i$ &  XY, YX  & \cellcolor[gray]{0.85}\hspace{-3mm} $\pi/4$ \hspace{-3mm} &  $0,1,i,0$  & -XY, YX & \hspace{-3mm} $\perp$ \hspace{-3mm} & $0,1,-i,0$ &  XY, -YX & \hspace{-3mm} $\perp$ \hspace{-3mm} & $1,0,0,-i$ &  -XY, -YX & \cellcolor[gray]{0.85}\hspace{-3mm} $\pi/4$ \hspace{-3mm} \\ 
    		
    		& \cellcolor[gray]{0.7} \hspace{-3mm} $1,1,1,-1$ \hspace{-3mm} &  XZ, ZX  & \hspace{-3mm} $\pi/3$ \hspace{-3mm} &  \cellcolor[gray]{0.7} $1,1,-1,1$ & -XZ, ZX   & \hspace{-3mm} $\pi/3$ \hspace{-3mm} &  \cellcolor[gray]{0.7} $1,-1,1,1$  &  XZ, -ZX  & \hspace{-3mm} $\pi/3$ \hspace{-3mm} &  \cellcolor[gray]{0.7} \hspace{-3mm} $1,-1,-1,-1$ \hspace{-3mm} &  -XZ, -ZX & \hspace{-3mm} $\pi/3$ \hspace{-3mm} \\ 
            & \cellcolor[gray]{0.7} $1,i,1,-i$ &  XZ, ZY  & \hspace{-3mm} $\pi/3$ \hspace{-3mm} &  \cellcolor[gray]{0.7} $1,i,-1,i$ & -XZ, ZY & \hspace{-3mm} $\pi/3$ \hspace{-3mm} & \cellcolor[gray]{0.7} $1,-i,1,i$  &  XZ, -ZY  & \hspace{-3mm} $\pi/3$ \hspace{-3mm} &  \cellcolor[gray]{0.7} $1,-i,-1,-i$  &  -XZ, -ZY & \hspace{-3mm} $\pi/3$ \hspace{-3mm} \\ 

            & \cellcolor[gray]{0.7} $1,1,i,-i$ &  YZ, ZX  & \hspace{-3mm} $\pi/3$ \hspace{-3mm} &  \cellcolor[gray]{0.7} $1,1,-i,i$  &  -YZ, ZX  & \hspace{-3mm} $\pi/3$ \hspace{-3mm} &  \cellcolor[gray]{0.7} $1,-1,i,i$ & YZ, -ZX  & \hspace{-3mm} $\pi/3$ \hspace{-3mm} & \cellcolor[gray]{0.7} $1,-1,-i,-i$  &  -YZ, -ZX & \hspace{-3mm} $\pi/3$ \hspace{-3mm} \\ 
            & \cellcolor[gray]{0.7} $1,i,i,1$  &  YZ, ZY  & \hspace{-3mm} $\pi/3$ \hspace{-3mm} &  \cellcolor[gray]{0.7} $1,i,-i,-1$ &  -YZ, ZY  & \hspace{-3mm} $\pi/3$ \hspace{-3mm} &  \cellcolor[gray]{0.7} $1,-i,i,-1$  & YZ, -ZY  & \hspace{-3mm} $\pi/3$ \hspace{-3mm} &  \cellcolor[gray]{0.7} $1,-i,-i,1$   &  -YZ, -ZY & \hspace{-3mm} $\pi/3$ \hspace{-3mm} \\
            \hline
        \end{tabular}
        }
    \end{table} 

    \begin{proposition} \label{prop:stab_tensor}
    		The tensor product $\ket{\psi} \otimes \ket{\varphi}$ of two stabilizer states
        is a stabilizer state.
    \end{proposition}
    \noindent
	{\bf Proof.} Consider two Pauli operators, $P$ and $Q$, which stabilize
    		the $n$-qubit state $\ket{\psi}$ and the $m$-qubit state
    		$\ket{\varphi}$, respectively. The $(n+m)$-qubit state
    		$\ket{\psi}\otimes\ket{\varphi}$ is stabilized by $P\otimes Q = PQ$ since
    		$PQ\ket{\psi}\ket{\varphi} = P\ket{\psi}Q\ket{\varphi} = \ket{\psi}\ket{\varphi}$.
   		Let $\{P_1,\ldots, P_n\}$ and $\{Q_1,...,Q_m\}$ be sets of generators
   		for $S(\ket{\psi})$ and $S(\ket{\varphi})$, respectively.
        We create an $(n+m)$-element set of Pauli operators
        as follows, $P_j\otimes I^{\otimes m}$, $j \in \{1,...,n\}$ and
        $I^{\otimes n}\otimes Q_k$, $k\in\{1,...,m\}$.
        This is a generator set for $S(\ket{\psi}\ket{\varphi})$ because
        each of the operators in the set stabilizes $\ket{\psi}\ket{\varphi}$ and
        the set generates all tensor products $P_jQ_k$.
    \square
\newpage
    \begin{observation} \label{obs:stabst_amps}
    		Consider a stabilizer state $\ket{\psi}$ represented by a set of generators of its
    		stabilizer group $S(\ket{\psi})$. Recall from the proof of Theorem \ref{th:gen_commute}
    		that, since $S(\ket{\psi}) \cong {\mathbb Z}_2^n$,
   	 	each generator imposes a linear constraint on $\ket{\psi}$. Therefore, the set
        of generators can be viewed as a system of linear equations whose solution
        yields the $2^k$ (for some $k$ between $0$ and $n$) non-zero computational 
        basis amplitudes that make up $\ket{\psi}$. Thus,
        one needs to perform Gaussian elimination to obtain such basis
        amplitudes from a generator set. 
    \end{observation} 

\ \\\noindent
{\bf Canonical stabilizer matrices}.
Although stabilizer states are uniquely determined by their stabilizer group, the set of 
generators may be selected in different ways. For example, the state 
$\ket{\psi} = (\ket{00} + \ket{11})/\sqrt{2}$ is uniquely specified by any of 
the following:

	\begin{center}
		\begin{tabular}{c|l|cc|l|cc|l|}
		\multirow{2}{8mm}{$\cM_1 =$} & $XX$ & & \multirow{2}{8mm}{$\cM_2 =$} & $XX$ & &
		\multirow{2}{8mm}{$\cM_3 =$} & -$YY$ \\
		& $ZZ$ & & & -$YY$ & & & $ZZ$ \\
		\end{tabular}
	\end{center} 
	
\begin{figure}[!b]
	\centering
	\begin{tabular}{ll}
	\includegraphics[scale=.30]{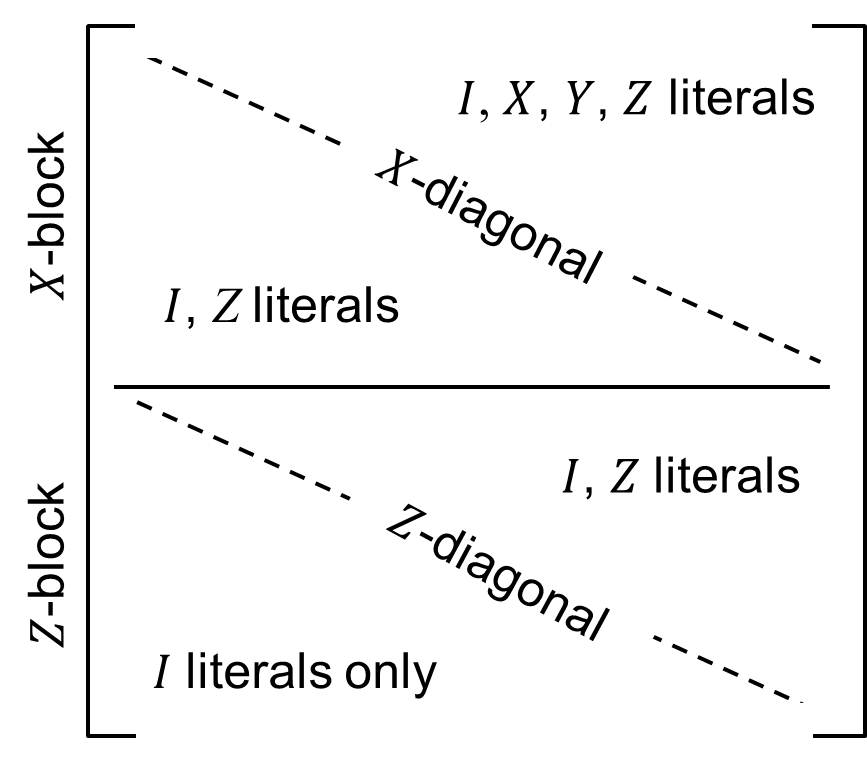} 
	&
	\raisebox{55pt}{
	\myfcaption{\label{fig:sminv} 
	Canonical (row-reduced echelon) form for stabilizer matrices. 
	The $X$-block contains a {\em minimal} set of rows with $X/Y$ literals. 
	The rows with $Z$ literals only appear in the $Z$-block. Each block 
	is arranged so that the leading non-$I$ literal of each row is strictly 
	to the right of the leading non-$I$ literal in the row above. The number
	of Pauli (non-$I$) literals in each block is minimal.}
	}
	\end{tabular}
	\end{figure}

\noindent
One obtains $\cM_2$ from $\cM_1$ by left-multiplying the second row
by the first. Similarly, one can also obtain $\cM_3$ from $\cM_1$ 
or $\cM_2$ via row multiplication. Observe that, multiplying any row by itself yields
$II$, which stabilizes $\ket{\psi}$. However, $II$ cannot
be used as a stabilizer generator because it is redundant
and carries no information about the structure of $\ket{\psi}$. 
This also holds true in general for $\cM$ of any size.
Any stabilizer matrix can be rearranged
by applying sequences of elementary row operations in order to obtain a particular
matrix structure. Such operations do not modify the stabilizer state.
The elementary row operations that can be performed on a stabilizer matrix are
transposition, which swaps two rows of the matrix, and multiplication,
which left-multiplies one row with another. Such operations allow one
to rearrange the stabilizer matrix in a series of steps that resemble
Gauss-Jordan elimination.\footnote{\scriptsize Since Gaussian elimination
essentially inverts the $n\times 2n$ matrix, this
could be sped up to $O(n^{2.376})$ time by using
fast matrix inversion algorithms. However, $O(n^3)$-time 
Gaussian elimination seems more practical.} ~Given an $n\times n$ stabilizer matrix, 
row transpositions are performed in constant time\footnote{\scriptsize Storing pointers to rows
facilitates $O(1)$-time row transpositions~---~one simply swaps 
relevant pointers.} ~while row multiplications require $\Theta(n)$ time.
Algorithm~\ref{alg:gauss_min} rearranges a stabilizer matrix into
a {\em row-reduced echelon form} that contains: ({\em i}) a {\em minimum 
set} of generators with $X$ and $Y$ literals appearing at the top, 
and ({\em ii}) generators containing a {\em minimum set} of $Z$ literals 
only appearing at the bottom of the matrix.
This particular stabilizer-matrix structure, shown in Figure~\ref{fig:sminv},  
defines a canonical representation for stabilizer states~\cite{Djor, Gottes98}.
Several row-echelon (standard) forms for stabilizer generators along 
with relevant algorithms to obtain them have been introduced in 
the literature~\cite{Audenaert, Gottes98, NielChu}. However, 
such standard forms are not always canonical, e.g, 
the row-echelon form described in~\cite{Audenaert} does not guarantee
a minimum set of $Z$ literals. 
Since most of our algorithms manipulate canonical stabilizer matrices, 
we will describe in detail our Gaussian-elimination procedure for 
obtaining the canonical structure depicted in Figure~\ref{fig:sminv}.
The algorithm iteratively determines which 
row operations to apply based on the Pauli (non-$I$) literals
contained in the first row and column of an increasingly smaller submatrix
of the full stabilizer matrix. Initially, the submatrix considered is the full stabilizer matrix.
After the proper row operations are applied, the dimensions of the submatrix
decrease by one until the size of the submatrix reaches one. 
The algorithm performs this process twice, once to position the
rows with $X$($Y$) literals at the top, and then again to position
the remaining rows containing $Z$ literals only at the bottom. 
Let $i\in\{1,\ldots,n\}$ and $j\in\{1,\ldots,n\}$ be the index
of the first row and first column, respectively, of submatrix
$\cA$. The steps to construct the upper-triangular portion of the row-echelon
form shown in Figure~\ref{fig:sminv} are as follows.  

	\begin{itemize}
		\item[{\bf 1.}] Let $k$ be a row in $\cA$ whose $j^{th}$ literal
		is $X$($Y$). Swap rows $k$ and $i$ such that $k$ is the first row of $\cA$.
		Decrease the height of $\cA$ by one (i.e., increase $i$).
		\item[{\bf 2.}] For each row $m \in \{0,\ldots,n\}, m \neq i$ that has
		an $X$($Y$) in column $j$, use row multiplication to set the $j^{th}$
		literal in row $m$ to $I$ or $Z$.
		\item[{\bf 3.}] Decrease the width of $\cA$ by one (i.e., increase $j$).
	\end{itemize}
	
\newcommand{\rswap}{\mathrm{\tt ROWSWAP}}
\newcommand{\rmult}{\mathrm{\tt ROWMULT}}

    \begin{algorithm}[!t]
        \caption{Canonical form reduction for stabilizer matrices}
        \label{alg:gauss_min}
        \footnotesize
        \begin{algorithmic}[1]
            \Require Stabilizer matrix $\cM$ for $S(\ket{\psi})$ with rows $R_1,\ldots,R_n$
            \Ensure $\cM$ is reduced to row-echelon form
            \Statex \hspace{-5mm} $\Rightarrow$ $\rswap(\cM, i, j)$ swaps rows $R_i$ and $R_j$ of $\cM$
            \Statex \hspace{-5mm} $\Rightarrow$ $\rmult(\cM, i, j)$ left-multiplies rows $R_i$ 
            and $R_j$, returns updated $R_i$
            \Statex
            \State $i \leftarrow 1$
            \For{$j \in \{1, \dots, n\}$} \Comment{Setup $X$ block}
            	\State $k \leftarrow$ index of row $R_{k \in \{i,\ldots, n\}}$ 
            	with $j^{th}$ literal set to $X$($Y$)
            	\If{$k$ {\bf exists}}
            		\State $\rswap(\cM, i, k)$
            		\For{$m \in \{0,\ldots, n\}$} 
            			\If{$j^{th}$ literal of $R_m$ is $X$ or $Y$ and $m\neq i$}
            				\State $R_m = \rmult(\cM, R_i, R_m)$ \Comment{Gauss-Jordan elimination step}
            			\EndIf
            		\EndFor
            		\State $i \leftarrow i + 1$
            	\EndIf
            \EndFor
           	\For{$j \in \{1, \dots, n\}$} \Comment{Setup $Z$ block}
            	\State $k \leftarrow$ index of row $R_{k \in \{i,\ldots, n\}}$ 
            	with $j^{th}$ literal set to $Z$
            	\If{$k$ {\bf exists}}
            		\State $\rswap(\cM, i, k)$
            		\For{$m \in \{0,\ldots, n\}$}
            			\If{$j^{th}$ literal of $R_m$ is $Z$ or $Y$ and $m\neq i$}
            				\State $R_m = \rmult(\cM, R_i, R_m)$ \Comment{Gauss-Jordan elimination step}
            			\EndIf
            		\EndFor
            		\State $i \leftarrow i + 1$
           	\EndIf
            \EndFor
        \end{algorithmic}
    \end{algorithm}

To bring the matrix to its lower-triangular form, one executes the same procedure
with the following difference: ($i$) step~1 looks for rows that have
a $Z$ literal (instead of $X$ or $Y$) in column $j$, and ($ii$) step~2
looks for rows that have $Z$ or $Y$ literals (instead of $X$ or $Y$) in column $j$.
Observe that Algorithm~\ref{alg:gauss_min} ensures that the columns
in $\cM$ have at most two distinct types of non-$I$ literals.
Since Algorithm~\ref{alg:gauss_min} 
inspects all $n^2$ entries in the matrix and performs a constant number of row
multiplications each time, its runtime is $O(n^3)$. 

 	\begin{definition} \label{def:mtxeq} {\em 
    		Two stabilizer matrices are {\em similar} if they contain the 
    		same Pauli generators with different $\pm 1$ phases, i.e., the
    		matrices are equivalent up to a phase-vector permutation. 
    		Otherwise, the matrices are called {\em dissimilar}. }
    \end{definition}
    
    \begin{observation}
    		The number of dissimilar $n$-qubit canonical stabilizer matrices is
    		$\mathcal{N}(n)/2^n$.
    \end{observation}
    
\ \\\noindent
{\bf Stabilizer-circuit simulation}.
Observe that the computational-basis states can be characterized as stabilizer states 
with the following stabilizer-matrix structure.

	\begin{definition} \label{def:basis_form} {\em 
		A stabilizer matrix is in its {\em \normform} if it has the following structure.
		\begin{equation*}
			\begin{array}{r}
				\pm \\
				\pm \\
				\vdots \\
				\pm
			\end{array}
			\left[
			\begin{array}{cccc}
				Z & I & \cdots & I  \\
				I & Z & \cdots & I   \\
				\vdots & \vdots & \ddots & \vdots \\
				I & I & \cdots & Z
			\end{array}
 			\right]
		\end{equation*} }
	\end{definition}
	
In this matrix form, the $\pm$ sign of each row along with 
its corresponding $Z_j$-literal designates whether 
the state of the $j^{th}$ qubit is $\ket{0}$ ($+$) or $\ket{1}$ ($-$). Suppose we 
want to simulate circuit $\cC$. Stabilizer-based
simulation first initializes $\cM$ to specify some basis 
state $\ket{\psi}$. To simulate the action of each gate $U \in \cC$,
we conjugate each row $g_i$ of $\cM$ by $U$.\footnote{\scriptsize
Since $g_i\ket{\psi} = \ket{\psi}$, 
the resulting state $U\ket{\psi}$ is stabilized by $Ug_iU^\dag$ 
because $(Ug_iU^\dag) U\ket{\psi} = Ug_i\ket{\psi} = U\ket{\psi}$.}
~We require that $Ug_iU^\dag$ maps to another
string of Pauli literals so that the resulting stabilizer
matrix $\cM'$ is well-formed. It turns out that the Hadamard,
Phase and CNOT gates (Figure \ref{fig:chp_pauli}a) as well as 
the Controlled-$Z$ and Controlled-$Y$ gates (Figure \ref{fig:ctrlgates}) 
have such mappings, i.e., these gates conjugate the Pauli group 
onto itself \cite{Gottes98, NielChu}. Since these gates are directly 
simulated using stabilizers, they are commonly called \emph{stabilizer gates}.  
Table~\ref{tab:cliff_mult} lists the transformation properties of Pauli literals
under conjugation by the Hadamard, Phase and CNOT gates. Figure~\ref{fig:ctrlgates} 
shows that the Controlled-$Z$ and Controlled-$Y$ operations can be decomposed into 
Pauli, Hadamard and CNOT gates and thus are also stabilizer gates.

	\begin{example}  {\em 
		Suppose we simulate a CNOT operation on 
		$\ket{\psi} = (\ket{00} + \ket{11})/\sqrt{2}$ using $\cM$,
		\begin{center}
			\begin{tabular}{c|l|cc|l|}
				\multirow{2}{6mm}{$\cM =$}   & $XX$ & \multirow{2}{8mm}{$\xrightarrow{CNOT}$} 
				&\multirow{2}{7mm}{$\cM' =$} & $XI$ \\
				& $ZZ$ &   						&                         &  $IZ$ \\
			\end{tabular}
		\end{center}
		\noindent
		One can verify that the rows of $\cM'$ stabilize
		$\ket{\psi}\xrightarrow{CNOT}(\ket{00} + \ket{10})/\sqrt{2}$ as required. }
	\end{example}

	\begin{table}[!t]
 		\tcaption{\label{tab:cliff_mult} \centering Transformation of Pauli-group 
 		elements under conjugation by the stabilizer gates \cite{NielChu}. \\ For 
 		the CNOT case, subscript $1$ indicates the control and $2$ the target.}	
        \centering \footnotesize
        \begin{tabular}{cc}
        \begin{tabular}{|c||c|c|}
            \hline
                \sc Gate & \sc Input  & \sc Output \\ \hline\hline
                          & $X$ & $Z$ \\
                $H$       & $Y$ & -$Y$ \\
                          & $Z$ & $X$  \\ \hline
                          & $X$ & $Y$ \\
                $P$       & $Y$ & -$X$ \\
                          & $Z$ & $Z$   \\
            \hline
        \end{tabular}
        & \hspace{-5pt}
         \begin{tabular}{|c||c|c|}
            \hline
            \sc Gate & \sc Input  & \sc Output  \\ \hline\hline
            \multirow{6}{10mm}{$CNOT$} & $I_1X_2$ & $I_1X_2$   \\
                      & $X_1I_2$ & $X_1X_2$    \\
                      & $I_1Y_2$ & $Z_1Y_2$      \\
                      & $Y_1I_2$ & $Y_1X_2$     \\
                      & $I_1Z_2$ & $Z_1Z_2$   \\
                      & $Z_1I_2$ & $Z_1I_2$    \\
            \hline
        \end{tabular} \\ \\
		\end{tabular}
		\vspace{-.5cm}
 	\end{table}
 	
The Hadamard, Phase and CNOT gates are also called \emph{Clifford gates} 
because they generate the Clifford group of unitary operators. We use 
these names interchangeably. Any circuit composed exclusively of stabilizer
gates is called a \emph{unitary stabilizer circuit}. 
Table~\ref{tab:cliff_mult} shows that at most two columns of $\cM$ 
are updated when one simulates a stabilizer gate. Thus, such gates are 
simulated in $\Theta(n)$ time. 

	\begin{figure}[!b]
 		\centering
 		\begin{tabular}{cc}
 		$
    		\Qcircuit @C=1.0em @R=1.0em {
    			& \\
    			& \ctrl{2} & \qw & & & \qw 	   & \ctrl{2} & \qw & \qw \\
    			& & & \equiv & & & \\
			& \gate{Z} & \qw & & & \gate{H}  & \targ    & \gate{H}  & \qw \\
			\\
   	 	}$
     	& \hspace{1cm}
   	 	$
    		\Qcircuit @C=1.0em @R=1.0em {
 			& \\
    			& \ctrl{2} & \qw  &  &	   & \qw & \qw       & \ctrl{2} & \qw        & \qw      & \qw \\
    			& & & \equiv & & & \\
			& \gate{Y} & \qw & & & \gate{Y}  & \gate{P}  & \targ    & \gate{P^3} & \gate{Y} & \qw
			\\
    		}$ \\
   	 	\footnotesize {\bf (a)} & \footnotesize {\bf (b)}
 		\end{tabular}
 		\fcaption{\label{fig:ctrlgates} Implementation of the {\bf (a)} Controlled-$Z$
 		and {\bf (b)} Controlled-$Y$ operations using Pauli and Clifford gates. These gates 
 		can be simulated directly on the stabilizer or using the equivalence shown here.}
 		\vspace{-.25cm}
 	\end{figure}

    \begin{theorem} \label{th:stabst}
        An $n$-qubit stabilizer state $\ket{\psi}$ can be obtained
        by applying a stabilizer circuit to the $\ket{0}^{\otimes n}$
        computational-basis state. 
    \end{theorem}
    \noindent
	{\bf Proof.} 
    		The work in \cite{AaronGottes} represents the generators using a tableau, 
    		and then shows how to construct a unitary stabilizer circuit from the 
    		tableau. We refer the reader to \cite[Theorem 8]{AaronGottes} for 
    		details of the proof.
    \square

    \begin{corollary} \label{cor:stab_allzeros}
	   	An $n$-qubit stabilizer state $\ket{\psi}$ can be transformed
  		by stabilizer gates into the $\ket{00\ldots 0}$ computational-basis state.
   	\end{corollary}
   	\noindent
	{\bf Proof.} 
        Since every stabilizer state can be produced by applying some unitary
        stabilizer circuit $\cC$ to the $\ket{0}^{\otimes n}$ state, it suffices to reverse
        $\cC$ to perform the inverse transformation. To reverse a stabilizer
        circuit, reverse the order of gates and replace every $P$ gate with $PPP$.
    \square
   
    \begin{definition} \label{def:biased} {\em 
        A pure state $\ket{\psi}$ with computational-basis decomposition
        $\sum_{k=0}^n \lambda_k\ket{k}$ is said to be {\em unbiased}
        if for all $\lambda_i \neq 0$ and $\lambda_j\neq 0$,
        $\mid\lambda_i\mid^2 = \mid\lambda_j\mid^2$. Otherwise,
        the state is {\em biased}.}
   \end{definition}
	
The stabilizer formalism also admits one-qubit measurements 
in the computational basis. 
However, the updates to $\cM$ for such gates are not as efficient
as for stabilizer gates. Note that any qubit
$j$ in a stabilizer state is either in a $\ket{0}$ ($\ket{1}$) state or 
in an unbiased (Definition~\ref{def:biased}) superposition 
of both \cite{Gottes98, NielChu, Vanden}. The former case is called a {\em deterministic
outcome} and the latter a {\em random outcome}. We can tell these 
cases apart in $\Theta(n)$ time by searching for $X$ or $Y$ literals in
the $j^{th}$ column of $\cM$. If such literals are found, the qubit must be in 
a superposition and the outcome is random with equal probability
($p(0) = p(1) = .5$); otherwise the outcome is deterministic
($p(0) = 1$ or $p(1) = 1$). 

	{\em Random case}: one flips an unbiased coin to decide the outcome 
	and then updates $\cM$ to make it consistent with the 
    outcome obtained. This requires at most $n$ row multiplications 
    leading to $O(n^2)$ runtime \cite{AaronGottes, NielChu}. 
	
	{\em Deterministic case}: no updates to 
	$\cM$ are necessary but we need to figure out whether the state 
	of the qubit is $\ket{0}$ or $\ket{1}$, i.e., whether the qubit is stabilized by $Z$ or -$Z$,
	respectively. One approach is to transform $\cM$ into  
	row-echelon form using Algorithm \ref{alg:gauss_min}.
	This removes redundant literals from $\cM$ in order
	to identify the row containing a $Z$ in its $j^{th}$ position
	and $I$ everywhere else. The $\pm$ phase of this row 
	decides the outcome of the measurement. Since this approach
	is a form of Gaussian elimination, it takes $O(n^3)$ time in 
	practice (and can be completed asymptotically faster in theory).   

Aaronson and Gottesman \cite{AaronGottes} improved the runtime of deterministic
measurements by doubling the size of $\cM$ to include $n$ {\em destabilizer generators} 
in addition to the $n$ stabilizer generators. Such destabilizer generators help 
identify exactly which row multiplications to compute in order 
to decide the measurement outcome. This approach avoids Gaussian elimination,
and thus deterministic measurements are computed in $O(n^2)$ time.

When dealing with single-qubit $Z$ measurements of the form
$P=Z\otimes I^{\otimes (n-1)}$, we call the post-measurement
states {\em cofactors}. Such states are separable stabilizer
states of the form $\ket{0}\ket{\alpha_0}$ and $\ket{1}\ket{\alpha_1}$.
We denote the $\ket{0}$- and $\ket{1}$-cofactor by $\ket{\psi_{j=0}}$ and $\ket{\psi_{j=1}}$,
respectively, where $j$ is the index of the measured qubit. The states $\ket{\alpha_0}$
and $\ket{\alpha_1}$ are called {\em reduced cofactors}, and are also
stabilizer states. The generators of the reduced cofactors are obtained from the generators
of the respective cofactors by dropping the generator
added during measurement and removing from each remaining generator
the $Z$ literal at the $j^{th}$ position.

	\begin{definition} {\em \label{def:cofactors}
		Let $\ket{\psi} = \sum_{k=0}^{2^n - 1}\alpha_k\ket{k}$ and $a\in \{0,1\}$.
		Furthermore, let $C_{j = a}$ be the set of computational-basis states in 
		$\ket{\psi}$ with the $j^{th}$ qubit set to $a$ and $\alpha_k\neq 0$.
		The {\em support} of $\ket{\psi_{j=a}}$,
		denoted by $\mid \psi_{j=a}\mid$, is $\mid C_{j=a} \mid$.
	 }
	\end{definition}

    \begin{observation} \label{obs:cofactors}
        The $\ket{0}$-cofactor of $\ket{\psi}$, with respect to qubit $j$,
        is $\ket{\psi_{j=0}} = \sum_{\ket{k}\in C_{j=0}}\alpha_k\ket{k}$.
        Similarly, the $\ket{1}$-cofactor of $\ket{\psi}$,
        with respect to $j$, is $\ket{\psi_{j=1}} = \sum_{\ket{k}\in C_{j=1}}\alpha_k\ket{k}$.
    \end{observation}

One can also consider {\em iterated cofactors}, such as
{\em double cofactors} $|\psi_{qr=00} \rangle\ldots |\psi_{qr=11}\rangle$.
Cofactoring with respect to all qubits produces the individual amplitudes 
of the computational-basis states. 
The work in \cite{Vanden} establishes an alternative representation for 
stabilizer states that tracks the non-zero amplitudes 
of a stabilizer state using linear transformations and bilinear forms 
over the two-element field.   

\newpage

\newcommand{\zfield}{\mathbb{Z}}
\newcommand{\by}{\mathbf y}
\newcommand{\bt}{\mathbf t}
\newcommand{\bx}{\mathbf x}
\newcommand{\bc}{\mathbf c}
\newcommand{\bQ}{\mathbf Q}
\newcommand{\bR}{\mathbf R}
\newcommand{\tok}{\frac{1}{2^{k/2}}}

	\begin{theorem}\cite[Appendix A]{Vanden} \label{th:stab_form}
		For an $n$-qubit stabilizer state $\ket{\psi}$, $\exists\ \bc,\bt\in\zfield_2^n$,
		$\bR$ an $n\times k$ binary matrix with column rank $k\leq n$, and $n\times n$ 
		symmetric binary matrix $\bQ$, such that
			\begin{equation} \label{eq:stab_form}
				\ket{\psi} = \tok\sum_{\bx\in \zfield_2^k}i^{2\bc^\top\by+\by^\top\bQ\by}\ket{\by = \bR\bx + \bt}
			\end{equation}
		\noindent
	\end{theorem}

   \begin{corollary} \label{cor:stab_props}
   		The basis amplitudes of a stabilizer state $\ket{s}$
   		have the following properties:
        \begin{itemize}
                \item[({i})] The number of non-zero amplitudes (support) in $\ket{s}$ is a power of two.
                \item[({ii})] State $\ket{s}$ is unbiased,
                and every non-zero amplitude is $\frac{\pm 1}{\sqrt{|s|}}$ or $\frac{\pm i}{\sqrt{|s|}}$,
                where $|s|$ is the support of $\ket{s}$. 
                \item[({iii})] The number of imaginary amplitudes in $\ket{s}$
                is either zero or half the number of non-zero amplitudes.\footnote{
                Assuming the amplitudes are normalized such that 
                the first non-zero basis amplitude is $1.0$.}
                \item[({iv})] The number of negative amplitudes in $\ket{s}$
                is either zero or a power of two.\footnote{For normalized amplitudes,
                this might require multiplying by a $-1$ global phase.}
                \item[({v})] For each qubit $q$ of $\ket{s}$,
                the number of basis states with $q=0$ in
                the $\ket{0}$- and $\ket{1}$-cofactors is either zero or a power of two.
                Furthermore, if both $\ket{0}$- and $\ket{1}$-cofactors with respect to a given qubit
                are nontrivial, the supports of the cofactors must be equal and the norms
                of the cofactors must be equal.
        \end{itemize}
    \end{corollary}

Corollary~\ref{cor:stab_props} is illustrated in Table~\ref{tab:two_qbssts} 
and Appendix~A. Its further implications are discussed 
in Section~\ref{sec:stabneighbors}. Corollary~\ref{cor:stab_props}-$v$ 
is restated as Lemma~\ref{lem:dbl_cofactors} and proven in Section~\ref{sec:stab_evade}.

\section{Geometric Properties of Stabilizer States} 
\label{sec:stabneighbors}
 
Given $\braket{\psi}{\varphi} = re^{i\alpha}$, we
normalize the global phase of $\ket{\psi}$ to ensure, without loss
of generality, that $\braket{\psi}{\varphi} \in \mathbb{R}_+$.

	\begin{theorem} \label{th:stab_ortho}
		Let $S(\ket{\psi})$ and $S(\ket{\varphi})$ be the stabilizer
		groups for $\ket{\psi}$ and $\ket{\varphi}$, respectively.
		If there exist $P \in S(\ket{\psi})$ and $Q \in S(\ket{\varphi})$
		such that $P =$ -$Q$, then $\ket{\psi}\perp\ket{\varphi}$.
	\end{theorem}
	\noindent
	{\bf Proof.} 
		Since $\ket{\psi}$ is a $1$-eigenvector of $P$ and
		$\ket{\varphi}$ is a $(-1)$-eigenvector of $P$, they
		must be orthogonal.
	\square

	\begin{theorem}\label{th:inprod_aron}{\em\bf \cite{AaronGottes}}
		Let $\ket{\psi}$, $\ket{\varphi}$ be oblique
		stabilizer states. Let $s$ be the minimum, over all sets of
		generators $\{P_1,\ldots,P_n\}$ for $S(\ket{\psi})$ and
		$\{Q_1,\ldots,Q_n\}$ for $S(\ket{\varphi})$, of the number
		of different $i$ values for which $P_i \neq Q_i$. Then,
		$|\braket{\psi}{\varphi}|=2^{-s/2}$.
	\end{theorem}
	\noindent
	{\bf Proof.} 
		Since $\braket{\psi}{\varphi}$ is not affected by
		unitary transformations $U$,
		we choose a stabilizer circuit
		such that $U\ket{\psi} = \ket{b}$, where $\ket{b}$ is a basis state.
		For this state, select the stabilizer generators $\cM$ of the
		form $I\ldots IZI\ldots I$. Perform Gaussian elimination
		on $\cM$ to minimize the incidence of $P_i \neq Q_i$.
		Consider two cases. If $U\ket{\varphi} \neq \ket{b}$ and its generators
		contain only $I/Z$ literals, then $U\ket{\varphi}\perp U\ket{\psi}$,
		which contradicts the assumption that $\ket{\psi}$ and $\ket{\varphi}$
		are oblique. Otherwise, each generator of
		$U\ket{\varphi}$ containing $X/Y$ literals contributes 
		a factor of $1/\sqrt{2}$ to the inner product.
	\square
	
	\begin{corollary} \label{cor:minmaxip}
		Let $\ket{\psi}$ and $\ket{\phi}$ be oblique
		$n$-qubit stabilizer states such that $\ket{\psi}\neq e^{i\alpha}\ket{\phi}$. Then,
		$2^{-n/2}\leq |\braket{\psi}{\phi}| \leq 2^{-1/2}$.	
	\end{corollary}

\subsection{Inner products and $k$-neighbor stabilizer states} 
\label{sec:kneighbors}

    \begin{definition} {\em 
		Given an arbitrary state $\ket{\psi}$ with $||\psi|| = 1$, a stabilizer state
		$\ket{\varphi}$ is a {\em $k$-neighbor stabilizer state}
		of $\ket{\psi}$ if $|\braket{\psi}{\varphi}|=2^{-k/2}$.
		}
	\end{definition}

When two stabilizer states are $1$-neighbors we will also refer to them as 
{\em nearest neighbors} since, by Corollary~\ref{cor:minmaxip}, this is 
the maximal inner-product value $\neq 1$. Furthermore, the 
distance between a stabilizer state and any of its $k$-neighbors 
is $\sqrt{2-2^{1-k/2}}$. Therefore, the distance between
closest stabilizer states (nearest neighbors) is $\sqrt{2-\sqrt{2}}\approx 0.765$.

	\begin{proposition}
		Consider two orthogonal stabilizer states $\ket{\alpha}$ and $\ket{\beta}$
		whose unbiased superposition $\ket{\psi}$
		is also a stabilizer state. Then $\ket{\psi}$ is a
		nearest neighbor of $\ket{\alpha}$ and $\ket{\beta}$.
	\end{proposition}
	\noindent
	{\bf Proof.} 
		Since stabilizer states are unbiased, $|\braket{\psi}{\alpha}| =
		|\braket{\psi}{\beta}| = \frac{1}{\sqrt{2}}$. 
		Thus, $\ket{\psi}$ is a $1$-neighbor or nearest-neighbor 
		stabilizer state to $\ket{\alpha}$ and $\ket{\beta}$.
		Figure~\ref{fig:basis_angle} illustrates this case.
	\square

	\begin{lemma} \label{lem:num_stabsts}
		Any two stabilizer states have equal numbers of $k$-neighbor
        stabilizer states.
  	\end{lemma}
  	\noindent
	{\bf Proof.}  
		Any stabilizer state can be mapped to another stabilizer state by a
		stabilizer circuit (Corollary~\ref{cor:stab_allzeros}). Since such 
		circuits effect unitary operators, inner products are preserved.
	\square

	\begin{lemma}\label{lem:cross}
        Let $|\psi\rangle$ and $|\varphi\rangle$ be $n$-qubit stabilizer states
        such that $\braket{\psi}{\varphi}\neq 1$. Then $\frac{|\psi\rangle+i^l|\varphi\rangle}{\sqrt{2}}$,
        $l\in\{0,1,2,3\}$, is a stabilizer state iff $\braket{\psi}{\varphi} = 0$ 
        and $|\varphi\rangle=P|\psi\rangle$, where $P\in\mathcal{G}_n$.
    \end{lemma}
    \noindent
	{\bf Proof.} We first prove that, if $\braket{\psi}{\varphi} = 0$ and 
        $|\varphi\rangle=P|\psi\rangle$, an unbiased sum of such states
        is also a stabilizer state. Suppose $S(|\psi\rangle)=\langle g_k\rangle_{k=1,2,\dots,n}$
        is generated by elements $g_k$ of the $n$-qubit Pauli group.
        Let 
        	\[
        		f(k)=\left\{\begin{array}{cl} 0 & \textrm{if}\ [P,g_k]=0 \\
        									  1 & \textrm{otherwise}\end{array}\right.
        	\]
        \noindent
        and write $S(|\varphi\rangle)=\langle (-1)^{f(k)}g_k\rangle$.
        Conjugating each generator $g_k$ by $P$ we see that $\ket{\varphi}$
        is stabilized by $\langle (-1)^{f(k)} g_k\rangle$.
        Let $Z_k$ (respectively $X_k$) denote the Pauli operator $Z$ ($X$) acting
        on the $k^{th}$ qubit. By Corollary~\ref{cor:stab_allzeros}, there exists
        an element $\cC$ of the $n$-qubit Clifford group which maps $S(|\psi\rangle)$
        to \normform~(Definition~\ref{def:basis_form}) such that $\cC|\psi\rangle=\ket{0}^{\otimes n}$
        and $\cC|\varphi\rangle=(\cC P\cC^\dag)\cC|\psi\rangle=i^m|f(1)f(2)\dots f(n)\rangle$.
        The second equality follows from the fact that $\cC P\cC^\dag$ is an element of
        the Pauli group and can therefore be written as $i^mX(v)Z(u)$ for some
        $m \in \{0,1,2,3\}$ and $u,v \in {\mathbb Z}_2^k$. Therefore,
        	\begin{equation}
		        \frac{|\psi\rangle+i^l|\varphi\rangle}{\sqrt{2}}= \frac{\cC^\dag(\ket{0}^{\otimes n}
        		+ i^{t=(l+m)\text{mod}\ 4}\ket{f(1)f(2)\ldots f(n)})}{\sqrt{2}}
        	\end{equation}
        	\noindent
        The state in parentheses on the right-hand side 
        	is the product of an all-zeros state and a GHZ state. Therefore, the sum 
        	is stabilized by $S' = \cC^\dag\langle S_{zero}, S_{ghz}\rangle \cC$
        where $S_{zero} = \langle Z_i, i \in \{k|f(k)=0\}\rangle$ and $S_{ghz}$
        is supported on $\{k|f(k)=1\}$ and equals $\langle(-1)^{t/2}XX\ldots X,\forall i\ Z_iZ_{i+1}\rangle$
        if $t=0\mod 2$ or $\langle(-1)^{(t-1)/2}YY\ldots Y,\forall i\ Z_iZ_{i+1}\rangle$
        if $t=1\mod 2$.
     
        We now prove the opposite implication. Let $\ket{u} = \frac{|\psi\rangle+i^l|\varphi\rangle}{\sqrt{2}}$, 
        where $|\psi\rangle$ and $|\varphi\rangle$ are $n$-qubit stabilizer states
        and $\braket{\psi}{\varphi}\neq 1$. Let $\cC_1$ and $\cC_2$ be Clifford circuits
        such that $\ket{\psi}=\cC_1\ket{\mathbf{0}}$ and $\ket{\varphi}=\cC_2\ket{\mathbf{0}}$,
        where $\ket{\mathbf{0}} = \ket{0}^{\otimes n}$. Observe that $\cC_1 \neq \cC_2$ by
        our assumption that $\braket{\psi}{\varphi}\neq 1$. Therefore,
		        $\ket{u} 
		        =(\cC_1\ket{\mathbf{0}} + i^l\cC_2\ket{\mathbf{0}})/\sqrt{2}
		        =\cC_1(\ket{\mathbf{0}} + i^l\cC_1^\dag\cC_2\ket{\mathbf{0}})\sqrt{2}$, and
        		\begin{align*}
       		 	\cC_1^\dag\ket{u}&=\frac{\ket{\mathbf{0}} + i^l\cC_1^\dag\cC_2\ket{\mathbf{0}}}{\sqrt{2}}
       		\end{align*}
       	\noindent
        	Since  $\cC_1^\dag\cC_2 \neq I^{\otimes n}$ and $\cC_1^\dag\ket{u}$ is a stabilizer state, 
        	$\cC_1^\dag\cC_2\ket{\mathbf{0}}$ must simplify to a basis state 
        	$\ket{b}\neq\ket{\mathbf{0}}$ (otherwise, $\cC_1^\dag\ket{u}$ is
        	either biased or has $2^n-1$ basis states). It follows that, 
        	 $\braket{\psi}{\varphi} = \bra{\mathbf{0}}\cC_1^\dag\cC_2\ket{\mathbf{0}}
        	 = \braket{\mathbf{0}}{b} = 0$. Let 	$i^l\ket{b}=P\ket{\mathbf{0}}$, 
        	 where $P$ is an element of the Pauli group,
        		\begin{align*}
       		 	\cC_1^\dag\ket{u}&=\ket{\mathbf{0}} + P\ket{\mathbf{0}} \\
       		 	\ket{u}&=\cC_1\ket{\mathbf{0}} + \cC_1 P\ket{\mathbf{0}} 
       		 	= \cC_1\ket{\mathbf{0}} + (\cC_1 P\cC_1^\dag)\cC_1\ket{\mathbf{0}} \\
       		 	&= \cC_1\ket{\mathbf{0}} + P'\cC_1\ket{\mathbf{0}} = \ket{\psi}+P'\ket{\psi}
       		 	\hspace{.5cm}\square
       		\end{align*} 

	\begin{corollary} \label{cor:stabsum}
		Let $\frac{|\psi\rangle+i^l|\varphi\rangle}{\sqrt{2}}$ be a stabilizer state.
		Then the canonical stabilizer matrices for $\ket{\psi}$ and $\ket{\varphi}$ 
		are similar.
	\end{corollary}
	\noindent
	{\bf Proof.} 
		Let $\cM^\psi = \{R_1, R_2,\ldots, R_n\}$ be the canonical stabilizer matrix 
		for $\ket{\psi}$. Since $\ket{\varphi} = P\ket{\psi}$ by Lemma~\ref{lem:cross},
		$\cM^\varphi = \{(-1)^cR_1, (-1)^cR_2,\ldots,(-1)^cR_n\}$, 
		where $c=0$ if $P$ and $R_i$ commute, and $c=1$ otherwise.
	\square

	\begin{theorem} \label{th:num_near}
        For any $n$-qubit stabilizer state $\ket{\psi}$, there are $4(2^n - 1)$
        nearest-neighbor stabilizer states, and these states 
        can be produced as described in Lemma~\ref{lem:cross}.
	\end{theorem}
    \noindent
	{\bf Proof.} 
		The all-zeros basis amplitude of any stabilizer state $\ket{\psi}$
		that is a nearest neighbor to $\ket{0}^{\otimes n}$
    		must be $\propto 1/\sqrt{2}$. Therefore, $\ket{\psi}$ is an unbiased superposition of
        $\ket{0}^{\otimes n}$ and one of the other $2^n-1$ basis states,
        i.e., $\ket{\psi} = \frac{\ket{0}^{\otimes n} + P\ket{0}^{\otimes n}}{\sqrt{2}}$,
        where $P \in \mathcal{G}_n$ such that 
        $P\ket{0}^{\otimes n} \neq \alpha\ket{0}^{\otimes n}$.
        As in the proof of Lemma~\ref{lem:cross},
        we have $\ket{\psi} = \frac{\ket{0}^{\otimes n} + i^l\ket{\varphi}}{\sqrt{2}}$,
        where $\ket{\varphi}$ is a basis state and $l \in \{0,1,2,3\}$. Thus,
        there are $4$ possible unbiased superpositions, and a total of
        $4(2^n - 1)$ nearest-neighbor stabilizer states. Since $\ket{0}^{\otimes n}$ is a stabilizer
        state, all stabilizer states have the same number of nearest neighbors
        by Lemma~\ref{lem:num_stabsts}.
    \square
    
	\begin{corollary} \label{cor:basis_angle}
		For any $n$-qubit stabilizer state $\ket{\psi}$, there exists a set 
		of states $V_\psi=\{\ket{s_i}\}_{i=1}^{2^n-1}$ such that each $\ket{s_i}$ is
		a nearest-neighbor to $\ket{\psi}$ and $\braket{s_i}{s_j}=1/2$ for 
		$i\neq j$. This set $V_\psi$ together with $\ket{\psi}$ form a basis 
		in $\cH^{\otimes n}$.
	\end{corollary}
	\noindent
	{\bf Proof.} 
		Without loss of generality, assume that $\ket{\psi}=\ket{0}^{\otimes n}$. 
		Theorem~\ref{th:num_near} shows that, for any given stabilizer states, its 
		nearest neighbors come in groups of four.
		Taking any one representative from each group of four, we get a set of states 
		$V_\psi=\left\{\ket{s_i}=\frac{\ket{0}^{\otimes n}+i^l\ket{b_i}}{\sqrt{2}}\right\}_{i=1}^{2^n-1}$,
		where $l\in\{0,1,2,3\}$ and $\ket{b_i}$ are computational-basis states other than $\ket{\psi}$.
		Thus, for all $\ket{s_i}$ and $\ket{s_j}$ such that $i\neq j$,
			\begin{equation}
				\braket{s_i}{s_j}=\left(\frac{\bra{0^{\otimes n}}+\bra{b_i}}{\sqrt{2}}\right)
				\left(\frac{\ket{0^{\otimes n}}+\ket{b_j}}{\sqrt{2}}\right) = \frac{1}{2}
			\end{equation} 
		$V_\psi$ together with $\ket{\psi}$ form a 
		linearly independent set (one can subtract $\ket{\psi}$ from each 
		$\ket{s_i}$ to get an orthogonal set) and thus a basis. Figure~\ref{fig:basis_angle}
		illustrates this nearest-neighbor structure for a small set of states.
	\square

In Table~\ref{tab:two_qbssts}, one can find the twelve nearest-neighbor states of $\ket{00}$.
We computed the angles between all-pairs of $3$-qubit stabilizer
states and confirmed that each was surrounded by exactly $28$ nearest neighbors. 
The same procedure confirmed that the number of nearest neighbors for any 
$4$-qubit stabilizer state is $60$. 
In Section~\ref{sec:stab_ortho}, we describe an orthogonalization procedure
for linear combinations of stabilizer states that takes advantage of the 
nearest-neighbor structure described in Theorem~\ref{th:num_near}.

Alternatively, Theorem~\ref{th:num_near} can also be proven using a counting
argument. By Theorem~\ref{th:inprod_aron}, we know that any nearest-neighbor 
stabilizer state to $\ket{0^{\otimes n}}$ can be represented by a stabilizer matrix
that has exactly one generator (row) with at least one $X/Y$ literal. Therefore,
there are $2(4^n-2^n)$ choices for the first generator $P_1$. The factor of $2$
accounts for the possible signs of $P_1$.
The remaining (independent) generators, $P_2,\ldots,P_n$, are then selected 
such that they commute with $P_1$ and consist of $Z/I$ literals only. Observe that such 
generators can be selected arbitrarily since they generate the same
($Z/I$)-element subgroup. 

	\begin{example} {\em 
		Suppose $P_1=XII$, then $P_2$ and $P_3$ can be selected arbitrarily 
		from the set $\{IZI,IZZ,IIZ\}$. To account for stabilizer matrices 
		that describe the same state, consider that $P_1$ can be replaced 
		by $P_1Q$, where $Q$ is an arbitrary product of the $Z/I$ generators.
		}
	\end{example}
		
Therefore, the number of nearest-neighbor stabilizer states 
$\mathcal{L}_n(1)$ is given by,

	\begin{equation} \label{eq:near2}
        	\mathcal{L}_n(1) = \frac{2(4^n-2^n)}{2^{n-1}} = 4(2^n-1)
    \end{equation}
   
\noindent
The following theorem generalizes Equation~\ref{eq:near2}
to the case of $k$-neighbor stabilizer states.

	\begin{theorem} \label{th:near_count}
		For any $n$-qubit stabilizer state, the number of
        $k$-neighbor stabilizer states is,
        \begin{equation}
        		\mathcal{L}_n(k) = 2^{k(k+1-n)}\prod_{j=0}^{k-1}\frac{4^{n}/2^j-2^{n}}{2^{k}-2^{j}}
        \end{equation}
	\end{theorem}
	\noindent
	{\bf Proof.} Without loss of generality, we count the $k$-neighbors of $\ket{0^{\otimes n}}$
	(Lemma~\ref{lem:num_stabsts}). By Theorem~\ref{th:inprod_aron}, we know that such 
	states are represented by an $n$-qubit stabilizer matrix with $k$ independent 
	$X/Y$ generators and $n-k$ independent $Z/I$ generators. Assume, for $j\leq k$, 
	that the $X/Y$ generators $P_1,\ldots,P_{j-1}$ have been chosen, and let 
	$Q_j,\ldots, Q_n$ be $Z/I$ generators that commute with them. Given
	this generating set, we count the possible choices for $P_j$ 
	to replace any one of the $Z/I$ generators. Observe that, $P_j$ must commute with 
	$P_1,\ldots,P_{j-1}$ and	cannot be an element of the subgroup generated 
	by $P_1,\ldots,P_{j-1},Q_{j},\ldots,Q_{n}$. 
	Thus, there are $2(4^n/2^{j-1}-2^n)$ choices for $P_j$. The factor of $2$
	accounts for the choice of sign.
	We need to account for choices of $X/Y$ generators that describe 
	the same state. Consider the subgroup generated by $P_{i\in \{1,\ldots, k\}}$.
	There are $2^k-1$ choices for $P_1$, $2^k-2$ for $P_2$, $2^k-4$ for $P_3$, 
	and so on. This gives a factor of $\prod_{j=1}^{k}(2^k-2^{j-1})$.
	Furthermore, we can replace $P_i$ by $P_iQ$, where $Q$ is an arbitrary product 
	of the $Z/I$ generators. This gives $k$ factors of $2^{n-k}$.
		\begin{equation*} \label{eq:nnear}
			\mathcal{L}_n(k) = \prod_{j=1}^{k}\frac{2}{2^{n-k}}\frac{4^n/2^{j-1}-2^n}{2^k-2^{j-1}} 
			\hspace{2cm}\square
		\end{equation*}

	\begin{corollary} \label{cor:stab_ortho}
		For an arbitrary $n$-qubit stabilizer state $\ket{\psi}$, the number of
        stabilizer states orthogonal to $\ket{\psi}$ is,
        		\begin{equation}
        			\mathcal{L}_{n}(\perp) = 
        			\mathcal{N}(n) - \sum_{k=1}^n\mathcal{L}_n(k) - 1 
        			= \frac{\mathcal{N}(n)(2^n-1)}{3\cdot 2^n}
        		\end{equation}
        	\noindent
        	where $\mathcal{N}(n)$ is the number of $n$-qubit stabilizer 
        	states (Proposition~\ref{prop:stab_count}). In other words, 
        	for any $n$-qubit stabilizer state $\ket{\psi}$ with a sufficiently 
        	large $n$, almost $1/3$ of remaining stabilizer states are orthogonal to $\ket{\psi}$. 
	\end{corollary}

As an illustration of Theorem~\ref{th:near_count} and Corollary~\ref{cor:stab_ortho}, 
Table~\ref{tab:near} numerically describes the distribution of inner products 
between any one $n$-qubit stabilizer state and all other stabilizer states 
for $n\in\{1,\ldots,7\}$. 
This table shows that the number of nearest-neighbor
stabilizer states as a fraction of all $n$-qubit stabilizer states approaches 
zero as $n$ increases. We now formalize other trends gleaned from Table~\ref{tab:near}.

	\begin{table}[!b]
		\tcaption{\centering \label{tab:near} Distribution of inner products (angles) 
		between any one $n$-qubit stabilizer state and all other stabilizer states 
		for $n\in\{1,\ldots,7\}$. The last column indicates the ratio 
		of orthogonal ($\perp$) states. 
		}
		\centering\footnotesize
		\begin{tabular}{|c|c||c|c|c|c|c|c|c|c|} \hline
		\multirow{3}{2mm}{$n$} & \multirow{3}{8mm}{$\mathcal{N}(n)$} 
		& \multicolumn{8}{c|}{$\mathcal{L}_n(k)/(\mathcal{N}(n)-1)$} \\ \cline{3-10}
								& & $k=1$ & $k=2$ & $k=3$ & $k=4$ & $k=5$ & $k=6$ & $k=7$ & $\perp$  \\
		    					    & & \hspace{-8pt} ($45.00^\circ$) \hspace{-8pt} 
		    					      & \hspace{-8pt} ($60.00^\circ$) \hspace{-8pt} 
		    					      & \hspace{-8pt} ($69.30^\circ$) \hspace{-8pt} 
		    					      & \hspace{-8pt} ($75.52^\circ$) \hspace{-8pt}
		    					      & \hspace{-8pt} ($79.82^\circ$) \hspace{-8pt} 
		    					      & \hspace{-8pt} ($82.82^\circ$) \hspace{-8pt} 
		    					      & \hspace{-8pt} ($84.93^\circ$) \hspace{-8pt}
		    					      & \hspace{-8pt} ($90.00^\circ$) \hspace{-8pt}
		    					     \\ \hline\hline
		    					 $1$ & $6$ 		& $80\%$  & & & & & & & $20\%$\\
		    					 $2$ & $60$ 		& $20.34\%$ & $54.24\%$ & & & & & & $25.42\%$\\
		    					 $3$ & $1080$ 	& $2.59\%$ & $20.76\%$ & $47.45\%$ & & & & & $29.19\%$ \\
		    					 $4$ & $36720$ 	& $0.16\%$ & $3.05\%$ & $20.92\%$ & $44.62\%$ & & & & $31.25\%$ \\
		    					 $5$ & $2423520$ & $0.01\%$ & $0.20\%$ & $3.27\%$ & $20.96\%$ 
		    					 & $43.27\%$ & & & $32.29\%$ \\
		    					 $6$ & $315057600$ & $\approx 0\%$ & $0.01\%$ & $0.23\%$ & $3.39\%$ 
		    					 & $20.97\%$ & $42.60\%$ & & $32.81\%$ \\ 
		    					 $7$ & $81284860800$ & $\approx 0\%$ & $\approx 0\%$ & $0.01\%$ & $0.24\%$ 
		    					 & $3.44\%$ & $20.97\%$ & $42.27\%$ & $33.07\%$ \\ \hline
		\end{tabular}
	\end{table}

    \begin{lemma} \label{lem:stab_knear}
		For an arbitrary $n$-qubit stabilizer state, consider the quantity 
		$a_{n, k}=\frac{{\mathcal L}_n(k)}{{\mathcal N}(n)}$, where $0<k\leq n$. 
		Then, for fixed $m$, the sequence $\{a_{n, n-m}\}^{n\rightarrow\infty}$ is monotonically 
		convergent.
	\end{lemma}
	\noindent
	{\bf Proof.} Using Theorem~\ref{th:near_count} and the 
	recurrence relation 	of Equation~\ref{eq:stab_count} we obtain,
		\begin{align} \label{eq:knear_ratio1}
			\frac{a_{n, k}}{a_{n+1, k+1}} &= \frac{\mathcal{L}_n(k)}{\mathcal{N}(n)}\cdot
			\frac{2(2^{n+1}+1)\mathcal{N}(n)}{\mathcal{L}_{n+1}(k+1)} 
			= 2(2^{n+1}+1)\frac{\mathcal{L}_n(k)}{\mathcal{L}_{n+1}(k+1)}
		\end{align}	
	\noindent where
		\begin{align} \label{eq:knear_ratio2}
			\frac{\mathcal{L}_n(k)}{\mathcal{L}_{n+1}(k+1)} &=2^{n-k-1}
			\prod_{j=0}^{k-1}\left(\frac{4^n/2^j-2^n}{2^k-2^j}\right)
			\prod_{j=0}^{k}\left(2\frac{2^{k}-2^{j-1}}{4^{n+1}/2^j-2^{n+1}}\right)  \notag \\
			&=2^n\prod_{j=1}^{k}\left(\frac{4^n/2^{j-1}-2^n}{2^k-2^{j-1}}\right)
			\prod_{j=0}^{k}\left(\frac{2^{k}-2^{j-1}}{4^{n+1}/2^{j}-2^{n+1}}\right)  \notag \\
			&=2^n\left(\frac{2^k-1/2}{4^{n+1}-2^{n+1}}\right)
			\prod_{j=1}^{k}\left(\frac{1}{2}\frac{4^n/2^{j-1}-2^n}{2(4^n/2^{j})-2^n}\right) 
			= \frac{2^{k+1}-1}{2^k(2^{n+3}-4)} \notag
		\end{align}
	\noindent Inserting the above result in Equation~\ref{eq:knear_ratio1} gives,
		\begin{align}
			\frac{a_{n, k}}{a_{n+1, k+1}} &= 2(2^{n+1}+1)\frac{2^{k+1}-1}{2^k(2^{n+3}-4)}
			= \frac{2^{n+k+3}+2^{k+2}-2^{n+2}-2}{2^{n+k+3}-2^{k+2}}
		\end{align}	
	\noindent Consider the case $0 < k < n$, 
		\begin{align*}
			\frac{2^{n+k+3}+2^{k+2}-2^{n+2}-2}{2^{n+k+3}-2^{k+2}} < 1
			\quad\quad\text{and thus}\quad\quad\frac{2^{k+2}}{ 2^{n+1}+1} < 1
		\end{align*}
	Therefore, for fixed $0 < m < n$, $\{a_{n, n-m}\}^{n\rightarrow\infty}$ is monotonically 
	increasing. It converges because $0 < a_{n, k} < 1$. For the case $k=n$,
		\begin{align}
			\frac{a_{n, n}}{a_{n+1, n+1}} &= \frac{2^{2n+3}-2}{2^{2n+3}-2^{n+2}}
			= \frac{2^{n+1}-2^{-n-1}}{2^{n+1}-1} = 1+\frac{1}{2^{n+1}} > 1
		\end{align}
	\noindent
	Therefore, the sequence $\{a_{n,n}\}^{n\rightarrow\infty}$ is monotonically decreasing and convergent.
	\square

	\begin{theorem} \label{th:stab_knear}
		For an arbitrary $n$-qubit stabilizer state, the limit of the sequence
		$\{a_{n,n-k}\}^{n\rightarrow\infty}$, $0\leq k < n$, lies in the 
		interval $[l(k), u(k)]$, where
		\begin{align*}
			l(k) = \frac{M_5}{2^{k(k+5)/2}}\cdot
			\exp\left(\frac{1}{32}-\frac{2}{2^{k+5}-1}\right) \quad\quad
			u(k) = \frac{M_5}{2^{k(k+3)/2}}\cdot
			\exp\left(\frac{1}{15}-\frac{2}{2^{k+5}+1}\right)
		\end{align*}
		and
		\[
			M_5 = \prod_{j=1}^5\left(\frac{1}{1-2^{-j}}\right)
			\prod_{j=1}^5\left(1-\frac{2}{2^{j+k}+1}\right)
		\]
	\end{theorem}
	\noindent
	{\bf Proof.} From the proof of Lemma~\ref{lem:stab_knear} we know $0 < l(k) < u(k) < 1$. 
	We now derive the sharper bounds, $l(k)$ and $u(k)$, for $\{a_{n,n-k}\}^{n\rightarrow\infty}$. 
	Using Theorem~\ref{th:near_count} and Proposition~\ref{prop:stab_count} we obtain,
		\begin{equation}
        		a_{n, n-k}
        		= 2^{-k}
        		\underbrace{\prod_{j=1}^{n-k}\left(\frac{1}{1-2^{-j}}\right)}_{A_{n-k}}
        		\underbrace{\prod_{j=1}^{n-k}\left(1-\frac{2}{2^{j+k}+1}\right)}_{B_{n-k}}
        		\underbrace{\prod_{j=1}^{k}\left(\frac{1}{1+2^{j}}\right)}_{C_k}
       	\end{equation}
    \noindent We now derive upper and lower bounds for the products $A_{n-k}$, $B_{n-k}$ and $C_k$.
    		\begin{align*}
    			\prod_{j=1}^{n-k}\exp\left(\frac{1}{2^{j}}\right) < 
    			&\ A_{n-k} < \prod_{j=1}^{n-k}\exp\left(\frac{1}{2^j-1}\right) \\
    			\prod_{j=1}^{n-k}\exp\left(\frac{-2}{2^{j+k}-1}\right) < 
    			&\ B_{n-k} < \prod_{j=1}^{n-k}\exp\left(\frac{-2}{2^{j+k}+1}\right)\\
    			\frac{1}{2^{k(k+5)/2}} = \prod_{j=1}^k\frac{1}{2^{j+1}} <
    			&\ C_k < \prod_{j=1}^k\frac{1}{2^j} = \frac{1}{2^{k(k+1)/2}}
    		\end{align*}
    	Therefore, as $n\rightarrow\infty$, 
    		\begin{align*}
    			A_5\cdot\exp\left(\frac{1}{32}\right)
    			< &\ A_{\infty} < 
    			A_5\cdot\exp\left(\frac{1}{15}\right) \\
    			B_5\cdot\exp\left(\frac{-2}{2^{j+k}-1}\right)
    			< &\ B_{\infty} < B_5\cdot\exp\left(\frac{-2}{2^{j+k}+1}\right)
    			\hspace{5pt}\square
    		\end{align*}
	
\begin{figure}[!t]
	\centering
	\hspace{-4cm}
	\begin{tabular}{cc}
	\begin{tabular}{|c|c|} \hline
	$k$ & \small $\displaystyle\lim_{n\rightarrow\infty}\frac{\mathcal{L}_n(n-k)}{\mathcal{N}(n)}$ \\\hline\hline
	$0$ & $41.9422\%$ \\ \hline
	$1$ & $20.9712\%$ \\ \hline
	$2$ & $3.4952\%$ \\ \hline
	$3$ & $.2497\%$ \\ \hline
	$4$ & $.0083\%$ \\ \hline
	$5$ & $\approx 0\%$ \\ \hline
	$\vdots$ & $\vdots$ \\ \hline
	$n-1$ & $\approx 0\%$ \\ 
	\hline
	\multicolumn{2}{c}{ } \\
	\end{tabular} 
	&
	\begin{tabular}{cc} 
		\includegraphics[scale=.6]{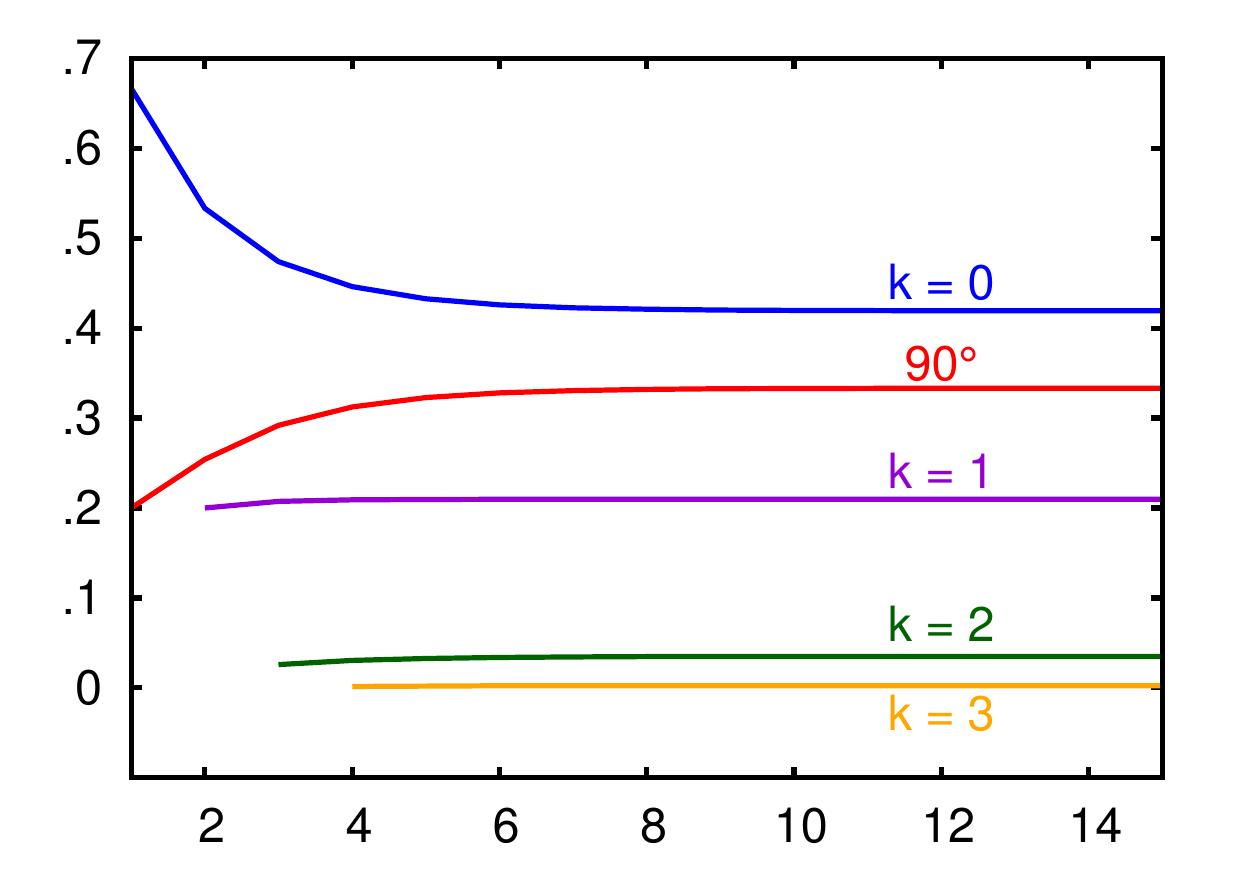}
		\vspace{-5pt}
		\hspace{-5cm}
		\raisebox{4cm}{\Large$\frac{\mathcal{L}_n(n-k)}{\mathcal{N}(n)}$}
		\\
		\hspace{4cm} \large $n$
	\end{tabular} 
	\\
	{\bf (a)} & \hspace{4cm} {\bf (b)}
	\end{tabular}
	\vspace{.25cm}
	\fcaption{\centering\label{fig:nnearconv} {\bf (a)} For an $n$-qubit stabilizer state $\ket{\psi}$,
	the fraction ($\approx 2/3$) of all stabilizer states that are oblique to $\ket{\psi}$ is 
	dominated by its $(n-k)$-neighbors, where $k \in \{0,1,2,3,4\}$. {\bf (b)} 
	The 	stabilizer states orthogonal to $\ket{\psi}$
	together with its $(n-k)$-neighbors ($k < 3$) account for $\approx 99\%$ of all states.
	}
	\vspace{-.25cm}
\end{figure}

In particular, observe that $u(k) \approx 0$ for $k>4$. Thus, the fraction of all stabilizer
states that are oblique to some $n$-qubit stabilizer state $\ket{\psi}$ is dominated
by its $(n-k)$-neighbors, where $k \in \{0,1,2,3,4\}$. Limit values 
for $\mathcal{L}_{n}(n-k)/\mathcal{N}(n)$
are approximated in Figure~\ref{fig:nnearconv}-a.
	
	\begin{theorem} \label{th:ipexpect}
		For two $n$-qubit stabilizer states $\ket{\psi}$ and $\ket{\phi}$ 
		drawn independently from a uniform distribution, the expected value 
		$E[\braket{\psi}{\phi}] \rightarrow 0$ as $n\rightarrow\infty$.
	\end{theorem}
	\noindent
	{\bf Proof.} Let $\ket{\varphi_{n-k}}$ be an $(n-k)$-neighbor of $\ket{\psi}$. 
	By Theorems~\ref{th:inprod_aron} and \ref{th:stab_knear},
		\begin{align}
		E[\braket{\psi}{\phi}]
		= \sum_{k=0}^{n-1} a_{n, n-k}|\braket{\psi}{\varphi_{n-k}}| 
		< \sum_{k=0}^{n-1} u(k)|\braket{\psi}{\varphi_{n-k}}| \approx \sum_{k=0}^{4} u(k) 2^{(k-n)/2}
		\end{align}
	Therefore,
		$
		E_\infty[\braket{\psi}{\phi}] < \sum_{k=0}^{4} u(k) 2^{(k-n)/2}
		$,
	which approaches zero as $n\rightarrow\infty$. 
	\square

\noindent

Theorem~\ref{th:ipexpect} suggests that stabilizer states are distributed
in a way similar to random quantum states \cite{DiVince, Klapp, Montanaro}.

\subsection{Exterior products and stabilizer bivectors} 
\label{sec:bivectors}

In this section, we study pairs of stabilizer states and the 
parallelograms they form, which lie in a $2n$-qubit Hilbert
space. Such pairs of states are called {\em bivectors} and 
are obtained by computing the wedge product $\ket{\psi\wedge\phi} = \ket{\psi}\otimes\ket{\phi} 
- \ket{\phi}\otimes\ket{\psi}$ between stabilizer states.
The wedge product is antisymmetric and yields zero for parallel vectors.
The norm of a stabilizer bivector can be interpreted as
the {\em signed area} of the parallelogram defined by non-parallel 
stabilizer states.

	\begin{definition} \label{def:bivector} {\em
		A {\em stabilizer bivector} $\ket{\varphi}$ is a $2n$-qubit stabilizer state
		such that $\ket{\varphi}\propto\ket{\psi\wedge\phi}$, where $\ket{\psi}$ and 
		$\ket{\phi}$ are $n$-qubit stabilizer states. }
	\end{definition}

	\begin{example} {\em 
		The wedge product of the stabilizer states 
		$\ket{\psi} = \ket{00}+\ket{11}$ and $\ket{\phi} = \ket{00}-\ket{11}$
		produces the stabilizer bivector $\ket{\psi\wedge\phi}=\ket{1100} + \ket{0011}$ 
		(up to a phase). In contrast, given
		the states $\ket{\psi} = \ket{00}+\ket{11}$ and $\ket{\phi} =\ket{10}$,
		$\ket{\psi\wedge\phi}=\ket{0010}+\ket{1100}+\ket{1110}-\ket{0011}-\ket{1000}-\ket{1011}$ 
		is not a stabilizer bivector because the number of  constituent basis states is
		not a power of two (Corollary~\ref{cor:stab_props}-$ii$). 
	}
	\end{example}

The work in \cite{Emary, Heydari} derives measures of entanglement for 
general pure bipartite states based on the alternating tensor product 
space defined by the wedge product of two states. Furthermore, 
Hayashi et al. \cite{Hayashi} conclude that the antisymmetric 
basis states (the wedge product of basis states) are 
more entangled than any symmetric basis states. Therefore,
stabilizer bivectors are potential candidates for developing
new entanglement monotones for quantifying quantum resources.
In Section~\ref{sec:bivec_comp}, we describe how to efficiently 
compute a generator set for stabilizer bivectors.


	\begin{theorem} \label{th:num_wedge}
        For any $n$-qubit stabilizer state $\ket{\psi}$, there are at least $5(2^n - 1)$
        distinct stabilizer states $\ket{\phi}$ such that the wedge product 
        $\ket{\psi\wedge\phi}$ is a stabilizer bivector.
	\end{theorem}
    \noindent
	{\bf Proof.} 
		By Proposition~\ref{prop:stab_tensor}, $\ket{\psi}\ket{\phi}$ and 
		$\ket{\phi}\ket{\psi}$ are both stabilizer states. Consider two cases.
	 
		{\em Case 1}: $\ket{\psi}$ and $\ket{\phi}$ are orthogonal. 
		From Lemma~\ref{lem:cross} and Corollary~\ref{cor:stabsum} we know that,
		if $\ket{\psi}\ket{\phi} - \ket{\phi}\ket{\psi}$ is a stabilizer state, the
	    canonical matrices of $\ket{\psi}$ and $\ket{\phi}$ must be {\em similar} (Definition~\ref{def:mtxeq}). 			Since each {\em dissimilar} matrix (Definition~\ref{def:mtxeq}) can 
		be used to represent $2^n$ states (the number of possible phase-vector permutations) 
		and $\ket{\psi\wedge\psi} = 0$, there are $2^n-1$ such wedge products that produce
		stabilizer bivectors. 
		
		{\em Case 2}: $\ket{\psi}$ and $\ket{\phi}$ are $1$-neighbors. 
		By Theorem~\ref{th:num_near}, $\ket{\psi}=\frac{\ket{\phi}+P\ket{\phi}}{\sqrt{2}}$
		and $\ket{\phi}=\frac{\ket{\psi}+P'\ket{\psi}}{\sqrt{2}}$, where $P$ and $P'$ are
		Pauli operators. Therefore, 
			\begin{equation}\label{eq:bivector_1near}
				\ket{\psi\wedge\phi} = \left(\frac{\ket{\phi}+P\ket{\phi}}{\sqrt{2}}\right)\wedge\ket{\phi}
				=  \frac{(P\ket{\phi})\wedge\ket{\phi}}{\sqrt{2}}
			\end{equation}
		Since $\ket{\phi}$ and $P\ket{\phi}$ are orthogonal with similar stabilizer 
		matrices (Case 1), $\ket{\psi\wedge\phi}$ is a stabilizer bivector
		(up to a normalizing factor), and there are $4(2^n-1)$ such wedge products.
		
		We now show that for $k$-neighbors $\ket{\psi}$ and $\ket{\phi}$, $k > 1$, 
		$\ket{\psi\wedge\phi}$ is not a stabilizer bivector. Without loss of generality,
		let $\ket{\psi}=\ket{0}$ and $\ket{\phi}=\sum_{i=0}^{2^k-1}\alpha_i\ket{b_i}$,
		where the $\ket{b_i}$ are computational-basis states. 
			\begin{align} \label{eq:nowedge}
				\ket{\psi\wedge\phi} 
				&= \ket{0}\wedge\left(\sum_{i=0}^{2^k-1}\alpha_i\ket{b_i}\right)
				= \sum_{i=1}^{2^k-1} \ket{0\wedge b_i}
			\end{align}
		Therefore, $\ket{\psi\wedge\phi}$ has $2^{k+1}-2$ computational-basis
		states, which is not a power of $2$ for $k>1$. By Corollary~\ref{cor:stab_props}-$i$,
		$\ket{\psi\wedge\phi}$ is not a stabilizer bivector for $(k>1)$-neighbors.
		\square

	\begin{proposition} \label{prop:bivec_norm}
		Given any two $n$-qubit stabilizer states $\ket{\psi}$ and $\ket{\phi}$, 
		the area of the parallelogram formed by these states is 
		$\|\psi\wedge\phi\| = \sqrt{1-2^{-k}}$, $0\leq k\leq n$.
	\end{proposition}
	\noindent
	{\bf Proof.} 
		Consider the determinant of the {\em Gramian matrix}
		of the states,
		\begin{equation*} \label{eq:wedge_norm}
			\text{det}\left|
			\begin{array}{cc}
				1 & \braket{\psi}{\phi} \\
				\braket{\phi}{\psi} & 1
			\end{array}
			\right| 
			= \|\psi\wedge\phi\|^2 = 1 - |\braket{\psi}{\phi}|^2
		\end{equation*}
		\noindent
		Thus, by Theorem~\ref{th:inprod_aron}, $\|\psi\wedge\phi\| = \sqrt{1-2^{-k}}$.
	\square

\subsection{Linear dependence of stabilizer states}
\label{sec:lindepstab}

We now characterize linearly-dependent triplets of stabilizer
states and discuss properties of minimally-dependent sets. 
Such properties can help in exploring succinct representations 
of arbitrary pure states using linear combinations of 
stabilizer states. 

	\begin{corollary}[of Theorem~\ref{th:num_near}] \label{cor:stabtriplets}
		Every linearly-dependent triplet $\{\ket{s_1},\ket{s_2},\ket{s_3}\}$ 
		of stabilizer states that are non-parallel to 
		each other includes two pairs of nearest neighbors and one pair of 
		orthogonal states. Furthermore, every nearest-neighbor pair 
		gives rise to a triplet of linearly-dependent stabilizer 
		states with two pairs of nearest neighbors and one pair 
		of orthogonal states. 
	\end{corollary}
	\noindent
	{\bf Proof.} Without loss of generality, assume that $\ket{s_1} = \alpha\ket{s_2} + \beta\ket{s_3}$  
		and $\ket{s_1} = \ket{00\ldots 0}$. 
		Recall from Corollary~\ref{cor:stab_props} that the non-zero amplitudes of a 
		stabilizer state $\ket{s}$ are 
		$\pm 1/\sqrt{|s|}$ or $\pm i/\sqrt{|s|}$
		and its support $|s|$ is a power of two.		
		By our assumptions, the supports of $\ket{s_2}$ and $\ket{s_3}$ cannot differ 
		by more than one.
		
		{\em Case 1}: $|s_2|\neq |s_3|$. Then one of $\ket{s_2}$ and $\ket{s_3}$ has 
		support one, while the other has support two. Suppose $|s_2| = 2$. 
		By Lemma~\ref{lem:cross}, 
		$\ket{s_2} = \frac{\ket{00\ldots 0}+i^l\ket{b}}{\sqrt{2}}$, 
		where $l\in\{0,1,2,3\}$ and $\ket{b}$ is computational-basis state 
		other than $\ket{00\ldots 0}$. Thus, $\ket{s_1}$ and $\ket{s_2}$
		are nearest neighbors. Since $|s_3|=1$ and the triplet
		is linearly dependent, $\ket{s_3}=\sqrt{2}\ket{s_2}-\ket{s_1}=i^l\ket{b}$.
		Therefore, $\ket{s_3}$ is orthogonal to $\ket{s_1}$ and 
		a nearest neighbor to $\ket{s_2}$.
		
		{\em Case 2}: $|s_2|=|s_3|$. At least one of $\ket{s_2}$ and $\ket{s_3}$ 
		must have a non-zero amplitude at $\ket{00\ldots 0}$. 
		If only one does, say $\ket{s_2}$, then $\ket{s_3}$ will have a
		non-zero amplitude at some other basis state where $\ket{s_2}$ has zero amplitude, 
		preventing $\ket{s_1} = \alpha\ket{s_2} + \beta\ket{s_3}$ from being $\ket{00\ldots 0}$. 
		Thus, both $\ket{s_2}$ and $\ket{s_3}$ must have a non-zero amplitude at $\ket{00\ldots 0}$
		and, more generally, their non-zero amplitudes must all be at the same basis elements.
		
		As non-zero amplitudes of stabilizer states must be either real or imaginary,
		we normalize the global phases of $\ket{s_2}$ and $\ket{s_3}$ so that the 
		$\ket{00\ldots 0}$ amplitudes are real, which implies that $\alpha$ and 
		$\beta$ are real.
		Using additional multiplication by $\pm 1$, we ensure that $\alpha,\beta>0$.
		Considering the amplitudes at $\ket{00\ldots 0}$, linear dependence
		implies $\pm\frac{\alpha}{\sqrt{|s_2|}}\pm\frac{\beta}{\sqrt{|s_3|}}=1$.
		Furthermore, $\alpha,\beta > 0$ and $|s_2|=|s_3|$ leave three possibilities:
		$\alpha\pm\beta=\sqrt{|s_2|}$ and $\beta-\alpha=\sqrt{|s_2|}$.
		At any other basis element where $\ket{s_2}$ and $\ket{s_3}$ have 
		non-zero amplitudes, there is the additional constraint that 
		$\alpha=\beta$ since these amplitudes must cancel out while $\alpha, \beta > 0$.
		This eliminates two of the three possibilities above, implying
		$\alpha=\frac{1}{\sqrt{|s_2|}}$,
		$\beta=\frac{1}{\sqrt{|s_2|}}$ and $|s_2|=|s_3|=2$.
		By Lemma~\ref{lem:cross}, $\ket{s_2}$ and $\ket{s_3}$ are
		nearest neighbors of $\ket{s_1}$ and $\braket{s_2}{s_3}=0$.
	\square


\ \\\noindent
{\bf Minimally-dependent sets of stabilizer states.} Given $k>3$, we now 
consider sets of $k$ stabilizer states that are {\em linearly dependent}, 
but such that all of their proper subsets are {\em independent}.
One such possibility are sets that contain $k-1$ 
mutually orthogonal stabilizer states and their sum. 
Consider the computational basis and a superposition of basis 
states that is also a stabilizer state. Theorem~\ref{th:stab_form} and 
Corollary~\ref{cor:stab_props} suggest examples of such 
minimally-dependent sets with $k=2^m+1$ stabilizer states.
Suppose $k=2^m-d$ and $\ket{s}$ is a stabilizer state
in a superposition of all $2^m$ computational-basis states. 
To construct a minimally-dependent set of size $k$, 
replace $2d+2$ basis states with 
$d+1$ stabilizer states formed as half-sums of $d+1$ disjoint
pairs of basis states. Note that all stabilizer states 
except for $\ket{s}$ remain orthogonal and contribute to $\ket{s}$.

	\begin{example}{\em 
		Let $\ket{s}=\frac{1}{2\sqrt{2}}\sum_{i=0}^7\ket{b_i}$ 
		where each $\ket{b_i}$ is a computational-basis state.
		One can obtain a minimally-dependent set of size $k=5$ 
		as $\left\{(\ket{b_0}+\ket{b_1}),(\ket{b_2}+\ket{b_3}),(\ket{b_4}+\ket{b_5}),\right.$
		$\left.(\ket{b_6}+\ket{b_7}), \ket{s}\right\}$,
		where each state in parentheses is a half-sum of two
		basis states. 
	}
	\end{example}

We close this section by outlining how to test the linear (in)dependence 
of a finite set of stabilizer states $\mathbf{s} = \{\ket{s_1}, \ldots,\ket{s_k}\}$. 
Recall that the {\em Gramian} of $\mathbf{s}$ is a matrix whose entries are given by 
$\braket{s_j}{s_i}$, and that $\mathbf{s}$ is linearly dependent if and only if the 
determinant of its Gramian matrix is zero. In Section~\ref{sec:ipalgo}, we describe 
an inner-product algorithm for stabilizer states. Thus, to test the linear 
(in)dependence of $\mathbf{s}$, generate the Gramian matrix by computing 
pairwise inner products using our algorithm.\footnote{The performance of pairwise 
computation of inner products between stabilizer states using techniques in 
Section~\ref{sec:ipalgo} can be improved by pre-computing and storing a basis normalization 
circuit for each state, rather than computing it from scratch for each pair of states.} 
Then, compute the determinant of the Gramian matrix and compare it to zero.

\section{The Embedding of Stabilizer Geometry in Hilbert Space} 
\label{sec:embedding}

The geometric structure of stabilizer states described in 
Section~\ref{sec:kneighbors} suggests an equally-spaced 
embedding in the finite-dimensional Hilbert space 
that can be exploited to study arbitrary 
quantum states. This is further evidenced by two types of results.
First, the uniform distribution over stabilizer states
is close to the uniform distribution of arbitrary quantum
states in terms of their first two moments \cite{DiVince, Klapp, Montanaro}.
Second, the entanglement of stabilizer states is nearly maximal
and similar to that of random states \cite{Dahlsten, Smith}. 

Consider a linear combination of stabilizer states $\Sigma_{i=1}^k \alpha_i\ket{s_i}$ 
and the task of finding a closest stabilizer state:
	\begin{equation} \label{eq:geoentangle}
		\argmax_{\ket{\phi}\in \text{Stab}(\cH)} 
		|\Sigma_{i=1}^k\alpha_i\braket{\phi}{s_i}|
	\end{equation}
where Stab($\cH$) is the set of stabilizer states in the Hilbert space $\cH$.
Here, we can work with any representation of $\ket{\phi}$ that allows us to
compute inner products with stabilizer states, e.g., a succinct superposition
of stabilizer states. The large count of nearest-neighbor stabilizer states given by 
Theorem~\ref{th:near_count} and the rather uniform structure of stabilizer geometry 
(Theorem~\ref{th:ipexpect}, Corollary~\ref{cor:basis_angle}) motivate the use of 
local search to compute Formula~\ref{eq:geoentangle}. 
However, it turns out that greedy local search does not guarantee 
finding a closest stabilizer state.

\subsection{Approximating an arbitrary quantum state with a single stabilizer state}
\label{sec:locsrch}

We analyze a simple {\em greedy local-search} algorithm that starts at the 
stabilizer state $\ket{s_m}$ in $\ket{\psi}=\Sigma_{i=1}^k \alpha_i\ket{s_i}$ with 
the largest $|\alpha_i|$. At each iteration, this algorithm evaluates the
nearest neighbors $N(\ket{s_m})$ of the current state $\ket{s_m}$ by computing 
$\max_{\ket{\phi}\in N(\ket{s_m})} |\braket{\phi}{\psi}|$, and then moves
to a neighbor that satisfies this metric. The algorithm 
stops at a stabilizer state that is closer to $\ket{\psi}$ 
than any of its nearest neighbors. As an illustration, consider the state
	\begin{equation}
		\ket{\psi}= (1+\varepsilon)\ket{00}+\ket{01}+\ket{10}+\ket{11}
	\end{equation}
Given a sufficiently small $\varepsilon>0$, the unique 
closest stabilizer state to $\ket{\psi}$ is $\ket{00}+\ket{01}+\ket{10}+\ket{11}$.
We start local search at $\ket{s_m}=\ket{00}$ which contributes most to $\ket{\psi}$.
By Theorem~\ref{th:num_near}, all nearest neighbors of $\ket{00}$ have the form
	$
  		\frac{\ket{00}+J\ket{ab}}{\sqrt{2}}
	$
where $ab\in\{01,10,11\}$ and $J\in\{\pm 1, \pm i\}$.
Here we maintain global constants, as they make a difference. We also pick a representative 
value $ab=01$ and note that the inner product with $\ket{\psi}$ is maximized by $J=1$.
Let $\ket{r}=(\ket{00}+\ket{01})/\sqrt{2}$. When $\langle\psi\ket{00}=1+\varepsilon < \langle\psi\ket{r}=(2+\varepsilon)/\sqrt{2}$ (up to the same constant), the algorithm 
sets $\ket{s_m}=\ket{r}$.
Nearest neighbors of $\ket{s_m}=\frac{\ket{00}+\ket{01}}{\sqrt{2}}$ are its half-sums 
with its orthogonal stabilizer states, hence we arrive at $\frac{\ket{00}+\ket{01}+\ket{10}+\ket{11}}{2}$.
Consider the $n$-qubit state
	\begin{equation}
 		\ket{\psi_n}=(1+\varepsilon)\ket{0\ldots00}+\ket{0\ldots01}+\ldots+\ket{1\ldots1}
	\end{equation}
Its inner products with $\ket{0\ldots00}$ and the full-superposition state (up to the same constant) 
are $(1+\varepsilon)$ and $(2^n+\varepsilon)/2^{n/2}$, respectively. The full-superposition state 
will be closer as long as $\varepsilon<\frac{2^n-2^{n/2}}{2^{n/2}-1}=2^{n/2}$.
If we start local search at $\ket{0\ldots0}$, it will terminate there when
$\varepsilon > \frac{2-\sqrt{2}}{\sqrt{2}-1}=\sqrt{2}$, regardless of $n$.
Thus, local search stops at a suboptimal stabilizer state when 
$\sqrt{2}<\varepsilon<2^{n/2}$.

Even though constructive approximation by single stabilizer states appears difficult, 
it is important to know if good approximations exist. As we show next, the answer is 
negative for the more general case of approximating by stabilizer superpositions.

\subsection{Approximating arbitrary states with superpositions of stabilizer states} 
\label{sec:stab_evade}

The geometric structure of stabilizer states also motivates the
approximation of arbitrary $n$-qubit unbiased states using superpositions 
of poly$(n)$ stabilizer states. In particular, some optimism 
for obtaining quality approximations using {\em small} superpositions 
is motivated by the count of $n$-qubit stabilizer states from 
Equation \ref{eq:stab_count}~---~the $\Omega(2^{n^2/2})$
growth rate can be contrasted with the $2^n$ growth rate of 
the number of basis states.

To evaluate the quality of approximation by stabilizer states, we employ the quantities
	\begin{equation}
  		\Upsilon=\lim_{n\rightarrow\infty} \inf_{|\psi\rangle} 
  		\max_{|s\rangle \in{\mathcal S}_n} \langle s |\psi \rangle
 		\ \ \ \mathrm{and} \ \ \
  		\Upsilon^{poly}=\sup_{\mathrm{poly}\ p(n)} \lim_{n\rightarrow\infty} 
  		\inf_{|\psi\rangle} \sup_{|s\rangle\in{\mathcal S}_n^{p(n)}} \langle s |\psi \rangle
	\end{equation}
where ${\mathcal S}_n^{p(n)}$ contains superpositions of up to $p(n)$ states from the
set of all ${\mathcal S}_n$ for a given polynomial $p(n)$. For illustration, replace ${\mathcal S}_n$ 
with a set ${\mathcal B}_n$ of $2^n$ orthogonal basis states.
The state $|\psi\rangle=\frac{1}{2^{n/2}} \Sigma_{k=0}^{2^n-1} |k\rangle$ minimizes 
$\max_{s\in{\mathcal B}_n} \langle s |\psi \rangle$ with the value $\frac{1}{2^{n/2}}$. 
Because this value approaches zero at an exponential rate, taking polynomial-sized 
superpositions of basis states (rather than single basis states) will still produce 
the zero limit. Perhaps, this result is not surprising given that 
$\max_{|s_1\rangle \neq |s_2\rangle \in {\mathcal B}_n} \langle s_1 | s_2 \rangle=0$. 
However, since $\max_{|s_1\rangle \neq |s_2\rangle \in {\mathcal S}_n} \langle s_1 | s_2 \rangle=1/\sqrt{2}$, and each stabilizer state $|s_1\rangle \in {\mathcal S}_n$ 
has $2^{n+2}-4$ nearest neighbors $|s_2\rangle \in {\mathcal S}_n$ 
such that $\langle s_1 | s_2 \rangle=1/\sqrt{2}$, one might hope 
that $\Upsilon=1/\sqrt{2}$.

    \begin{lemma} \label{lem:dbl_cofactors}
        Given an $n$-qubit stabilizer state $|s\rangle$ and two qubits $q$ and $r$,
        supports and norms of all non-zero double cofactors $|s_{qr}\rangle$  are equal.
        In particular, since all supports are powers of two, the number of double cofactors
        with non-zero support cannot be three.
    \end{lemma}
    \noindent
	{\bf Proof.} 
        The claim is trivial when only one double cofactor is non-zero. When only two are 
        non-zero, they may originate from one single-qubit cofactor or from two different 
        single-qubit cofactors. By Corollary~\ref{cor:stab_props}-{\em iv}, 
        the supports and norms of cofactors must be equal. When all four double cofactors 
        are non-zero, Corollary~\ref{cor:stab_props}-{\em iv} implies that 
        their support must be one fourth of that of the initial state. Additionally, 
        the orthogonality of cofactors implies that their norm must be one half of 
        the original norm. Now we show that the case of exactly three non-zero double 
        cofactors is impossible. Without the loss of generality, assume that 
        $|\psi_{qr=11} \rangle=0$, but other cofactors are $\neq 0$. 
        Then single cofactors must have equal support. Therefore 
        $|\psi_{r=1}|=|\psi_{qr=01}|=|\psi_{qr=00}|+|\psi_{qr=10}|=|\psi_{r=0}|=2|\psi_{qr=10}|$, 
        but also $|\psi_{q=1}|=|\psi_{qr=10}|=|\psi_{qr=00}|+|\psi_{qr=01}|=|\psi_{q=0}|=2|\psi_{qr=01}|$. 
        Thus $|\psi_{qr=01}|=4|\psi_{qr=01}|$, which contradicts $|\psi_{qr=01}|\neq 0$.
    \square
    
Lemma~\ref{lem:dbl_cofactors} is illustrated in Table~\ref{tab:two_qbssts} and Appendix~A.

    \begin{theorem} \label{th:stab_evade}
    		$\Upsilon = \Upsilon^{poly} = 0$. 
    \end{theorem}
	\noindent
	{\bf Proof.} Consider the family of $2n$-qubit unbiased states 
    	$\mu^{2n}=\big(\frac{|00\rangle+|01\rangle+|10\rangle}{\sqrt{3}}\big)^{\otimes n}$, 
    	which are not stabilizer states per Lemma~\ref{lem:dbl_cofactors}.
	For an arbitrary stabilizer state $|s\rangle$, double cofactoring 
	over $q=2n-1$ and $r=2n-2$ yields
    \begin{equation}
        \langle s |\mu^{2n} \rangle = \langle s_{qr=00} |\mu^{2n}_{qr=00} \rangle 
        + \langle s_{qr=01} |\mu^{2n}_{qr=01} \rangle
        + \langle s_{qr=10} |\mu^{2n}_{qr=10} \rangle
    \end{equation}
	because $|\mu^{2n}_{qr=11}\rangle=0$. We can upper-bound this expression by over-estimating 
	each non-zero term with $\langle v | w \rangle \leq ||v||\cdot ||w||$. To this end, 
	$||\mu^{2n}_{qr}||=\frac{1}{\sqrt{3}}$, while the norm of (orthogonal) double cofactors 
	of $|s\rangle$ depends on how many of them are non-zeros~---~one, two or four. In these 
	cases, the norms are $1$, $\frac{\sqrt{2}}{2}$ and $\frac{1}{2}$, respectively,
	yielding upper bounds $\langle s |\mu^{2n} \rangle \leq \alpha=\frac{\sqrt{3}}{3}$, 
	$\frac{\sqrt{6}}{3}$ and $\frac{\sqrt{3}}{2}$. Therefore, cumulatively, 
	$\langle s |\mu^{2n} \rangle \leq \frac{\sqrt{3}}{2}$.

	More accurate bounds can be obtained by rewriting the same three terms using
	$$\langle v | w \rangle = ||v||\cdot ||w|| \frac{\braket{v}{w}}{||v||\cdot||w||},$$
	then observing that $ \langle v00 | w00 \rangle = \langle v | w \rangle $,
	$ \langle v01 | w01 \rangle = \langle v | w \rangle $, and 
	$ \langle v10 | w10 \rangle = \langle v | w \rangle $.
	In particular, $\sqrt{3}|\mu^{2n}_{qr=00}\rangle$,  $\sqrt{3}|\mu^{2n}_{qr=01}\rangle$ and  $\sqrt{3}|			\mu^{2n}_{qr=10}\rangle$ can be replaced by $|\mu^{2n-2} \rangle$, noting that
 	$|\mu^{2n}\rangle$ is separable. Then the cofactors of $|s\rangle$
	can be relaxed to a best-case $(2n-2)$-qubit stabilizer state $s'$ to obtain
    \begin{equation}
        \langle s |\mu^{2n} \rangle \leq \frac{\sqrt{3}}{2} \langle s' |\mu^{2n-2} \rangle
        \leq \left(\frac{\sqrt{3}}{2}\right)^n
    \end{equation}
	\noindent
	Therefore
    \begin{equation}
        \Upsilon=\lim_{n\rightarrow\infty} \inf_{|\psi\rangle} 
        \max_{|s\rangle \in{\mathcal S}_{2n}} \langle s |\psi \rangle=
        \lim_{n\rightarrow\infty} \left(\frac{\sqrt{3}}{2}\right)^n = 0
    \end{equation}
	\noindent
	For any polynomial $p(n)$,
    \begin{equation}
        \lim_{n\rightarrow\infty} \inf_{|\psi\rangle} \sup_{|s\rangle\in{\mathcal S}_{2n}^{p(n)}} 
        \langle s |\psi \rangle =\lim_{n\rightarrow\infty} p(n)\left(\frac{\sqrt{3}}{2}\right)^n = 0
    \end{equation}
	\noindent
	Therefore $\Upsilon^{poly}=0$ as well.
	\square

Theorem~\ref{th:stab_evade} is somewhat surprising because no $n$-qubit state can be 
orthogonal to the set of all stabilizer states, which contains many basis sets. 
However, they establish {\em asymptotic orthogonality}, which can be viewed as an 
infinite-dimensional phenomenon. Such families of states that are 
asymptotically orthogonal to all stabilizer states can be neither represented 
nor approximated by polynomial-sized superpositions of stabilizer states. 

Our construction of {\em stabilizer-evading} states can be modified to yield
other {\em separable} states with similar properties, e.g.,
$\bar\mu^{3n}=\big(\frac{|001\rangle+|010\rangle+|100\rangle}{\sqrt{3}}\big)^{\otimes n}$.
Superpositions such as $(\mu^{6n}+\bar\mu^{6n})/\sqrt{2}$ offer {\em entangled} states
that cannot be approximated by polynomial-sized superpositions of stabilizer states.

The above results along with the results from Section~\ref{sec:locsrch}
suggest that there are wide gaps between stabilizer states.  The
following proposition quantifies the size of such gaps.

	\begin{proposition} \label{prop:stabgap}
		Consider $2^n$-dimensional balls centered at a point
  		on the unit sphere that do not contain any $n$-qubit stabilizer states
  		in their interior.
  		The radius of such balls cannot exceed $\sqrt{2}$,
   		but approaches $\sqrt{2}$ as $n\rightarrow\infty$.
	\end{proposition}
	\noindent
	{\bf Proof sketch.} Consider an arbitrary ball $\mathbf{B}$ with radius $\sqrt{2}$ 
	and centered on the unit sphere as shown in Figure~\ref{fig:sphere}.
	$\mathbf{B}$ covers half of the unit sphere. Let $\ket{s}$ be 
	a stabilizer state that does not lie on the boundary of $\mathbf{B}$.
	Then, either $\ket{s}$ or -$\ket{s}$ is inside $\mathbf{B}$.
	Furthermore, observe that the intersection of the unit sphere and 
	the boundary of $B$ is contained in a hyperplane of dimension $n-1$. 
	If all stabilizer states were contained there, they would not have 
	included a single basis set. However, since stabilizer states
	contain the computational basis, they cannot all be contained in that intersection. 
	An asymptotic lower bound for the radius of $B$ is obtained
	using the family of {\em stabilizer-evading}
	states defined in Theorem~\ref{th:stab_evade}. These states are 
	asymptotically orthogonal to all $n$-qubit stabilizer states 
	($n\rightarrow\infty$). Therefore, the distance to their closest 
	stabilizer state approaches $\sqrt{2}$.
	\square

	\begin{figure}[!t]
		\centering\LARGE
		\scalebox{.42}[.42]{\tdplotsetmaincoords{60}{50}
\usetikzlibrary{3d}
\usetikzlibrary{calc}

\begin{tikzpicture}[scale=5,tdplot_main_coords]

\def\R{1}
\def\sqr2{1.41421}

\draw [dotted] (-\R,0,0) -- (1-\sqr2,0,0);
\draw [thick] (1-\sqr2,0,0) -- (0,0,0);
\draw [thick, blue] (0,0,0) -- (\R,0,0);
\draw [thick] (\R,0,0) -- (\R+\sqr2,0,0);
\draw [thick, blue] (0,-\R,0) -- (0,\R,0);
\draw [thick, blue] (0,0,0) -- (0,0,\R);
\draw [thick, blue] (0,0,\R) -- (\R,0,0);
\draw [thick, blue] (0,-\R,0) -- (\R,0,0);
\draw [thick, blue] (0,\R,0) -- (\R,0,0);
\draw [dotted] plot [domain=0:360, samples=60, variable=\i] 
    (\R*cos \i, \R*sin \i, 0) -- cycle;

\node (unitya) at (.4,-.1,0) {$\mathbf{1}$};
\node (unityb) at (-.1,0,.4) {$\mathbf{1}$};
\node (root2a) at (\R+\sqr2/2,-.15,0) {$\mathbf{\sqrt{2}}$};
\node (root2b) at (-.1,.5,.25) {$\mathbf{\sqrt{2}}$};
\node (usphere) at (0,-1,1.5) {{\bf Unit sphere}};
\node (ssphere) at (0,2.35,0) {$\mathbf{B}$};

\foreach \i in {90}
    \draw [thick, blue] plot [domain=-90:90, samples=60, variable=\j]        
        (\R*cos \i*sin \j,\R*sin \i*sin \j, \R*cos \j);

\foreach \i in {0, 30,...,150}
    \draw [dotted, thick, blue] plot [domain=-90:90, samples=60, variable=\j]        
        (\R*cos \i*sin \j,\R*sin \i*sin \j, \R*cos \j);

\foreach \j in {0, 15,...,90}
    \draw [dotted, thick, blue] plot [domain=0:360, samples=60, variable=\i]  
         (\R*cos \i*sin \j,\R*sin \i*sin \j, \R*cos \j);
        
\draw [thick] (\R,0,0) circle (\sqr2);
\shade[ball color=blue!10!white,opacity=0.20] (\R,0,0) circle (\sqr2);

\foreach \i in {0}
    \draw [thick] plot [domain=-90:90, samples=60, variable=\j]        
        (\sqr2*cos \i*sin \j + \R,\sqr2*sin \i*sin \j,\sqr2*cos \j);

\foreach \i in {45, 90, ..., 150}
    \draw [thick, dotted] plot [domain=-90:90, samples=60, variable=\j]        
        (\sqr2*cos \i*sin \j + \R,\sqr2*sin \i*sin \j,\sqr2*cos \j);
        

\end{tikzpicture}}
		\fcaption{\label{fig:sphere} A ball $\mathbf{B}$ with radius $\sqrt{2}$ 
		centered on the unit sphere covers half of the unit sphere. 
		Every such ball contains at least one stabilizer state 
		(in most cases, half of all stabilizer states). 
		\vspace{-.25cm}
		}
	\end{figure}

\section{Computational Geometry of Stabilizer States}  
\label{sec:inprod_stab}

Section~\ref{sec:embedding} illustrates how straightforward approaches to 
several natural geometric tasks fail. In this section, we focus on more specialized, 
but no less useful tasks and develop computational techniques that 
we successfully implemented in software and evaluate in Section~\ref{sec:empirical}. 
In Section~\ref{sec:synthesis}, we discuss our algorithm for synthesizing new 
{\em canonical stabilizer circuits}.
In Section~\ref{sec:ipalgo}, we turn our attention to the computation of inner products 
of stabilizer states and, by extension, their linear combinations. As Gram-Schmidt 
orthogonalization cannot be used directly with stabilizer states, we develop an 
alternative approach in Section~\ref{sec:stab_ortho}.
The efficient computation of generator sets for stabilizer bivectors is
discussed in Section~\ref{sec:bivec_comp}.

\subsection{Synthesis of canonical stabilizer circuits}
\label{sec:synthesis}

A crucial step in the inner-product computation for stabilizer states (Theorem \ref{th:inprod_aron}) 
is the synthesis of a stabilizer circuit that brings an $n$-qubit stabilizer state $\ket{\psi}$ to a computational basis state $\ket{b}$.
Consider a stabilizer matrix $\cM$ that uniquely identifies $\ket{\psi}$. $\cM$ is reduced 
to \normform\ (Definition~\ref{def:basis_form}) by applying a series of elementary row and column 
operations. Recall that row operations
(transposition and multiplication) do not modify the state, but column (Clifford) operations do.
Thus, the column operations involved in the reduction process constitute a unitary stabilizer
circuit $\cC$ such that $\cC\ket{\psi} = \ket{b}$, where $\ket{b}$ is a basis state. 
Algorithm~\ref{alg:inprod_circ} reduces an input matrix $\cM$ to \normform\
and returns such a circuit $\cC$. 

	\begin{definition} {\em 
		Given a finite sequence of quantum gates,
		a {\em circuit template} describes a segmentation
		of the circuit into blocks where each block
		uses only one gate type.
		The blocks must match the sequence and be
  		concatenated in that order. For example,
  		a circuit satisfying the $H$-$C$-$P$ template starts with
  		a block of Hadamard ($H$) gates, followed by a block of
  		CNOT ($C$) gates, followed by a block of Phase ($P$) gates. }
	\end{definition}

	\begin{definition} {\em 
		A circuit with a {\em template structure} consisting
		entirely of stabilizer-gate blocks is called
		a {\em canonical stabilizer circuit}.}
	\end{definition}

Canonical forms are useful for synthesizing stabilizer circuits
that minimize the number of gates and qubits required to produce a
particular computation. This is particularly important in the context
of quantum fault-tolerant architectures that are based on stabilizer codes. 
The work in \cite{AaronGottes} establishes 
a $7$-block\footnote{Theorem 8 in~\cite{AaronGottes}
actually describes an $11$-step procedure to obtain such circuits. However, the last four steps
are used to reduce destabilizer rows, which we do not consider here.}~
canonical-circuit template with the sequence $H$-$C$-$P$-$C$-$P$-$C$-$H$. 
Furthermore, the work in~\cite{Vanden} proves the existence of 
canonical circuits with the shorter sequence $H$-$C$-$X$-$P$-$CZ$, where the $X$
and $CZ$ blocks consist of NOT and Controlled-$Z$ (CPHASE) gates, respectively. 
However, the synthesis algorithm sketched in~\cite{Vanden} employs the 
$\zfield_2$-representation for states (Theorem~\ref{th:stab_form})
rather than the stabilizer formalism. Thus, no detailed algorithms are known 
for obtaining such canonical circuits {\em given an arbitrary generator set}.
Algorithm~\ref{alg:inprod_circ} takes as input a stabilizer matrix $\cM$ and synthesizes 
a $5$-block canonical circuit with template $H$-$C$-$CZ$-$P$-$H$ (Figure~\ref{fig:basiscirc}).
	\begin{figure}[!t]
		\centering
		\hspace{-2cm}
		\scalebox{.6}[.6]{
			\begin{minipage}[b]{.5\linewidth}
				\large \input{basiscirc}
			\end{minipage}
		} 
		\fcaption{\label{fig:basiscirc} Template structure for the basis-normalization 
		circuit synthesized by Algorithm~\ref{alg:inprod_circ}. The input is an arbitrary
		stabilizer state $\ket{\psi}$ while the output is a basis state 
		$\ket{b_1b_2\ldots b_n\in \{0,1\}^n}$.
		}
		\vspace{-.2cm}
    \end{figure}
We now describe the main steps in the algorithm.
The updates to the phase vector under row/column operations are 
left out of the discussion as such updates do not affect the overall execution 
of the algorithm.

	\begin{itemize}
		\vspace{-2mm}
		\item[{\bf 1.}] Reduce $\cM$ to canonical form.
		\vspace{-2mm}
		\item[{\bf 2.}] Use row transposition to diagonalize $\cM$. For $j \in \{1, \ldots, n\}$, 
		if the diagonal literal $\cM_{j, j} = Z$ and there are other Pauli (non-$I$) 
		literals in the row (qubit is entangled), conjugate $\cM$ by H$_j$. 
		Elements below the diagonal are $Z/I$ literals. 
		\vspace{-2mm}
		\item[{\bf 3.}] For each above-diagonal element $\cM_{j, k} = X/Y$, 
		conjugate by CNOT$_{j,k}$. Elements above the diagonal are now $I/Z$ literals.
		\vspace{-2mm}
		\item[{\bf 4.}] For each above-diagonal element $\cM_{j, k} = Z$, 
		conjugate by CPHASE$_{j,k}$. Elements above the diagonal are now $I$ literals.
		\vspace{-2mm}
		\item[{\bf 5.}] For each diagonal literal $\cM_{j, j} = Y$, conjugate by P$_j$.
		\vspace{-2mm}
		\item[{\bf 6.}] For each diagonal literal $\cM_{j, j} = X$, conjugate by H$_j$.
		\vspace{-2mm}
	\end{itemize}
	
\newcommand{\apply}{\mathrm{\tt CONJ}}
\newcommand{\gauss}{\mathrm{\tt GAUSS}}

    \begin{algorithm}[!t]
         \caption{Synthesis of a basis-normalization circuit}
         \footnotesize
         \label{alg:inprod_circ}
         \begin{algorithmic}[1]
            \Require Stabilizer matrix $\cM$ for $S(\ket{\psi})$ with rows $R_1,\ldots,R_n$
            \Ensure ({\em i}) Unitary stabilizer circuit $\cC$ such that $\cC\ket{\psi}$
            equals basis state $\ket{b}$, and ({\em ii}) reduce $\cM$ to \normform 
            \Statex \hspace{-5mm} $\Rightarrow$ $\gauss(\cM)$ reduces $\cM$ to canonical form 
            (Figure \ref{fig:sminv})
            \Statex \hspace{-5mm} $\Rightarrow$ $\rswap(\cM, i, j)$ swaps rows $R_i$ and $R_j$ of $\cM$
            \Statex \hspace{-5mm} $\Rightarrow$ $\rmult(\cM, i, j)$ left-multiplies rows $R_i$ 
            and $R_j$, returns updated $R_i$
            \Statex \hspace{-5mm} $\Rightarrow$ $\apply(\cM, \alpha_j)$ conjugates $j^{th}$ 
            column of $\cM$ by Clifford sequence $\alpha$
            \Statex
            \State $\gauss(\cM)$ \Comment{Set $\cM$ to canonical form}
            \State $\cC \leftarrow \emptyset$   
            \For{$j \in \{1, \dots, n\}$} 
            \Comment{Apply block of Hadamard gates}
            	\State $k \leftarrow$ index of row $R_{k \in \{j,\ldots, n\}}$ with $j^{th}$ 
            	literal set to $X$ or $Y$
            	\If{$k$ {\bf exists}}
            		\State $\rswap(\cM, j, k)$
            	\Else
            		\State $k_2 \leftarrow$ index of {\em last} row 
            		$R_{k_2 \in \{j,\ldots, n\}}$ with $j^{th}$ literal set to $Z$
            		\If{$k_2$ {\bf exists}}
            			\State $\rswap(\cM, j, k_2)$
            			\If{$R_{j}$ has $X$, $Y$ or $Z$ literals in columns $\{j+1, \ldots, n\}$}
            				\State $\apply(\cM, \text{H}_j)$
            				\State $\cC \leftarrow \cC \cup \text{H}_j$
            			\EndIf
           		\EndIf
           	\EndIf
            \EndFor
            \For{$j \in \{1, \dots, n\}$} \Comment{Apply block of CNOT gates}
            		 \For{$k \in \{j+1, \dots, n\}$} 
            		 	\If{$k^{th}$ literal of row $R_j$ is set to $X$ or $Y$}
            				\State $\apply(\cM, \text{CNOT}_{j, k})$
            				\State $\cC \leftarrow \cC \cup \text{CNOT}_{j, k}$
            			\EndIf
           		 \EndFor
            \EndFor
            \For{$j \in \{1, \dots, n\}$} 
            \Comment{Apply a block of Controlled-$Z$ gates}
            		\For{$k \in \{j+1, \dots, n\}$} 
            			\If{$k^{th}$ literal of row $R_j$ is set to $Z$}
            				\State $\apply(\cM, \text{CPHASE}_{j, k})$ 
            				\State $\cC \leftarrow \cC \cup \text{CPHASE}_{j, k}$
            			\EndIf
           		\EndFor
            \EndFor
            \For{$j \in \{1, \dots, n\}$} \Comment{Apply block of Phase gates}
            		\If{$j^{th}$ literal of row $R_j$ is set to $Y$}
            			\State $\apply(\cM, \text{P}_j)$
            			\State $\cC \leftarrow \cC \cup \text{P}_j$
            		\EndIf
            \EndFor
            \For{$j \in \{1, \dots, n\}$} \Comment{Apply block of Hadamard gates}
            		\If{$j^{th}$ literal of row $R_j$ is set to $X$}
            			\State $\apply(\cM, \text{H}_j)$
            			\State $\cC \leftarrow \cC \cup \text{H}_j$
            		\EndIf
            \EndFor
            \State \Return $\cC$
        \end{algorithmic}
    \end{algorithm}

	\begin{proposition} \label{prop:normform_circsize}
		For an $n\times n$ stabilizer matrix $\cM$, the number of 
		gates in the circuit $\cC$ returned by Algorithm~\ref{alg:inprod_circ} 
		is $O(n^2)$.
	\end{proposition}
	\noindent
	{\bf Proof.} 
		The number of gates in $\cC$ is dominated by the CPHASE block, 
		which consists of $O(n^2)$ gates. This agrees with previous
		results regarding the number of gates needed for an $n$-qubit 
		stabilizer circuit in the worst case \cite{Cleve, DehaeDemoor}.
	\square 
	
Observe that, for each gate added to $\cC$, the corresponding
column operation is applied to $\cM$. Since column 
operations run in $\Theta(n)$ time, it follows from Proposition~\ref{prop:normform_circsize}
that the runtime of Algorithm \ref{alg:inprod_circ} is $O(n^3)$.
Audenaert and Plenio~\cite{Audenaert} described an algorithm to 
compute the fidelity between two mixed stabilizer states. 
Similar to our use of basis-normalizing
circuits, the approach from \cite{Audenaert} relies on a stabilizer circuit
to map a stabilizer state to a normal form where all basis states
have non-zero amplitudes. 
However, the circuits generated by the Audenaert-Plenio algorithm 
exhibit two disadvantages: ($i$) they are not canonical 
and ($ii$) they have, on average, twice as many gates as 
our basis-normalization circuits (Section~\ref{sec:empirical}). 

We note that our canonical stabilizer circuits can be optimized using the 
techniques from~\cite{Moore}, which describes how to restructure circuits 
to facilitate parallel quantum computation. The authors show that such 
parallelization is possible for circuits consisting of H and CNOT gates, 
and for diagonal operators. Since CPHASE gates are diagonal, one can apply the techniques 
from~\cite{Moore} to parallelize our canonical circuits and produce equivalent circuits with 
$O(n^2)$ gates and parallel depth $O(\log n)$. 
Canonical stabilizer circuits that follow the $7$-block template structure from \cite{AaronGottes}
can be optimized to obtain a tighter bound on the number of gates. As in 
our approach, such circuits are dominated by the size of the CNOT blocks,
which contain $O(n^2)$ gates. The work in \cite{PatelMarkov} shows that 
any CNOT circuit has an equivalent CNOT circuit with $O(n^2/\log n)$ gates.
Thus, one simply applies such techniques to each of the CNOT blocks in the
canonical circuit. It is an open problem whether one can apply some variation
of the same techniques to CPHASE blocks, which would 
facilitate similar optimizations for our $5$-block canonical form. 

\subsection{An inner-product algorithm} 
\label{sec:ipalgo}

Let $\ket{\psi}$ and $\ket{\phi}$ be two stabilizer states represented by stabilizer matrices
$\cM^\psi$ and $\cM^\phi$, respectively. Our approach for computing the inner product between these two
states is shown in Algorithm~\ref{alg:inprod}. Following the proof of Theorem \ref{th:inprod_aron},
Algorithm \ref{alg:inprod_circ} is applied to $\cM^\psi$ in order to reduce it to \normform.
The stabilizer circuit generated by Algorithm~\ref{alg:inprod_circ} is then applied
to $\cM^\phi$ in order to preserve the inner product. Then, we minimize the number of $X$ 
and $Y$ literals in $\cM^\phi$ by applying Algorithm~\ref{alg:gauss_min}. 
Lastly, each generator in $\cM^\phi$ that anticommutes with 
$\cM^\psi$ (since $\cM^\psi$ is in \normform, we only
need to check which generators in $\cM^\phi$ have $X$
or $Y$ literals) contributes a factor of $1/\sqrt{2}$ 
to the inner product. If a generator in $\cM^\phi$, say $Q_i$, commutes with $\cM^\psi$, then we
check orthogonality by determining whether $Q_i$ is in the stabilizer group generated
by $\cM^\psi$. This is accomplished by multiplying the appropriate generators in $\cM^\psi$ such that
we create Pauli operator $R$, which has the same literals as $Q_i$,
and check whether $R$ has an opposite sign to $Q_i$. If this is the case, then, by
Theorem \ref{th:stab_ortho}, the states are orthogonal. 
The bottleneck of Algorithm \ref{alg:inprod} is
the call to Algorithm \ref{alg:inprod_circ}, and the overall runtime is $O(n^3)$.
As Section \ref{sec:empirical} shows, the performance of 
our algorithm is sensitive to the input stabilizer matrix
and takes $O(n^2)$ time in important cases.

\newcommand{\norm}{\mathrm{\tt BASISNORMCIRC}}
\newcommand{\lmult}{\mathrm{\tt LEFTMULT}}
	
        \begin{algorithm}[!t]
            \caption{Inner product for stabilizer states}
            \label{alg:inprod} \footnotesize
            \begin{algorithmic}[1]
                \Require Stabilizer matrices ({\em i}) $\cM^\psi$ for $\ket{\psi}$ with rows $P_1,\ldots,P_n$,
                and ({\em ii}) $\cM^\phi$ for $\ket{\phi}$ with rows $Q_1,\ldots,Q_n$
                \Ensure Inner product between $\ket{\psi}$ and $\ket{\phi}$
                \Statex \hspace{-5mm} $\Rightarrow$ $\norm(\cM^\psi)$ reduces $\cM$ to \normform, 
                i.e, $\cC\ket{\psi}=\ket{b}$, where $\ket{b}$ is a basis state, and returns $\cC$
                \Statex \hspace{-5mm} $\Rightarrow$ $\apply(\cM, \cC)$ conjugates $\cM$ 
                by Clifford circuit $\cC$
                \Statex \hspace{-5mm} $\Rightarrow$ $\gauss(\cM)$ reduces $\cM$ to canonical form 
                (Figure \ref{fig:sminv})
                \Statex \hspace{-5mm} $\Rightarrow$ $\lmult(P, Q)$ left-multiplies Pauli operators $P$ and $Q$, 
                and returns the updated $Q$
                \Statex
                \State $\cC \leftarrow \norm(\cM^\psi)$
                \Comment{Apply Algorithm \ref{alg:inprod_circ} to $\cM^\psi$}  
                	\State $\apply(\cM^\phi, \cC)$  \Comment{Compute $\cC\ket{\phi}$}
                \State $\gauss(\cM^\phi)$ \Comment{Set $\cM^\phi$ to canonical form}
                \State $k \leftarrow 0$
                \For{{\bf each} row $Q_i \in \cM^\phi$}
                	\If{$Q_i$ contains $X$ or $Y$ literals}
	              		\State $k \leftarrow k + 1$
	              	\Else\Comment{Check orthogonality, i.e., $Q_i \notin S(\ket{b})$}
	              		\State $R \leftarrow I^{\otimes n}$
	              		\For{{\bf each} $Z$ literal in $Q_i$ found at position $j$}
	              			\State $R \leftarrow \lmult(P_j, R)$
		              	\EndFor
		              	\If{$R = -Q_i$}            	
		              		\State\Return 0 \Comment{By Theorem \ref{th:stab_ortho}}		  
		              	\EndIf
	              	\EndIf
                \EndFor
                \State\Return $2^{-k/2}$ \Comment{By Theorem \ref{th:inprod_aron}}
            \end{algorithmic}
        \end{algorithm}

\vspace{-.2cm}\ \\
\noindent
{\bf Complex-valued inner product}. Recall that Algorithm~\ref{alg:inprod}
computes the {\em real-valued} (phase-normalized) inner product $r = e^{-i\theta}\braket{\psi}{\phi}$.
To compute the {\em complex-valued} inner product, one needs to additionally calculate
$e^{-i\theta}$. This is accomplished by modifying Algorithm~\ref{alg:inprod} 
such that it maintains the global phases generated when computing $\cC\ket{\psi}$ and $\cC\ket{\phi}$, 
where $\cC$ is the basis-normalization circuit from Algorithm \ref{alg:inprod_circ}.
Let $\alpha$ and $\beta$ be the global phases of $\cC\ket{\psi}$ and $\cC\ket{\phi}$,
respectively. The complex-valued inner product is computed as $\alpha^*\cdot\beta\cdot 2^{-k/2}$. 
Since stabilizer matrices represent states up to a global phase (i.e., states are
normalized such that the first non-zero basis amplitude is $1.0$), 
the functions $\norm(\cM)$ and $\apply(\cM, \cC)$ 
need to be modified to keep track of the global-phase factors generated when
each gate in $\cC$ is applied to the input state. Let $U$ be a stabilizer gate
applied to the state $\ket{\psi}$ represented by stabilizer matrix $\cM$. The following 
process computes the global phase of $U\ket{\psi}$:

	\begin{itemize}
		\vspace{-2mm}
		\item[{\bf 1.}] Use Gaussian elimination to obtain a basis state 
		$\ket{b}$ from $\cM$ (Observation~\ref{obs:stabst_amps}) and store 
		its non-zero amplitude $\alpha$. If $U$ is the Hadamard gate, it
		may be necessary to sample a sum of two non-zero (one real, one 
		imaginary) basis amplitudes (see Example~\ref{ex:stabgph}).
		\vspace{-2mm}
		\item[{\bf 2.}] Compute $\alpha U\ket{b}=\alpha'\ket{b'}$ directly 
		using the state-vector representation.
		\vspace{-2mm}
		\item[{\bf 3.}] Obtain $\ket{b'}$ from $U\cM U^\dag$ and 
		store its non-zero amplitude $\beta$. 
		\vspace{-2mm}
		\item[{\bf 4.}] The phase factor generated is $\alpha'/\beta$.
		\vspace{-2mm}
	\end{itemize}
	
	\begin{example} \label{ex:stabgph} {\em 
		Suppose $\ket{\psi} = \ket{00}+\ket{01}-i\ket{10}-i\ket{11}$,
		where we omit the normalization factor for clarity. A generator
		set for $\ket{\psi}$ is $\cM = \{-YI, IX\}$. We will compute the 
		global-phase factor generated when a Hadamard gate is applied to 
		the first qubit. Following Step~1, we obtain basis states $\ket{00}$ 
		and $i\ket{10}$ from $\cM$, and set $\alpha = 1$ (the amplitude of $\ket{00}$). 
		Next, we compute $H_1(\ket{00}+i\ket{10}) = \frac{1-i}{\sqrt{2}}\ket{00}+\frac{1+i}{\sqrt{2}}\ket{10}$ 
		and set $\alpha' = \frac{1-i}{\sqrt{2}}$ (Step 2). According to Step~3, we obtain 
		the $\ket{00}$ amplitude from $H_1\cM H_1^\dag = \{YI,IX\}$, which gives $\beta = 1$. 
		The global-phase factor is $\alpha'/\beta = \frac{1-i}{\sqrt{2}}$. One
		can obtain the factor generated when CNOT, Phase and measurements gates 
		are applied in a similar fashion. However, for such gates, only one 
		basis-state amplitude needs to be sampled in Step~1.
	}
	\end{example}

\subsection{Orthogonalization of stabilizer states} 
\label{sec:stab_ortho}

We now shift our focus to the task of orthogonalizing 
a linear combination of stabilizer states 
%
$\ket{\Psi} = \sum_{j=1}^N c_j\ket{\psi_j}$, where each $\ket{\psi_j}$ is represented
by its own stabilizer matrix. To simulate measurements of $\ket{\Psi}$, it is helpful
to transform the set of states that define $\ket{\Psi}$ into an orthogonal 
set. Since a linear combination of stabilizer states 
is usually not a stabilizer state, Gram-Schmidt orthogonalization cannot be 
used directly. Therefore, we develop an orthogonalization procedure 
that exploits the nearest-neighbor structure
of stabilizer states (Section~\ref{sec:kneighbors}) and their efficient manipulation 
via stabilizers.

	\begin{proposition} \label{prop:nneardecomp}
		Let $\ket{\psi}$ be a state represented by $\cM^\psi$,
		which contains at least one row with at least one $X$ or $Y$ literal. 
		Then $\ket{\psi}$ can be decomposed into a superposition
		$\frac{|\phi\rangle+i^l|\varphi\rangle}{\sqrt{2}}$,
		where $\ket{\phi}$ and $\ket{\varphi}$ are nearest neighbors 
		of $\ket{\psi}$ 	whose matrices are similar to each other.
	\end{proposition}
	\noindent
	{\bf Proof.} 
		Let $R_j$ be a row in $\cM^\psi$ with an $X/Y$ literal
		in its $j^{th}$ position, and let $Z_j$ be a Pauli operator with 
		a $Z$ literal in its $j^{th}$ position and $I$ everywhere else. 
		Observe that $R_j$ and $Z_j$ anticommute. If any other rows in $\cM^\psi$ 
		anticommute with $Z_j$, multiply them by $R_j$ to make them commute
		with $Z_j$. Let $\cM^\phi$ and $\cM^\varphi$ be the matrices obtained 
		by replacing row $R_j$ in $\cM^\psi$ with $Z_j$ and -$Z_j$, respectively. 
		This operation is equivalent to applying $(\pm Z_j)$-measurement projectors 
		to $\ket{\psi}$. Thus, $\ket{\phi}\equiv \frac{(I+Z_j)\ket{\psi}}{\sqrt{2}}$
		and $\ket{\varphi}\equiv \frac{(I-Z_j)\ket{\psi}}{\sqrt{2}}$,
		such that $|\braket{\psi}{\phi}|=|\braket{\psi}{\varphi}|=1/\sqrt{2}$. 
		The matrices $\cM^\phi$ and $\cM^\varphi$ are similar since $\cM^\varphi=X_j(\cM^\phi)X_j^\dag$ 
		(and $\cM^\phi=X_j(\cM^\varphi)X_j^\dag$), where $X_j$ is a Pauli 
		operator with an $X$ literal in its $j^{th}$ position 
		and $I$ everywhere else. This implies that the $j^{th}$ 
		qubit in $\ket{\phi}$ and $\ket{\varphi}$ is in the deterministic state 
		$\ket{0}$ and $\ket{1}$, respectively. ($\ket{\phi}$ and $\ket{\varphi}$
		are cofactors of $\ket{\psi}$ with respect to qubit $j$.)
		To produce $\frac{\ket{\phi}+i^l\ket{\varphi}}{\sqrt{2}}$, we need the 
		$i^l$ factor, which is not maintained by the stabilizer $\cM^\varphi$.
		This factor can be obtained by the same global-phase computation procedure 
		described in Section~\ref{sec:ipalgo} in the context of complex-valued 
		inner products.
	\square
	
\newcommand{\mlst}{\mathbf{M}}
\newcommand{\clst}{\mathbf{C}}

Algorithm~\ref{alg:stabortho} takes as input a linear combination of $n$-qubit states 
$\ket{\Psi}$ represented by a pair consisting of: ($i$)~a list of {\em canonical stabilizer
matrices} $\mlst=\{\cM^{1},\ldots,\cM^{N}\}$ and ($ii$)~a list of coefficients 
$\clst=\{c_1,\ldots,c_N\}$. The algorithm
iteratively applies the decomposition procedure described in the proof of 
Proposition~\ref{prop:nneardecomp} until all the matrices in $\mlst$ are {\em similar}
(Definition~\ref{def:mtxeq}) to each other. At each iteration, the algorithm
selects a pivot qubit based on the composition of Pauli literals in the corresponding
column. The states in the linear combination are decomposed with respect to the pivot
only if there exists a pair of matrices in $\mlst$ that contain different types of 
Pauli literals in the pivot column. Since similar matrices contain 
distinct phase vectors, the states represented by the modified list of matrices are 
mutually orthogonal (Theorem~\ref{th:stab_ortho}). The algorithm maintains 
the invariant that the sum of stabilizer states (represented by the pair $\mlst$ and $\clst$) 
equals the (non-stabilizer) vector $\ket{\Psi}$.
Observe that, for each pivot qubit that satisfies $b=1$, the size of $\mlst$
doubles. Let $m$ be the maximum value in $\{m_1,\ldots,m_N\}$, where 
$m_j$ is the number of rows with $X/Y$ literals in matrix $\cM^j$. 
An orthogonalization approach that obtains $\ket{\Psi}$ in the computational
basis (i.e., calculating computational-basis amplitudes of each $\ket{\psi_j}$ and
summing them with weights $c_j$) requires exactly $2^m$ terms. In contrast, Algorithm~\ref{alg:stabortho} 
expands $\mlst$ to $2^k$ terms, where $k\leq m$. In particular, if the
matrices in $\mlst$ are ``close'' to similar (Definition~\ref{def:mtxeq}), 
then $k<m$ and thus Algorithm~\ref{alg:stabortho} provides an advantage over
direct computation of basis amplitudes. 

	\begin{example}\label{ex:stabortho} {\em
		Let $\mlst=\left\{\cM^1, \cM^2\right\}$, where $\cM^1$ and $\cM^2$
		are $n$-qubit matrices. Suppose $\cM^1$ has $X$ literals along its 
		diagonal and $I$ literals everywhere else. Similarly, assume $\cM^2$
		follows the same diagonal structure, but replaces an $X$
		at diagonal position $p$ with $Y$. Algorithm~\ref{alg:stabortho}
		decomposes both matrices over the selected pivot qubit $p$ using
		Proposition~\ref{prop:nneardecomp}, which yields a list of two similar matrices.
		In contrast, calculating the basis amplitudes of the states
		represented by $\cM^1$ and $\cM^2$ would require summing over $2^n$ terms. 
 	}
	\end{example}

\newcommand{\paul}{\mathrm{\tt PAULI}}
\newcommand{\ins}{\mathrm{\tt INSERT}}
\newcommand{\rem}{\mathrm{\tt REMOVE}}
\newcommand{\decomp}{\mathrm{\tt DECOMPOSE}}

	\begin{algorithm}[!t]
            \caption{Orthogonalization procedure for a linear combination of stabilizer states}
            \label{alg:stabortho} \footnotesize
            \begin{algorithmic}[1]
                \Require Linear combination of $n$-qubit states $\ket{\Psi} = \sum_{j=1}^N c_j\ket{\psi_j}$
                represented by ($i$) a list of {\em canonical stabilizer matrices}
                $\mlst=\{\cM^{1},\ldots,\cM^{N}\}$
                and ($ii$) a list of coefficients $\clst=\{c_1,\ldots,c_N\}$ 
                \Ensure Modified lists $\mlst'$ and $\clst'$ representing a
                linear combination of {\em mutually orthogonal} states 
                \Statex \hspace{-5mm} $\Rightarrow$ $\paul(\cM, j)$ returns $0$ if the
                $j^{th}$ column in $\cM$ has $Z$ literals only (ignores $I$ literals), 
                $1$ if it has $X$ literals only, $2$ if it has $Y$ literals only,
                $3$ if it has $X/Z$ literals only, and $4$ if it has $Y/Z$ literals only
                \Statex \hspace{-5mm} $\Rightarrow$ $\rem(\mlst, \clst, j)$ removes
                the $j^{th}$ element in $\mlst$ and $\clst$
                \Statex \hspace{-5mm} $\Rightarrow$ $\ins(\mlst, \clst, \cM, c)$ appends
                $\cM$ to $\mlst$ and $c$ to $\clst$ if an {\em equivalent matrix} does not 
                exist in $\mlst$; otherwise, sets $c_j = c_j+c$, where $c_j$
                is the coefficient of the matrix in $\mlst$ that is equivalent to $\cM$
                \Statex \hspace{-5mm} $\Rightarrow$ $\decomp(\cM, j, a\in\{0,1\})$ implements the
                proof of Proposition~\ref{prop:nneardecomp} and returns
                the pair $[\cM_{j=a}, \alpha]$, where $\cM_{j=a}$ is the nearest-neighbor 
                canonical matrix with the $j^{th}$ qubit in state $\ket{a}$, 
                and $\alpha$ is the phase factor
                \Statex 
				\For{$j \in \{1, \dots, n\}$}
					\State $b\leftarrow 0$
					\State $l\leftarrow \paul(\cM^1, j)$
                		\For{{\bf each} $\cM^i$, $i\in\{2,\ldots, N\}$}
	              		\State $k\leftarrow \paul(\cM^i, j)$
	               		\If{$k\neq l$ {\bf or} the rows with Pauli literals in the $j^{th}$ 
	               		column are distinct}
	               			\State $b\leftarrow 1$
	               			\State {\bf break}
	               		\EndIf
                		\EndFor
					\If{$b=1$}
						\For{{\bf each} $\cM^i$, $i\in\{1,\ldots, N\}$}
							\If{$\paul(\cM^i, j)\neq 0$}
								\State $[\cM^i_{j=0}, \alpha]\leftarrow\decomp(\cM^i, j, 0)$ 
								\State $[\cM^i_{j=1}, \beta]\leftarrow\decomp(\cM^i, j, 1)$
								\State $\rem(\mlst, \clst, i)$
								\State $\ins(\mlst, \clst, \cM^i_{j=0}, \alpha/\sqrt{2})$
								\State $\ins(\mlst, \clst, \cM^i_{j=1}, \beta/\sqrt{2})$
							\EndIf
						\EndFor
					\EndIf
				\EndFor
            \end{algorithmic}
        \end{algorithm}

\subsection{Computation of stabilizer bivectors} 
\label{sec:bivec_comp}

In Section~\ref{sec:bivectors}, we defined the notion of a {\em stabilizer 
bivector}, which can be used to represent antisymmetric basis states \cite{Hayashi}
compactly on conventional computers. We now describe an algorithm  
to efficiently compute a generator set for stabilizer bivectors.

Let $\ket{\psi}$ and $\ket{\phi}$ be two 
stabilizer states represented by matrices $\cM^\psi$ and $\cM^\phi$, 
respectively. Algorithm~\ref{alg:exprod} shows our approach to computing the 
stabilizer bivector $\ket{\psi\wedge\phi}$. First, we compute $\braket{\psi}{\phi}$ 
and obtain the canonical matrices for $\cM^\psi$ and $\cM^\phi$. This allows us
to determine whether the conditions established by the proof of Theorem~\ref{th:num_wedge}
are satisfied. In the case that $\ket{\psi}$ and $\ket{\phi}$ are $1$-neighbors,
we modify the {\em dissimilar} matrices as per the proof of Theorem~\ref{th:num_wedge}.
This ensures that the input matrices are {\em similar} before computing the matrices
for the tensor products $\ket{\psi\otimes\phi}$ and $\ket{\phi\otimes\psi}$. 

\vspace{1mm}
	\begin{example} {\em 
		Suppose $\cM^\psi = \{ZI,IZ\}$ and $\cM^\phi = \{XI,IZ\}$,
		which represent a nearest-neighbor pair. Following Equation~\ref{eq:bivector_1near}, 
		we replace $\cM^\phi$ with $X_1\cM^\psi X_1^\dag=\{-ZI,IZ\}$ (or $\cM^\psi$ with 
		$X_1\cM^\phi X_1^\dag=\{-XI,IZ\}$) to make the input matrices similar.
		}
	\end{example}
\vspace{1mm}

The last step is to compute the difference $\ket{\psi\otimes\phi}-\ket{\phi\otimes\psi}$ 
as follows:

	\begin{itemize}
		\item[{\bf 1.}] Obtain the basis-normalization circuit $\cC$ for both 
		$\cM^{\psi\otimes\phi}$ and $\cM^{\phi\otimes\psi}$. Observe that its
		the same circuit since the matrices are similar.
		\item[{\bf 2.}] Conjugate both matrices by $\cC$ to obtain the
		matrices for basis states $\ket{b_1}$ and $\ket{b_2}$.
		\item[{\bf 3.}] Following the proof of Lemma~\ref{lem:cross}, obtain
		the matrix $\cM^{b_1-b_2}$ for the sum $\ket{b_1}-\ket{b_2}$.
		\item[{\bf 4.}] Conjugate $\cM^{b_1-b_2}$ by $\cC^\dag$ to obtain
		the matrix $\cM^{\psi\wedge\phi}$.
	\end{itemize}
   
\newcommand{\tensor}{\mathrm{\tt TENSOR}}
\newcommand{\inprod}{\mathrm{\tt INPROD}}
\newcommand{\stbsum}{\mathrm{\tt SUM}}

		\begin{algorithm}[!b]
            \caption{Computation of stabilizer bivectors}
            \label{alg:exprod} \footnotesize
            \begin{algorithmic}[1]
                \Require Stabilizer matrices $\cM^\psi$ and $\cM^\phi$ representing $\ket{\psi}$
                and $\ket{\phi}$, respectively
                \Ensure Stabilizer matrix $\cM^{\psi\wedge\phi}$ for bivector $\ket{\psi\wedge\phi}$, 
                if it exists
                \Statex \hspace{-5mm} $\Rightarrow$ $\norm(\cM^\psi)$ returns circuit $\cC$ 
                such that $\cC\ket{\psi}=\ket{b}$, where $\ket{b}$ is a basis state
                \Statex \hspace{-5mm} $\Rightarrow$  $\apply(\cM, \cC)$ conjugates $\cM$ 
                by Clifford or Pauli operator $\cC$ and returns the modified matrix
                \Statex \hspace{-5mm} $\Rightarrow$ $\gauss(\cM)$ reduces $\cM$ to canonical form 
                \Statex \hspace{-5mm} $\Rightarrow$ $\tensor(\cM^\psi, \cM^\phi)$ computes
                $\ket{\psi}\otimes\ket{\phi}$ and returns the resulting matrix 
                $\cM^{\psi\otimes\phi}$ (Proposition~\ref{prop:stab_tensor})
                \Statex \hspace{-5mm} $\Rightarrow$ $\inprod(\cM^\psi, \cM^\phi)$ computes
                and returns $\braket{\psi}{\phi}$
                \Statex \hspace{-5mm} $\Rightarrow$ $\stbsum(\cM^{b_1}, \cM^{b_2})$ computes
                $\ket{b_1}-\ket{b_2}$ and returns the resulting matrix $\cM^{b_1-b_2}$
                \Statex
                \State $\alpha\leftarrow \inprod(\cM^\psi, \cM^\phi)$
                \Comment{Apply Algorithm~\ref{alg:inprod}}
                \State $\gauss(\cM^\psi)$ \Comment{Set $\cM^\psi$ to canonical form}
                \State $\gauss(\cM^\phi)$ \Comment{Set $\cM^\phi$ to canonical form}
                \If{($\alpha = 0$ {\bf and} $\cM^\psi$ {\em dissimilar from} $\cM^\phi$)
                {\bf or} ($\alpha=1$) {\bf or} ($0 < \alpha < 1/\sqrt{2}$)} 
                		\State {\bf exit} \Comment{By Theorem~\ref{th:num_wedge}}
                	\EndIf
                	\If{$\cM^\psi$ {\em dissimilar from} $\cM^\phi$}
                	\Comment{$1$-neighbor case from Theorem~\ref{th:num_wedge}}
                			\State $\cM^{\phi}\leftarrow\apply(\cM^\psi, P)$ 
                			\Comment{$\cM^{\phi}$ now represents $P\ket{\psi}$, where $P$ is a Pauli operator}
                	\EndIf
                	\State $\cM^{\psi\otimes\phi} \leftarrow\tensor(\cM^{\psi}, \cM^{\phi})$
                	\State $\cM^{\phi\otimes\psi} \leftarrow\tensor(\cM^{\phi}, \cM^{\psi})$
                \State $\cC \leftarrow \norm(\cM^{\psi\otimes\phi})$
                \Comment{Obtains basis-normalization circuit}  
                	\State $\cM^{b_1}\leftarrow\apply(\cM^{\psi\otimes\phi}, \cC)$
                	\Comment{Maps $\cM^{\psi\otimes\phi}$ to basis state $\ket{b_1}$}
                	\State $\cM^{b_2}\leftarrow\apply(\cM^{\phi\otimes\psi}, \cC)$
                	\Comment{Maps $\cM^{\phi\otimes\psi}$ to basis state $\ket{b_2}$}
                	\State $\cM^{b_1-b_2}\leftarrow\stbsum(\cM^{b_1}, \cM^{b_2})$
                	\Comment{Obtains matrix for $\ket{b_1}-\ket{b_2}$ (proof of Lemma~\ref{lem:cross})}
                	\State $\cM^{\psi\wedge\phi}\leftarrow\apply(\cM^{b_1-b_2}, \cC^\dag)$
                \State\Return $\cM^{\psi\wedge\phi}$
            \end{algorithmic}
        \end{algorithm}
            
    \begin{proposition} \label{prop:normform_circsize}
		Let $\ket{\psi}$ and $\ket{\phi}$ be two $n$-qubit stabilizer states.
		The $2n$-qubit stabilizer bivector $\ket{\psi\wedge\phi}$ can be 
		computed in $O(n^3)$ time.
	\end{proposition}
	\noindent
	{\bf Proof.} 
		The bottleneck in Algorithm~\ref{alg:exprod} is the computation of 
		the basis-normalization circuit for the $2n\times 2n$ stabilizer
		matrix $\cM^{\psi\otimes\phi}$, which takes $O(n^3)$ time.
	\square

\subsection{Mixed stabilizer states}
\label{sec:mixedstab}

The work in \cite{AaronGottes} describes {\em mixed stabilizer states},
which are states that are {\em uniformly distributed} over all states in a subspace.
Since such mixed states can be written as the partial trace of a pure 
stabilizer state, they can be represented compactly using stabilizer matrices.
Recall from Theorem~\ref{th:gen_commute} that the set of Pauli operators (rows) 
in a stabilizer matrix that represents a pure state are linearly independent. 
In contrast, to represent mixed stabilizer states, one needs to maintain 
stabilizer matrices with linearly-dependent rows \cite{AaronGottes, Audenaert}.
This implies that a subset of the rows in the matrix can be reduced to the 
identity operator. Since Algorithm~\ref{alg:gauss_min} reduces a stabilizer
matrix to its Gauss-Jordan form, it can be used to find and eliminate 
linearly-dependent rows. Such rows will show up as $I$-literal only rows
at the bottom of the matrix after Algorithm~\ref{alg:gauss_min} is applied. 
Similarly, Algorithms~\ref{alg:inprod_circ} and \ref{alg:inprod},
generalize to the case of mixed stabilizer states
since one can identify linearly-dependent rows via Algorithm~\ref{alg:gauss_min}
and then ignore them throughout the rest of the computation. 
Therefore, if the input matrices represent mixed states, Algorithm~\ref{alg:inprod}
computes the fidelity between them. 

As outlined in \cite{AaronGottes} and proven in \cite{Audenaert}, 
one can also compute the partial trace over qubit $j$ as follows: ($i$) apply 
a modified version of Algorithm~\ref{alg:gauss_min} such 
that the $j^{th}$ column of the stabilizer matrix has at most
one $X/Y$ literal (if $X/Y$ literals exist in the column) 
and one $Z$ literal (if $Z$ literals exist in the column), and ($ii$) remove the 
rows and columns containing such literals. The resulting stabilizer
matrix represents the reduced mixed state. 



\section{Empirical Studies} 
\label{sec:empirical}

\begin{figure}[!b]
	\centering
	\begin{tabular}{cc} 
		\hspace{-.4cm}
		\includegraphics[scale=.55]{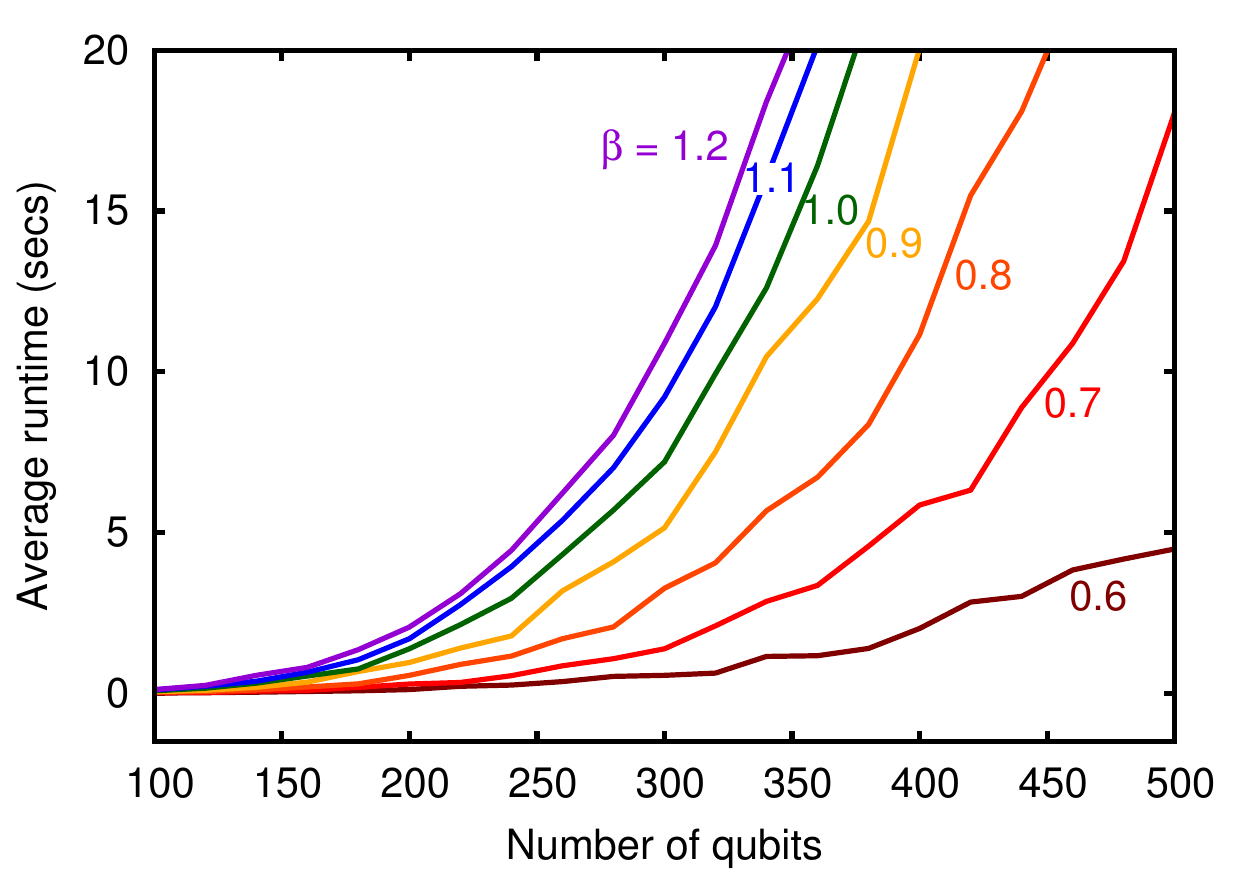} 
		&
		\includegraphics[scale=.55]{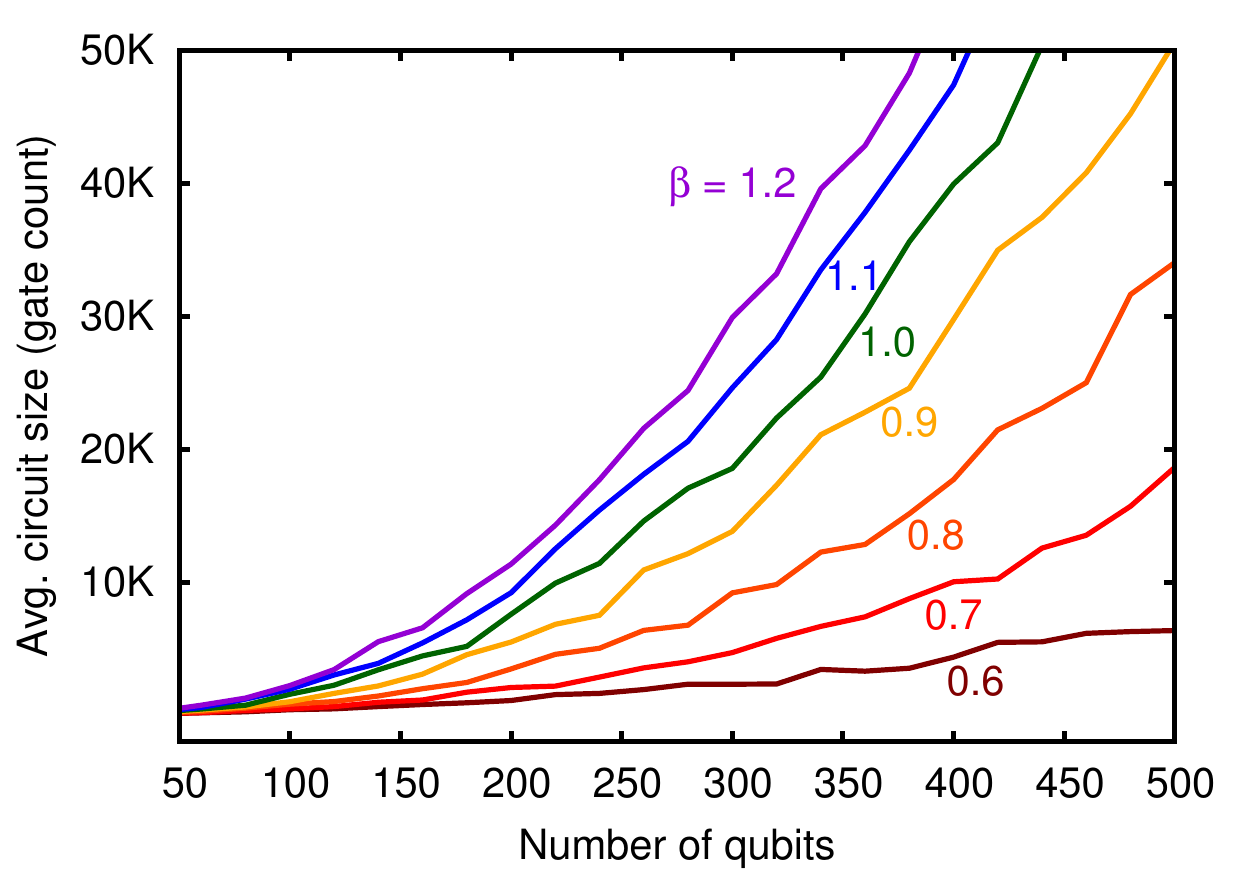} \\
		\hspace{1cm} {\bf\small (a)} & {\bf\small (b)}  \\
		&
		\vspace{-.25cm}
	\end{tabular}
	\fcaption{\label{fig:stabip}  {\bf\small (a)} Average runtime for Algorithm \ref{alg:inprod}
	to compute the inner product between two random $n$-qubit stabilizer states. The stabilizer
	matrices that represent the input states are generated by applying $\beta n\log_2 n$
	unitary stabilizer gates to $\ket{0^{\otimes n}}$. {\bf\small (b)} Average number of gates
	in the circuits produced by Algorithm \ref{alg:inprod_circ}.
	}
\end{figure}

\begin{figure}[!b]
	\centering
	\begin{tabular}{cc}
		\includegraphics[scale=.55]{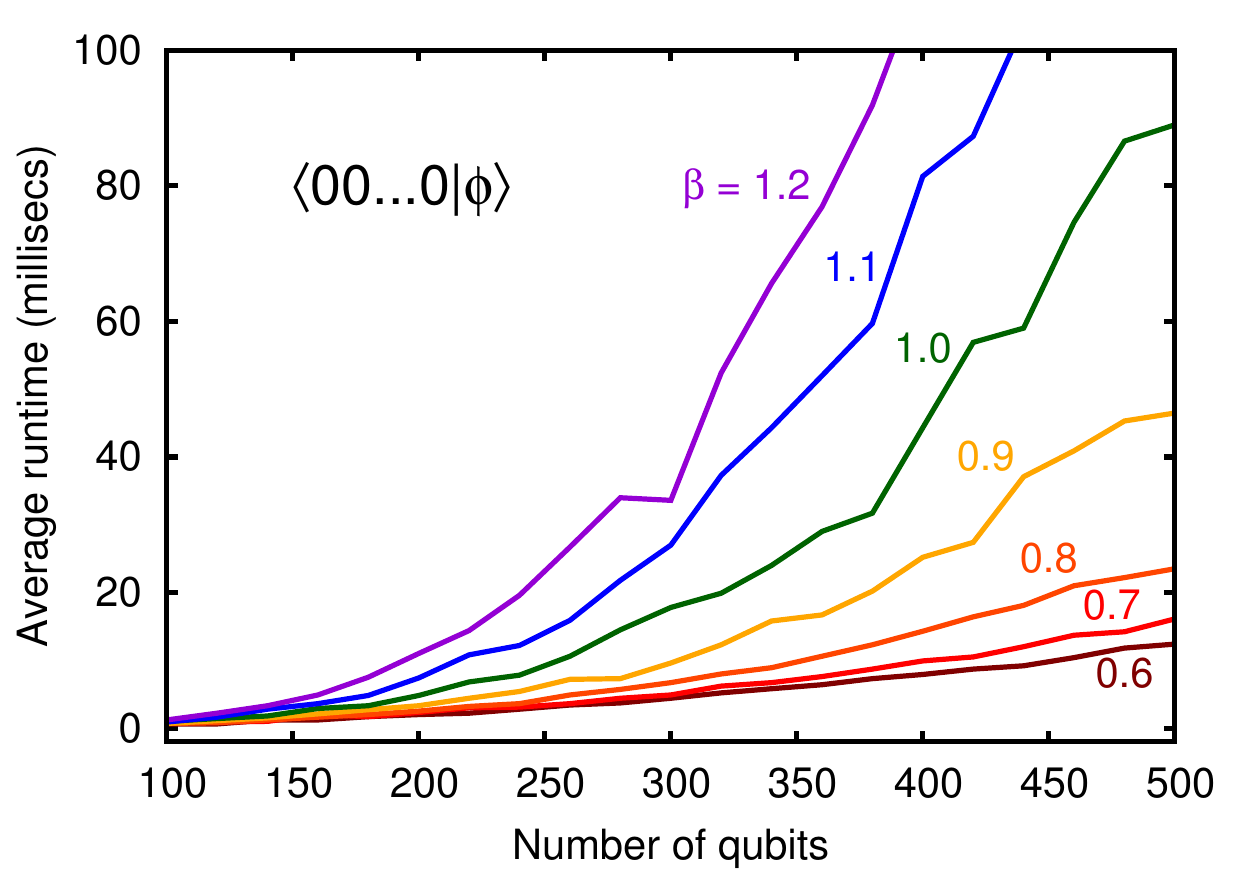} &
		\includegraphics[scale=.55]{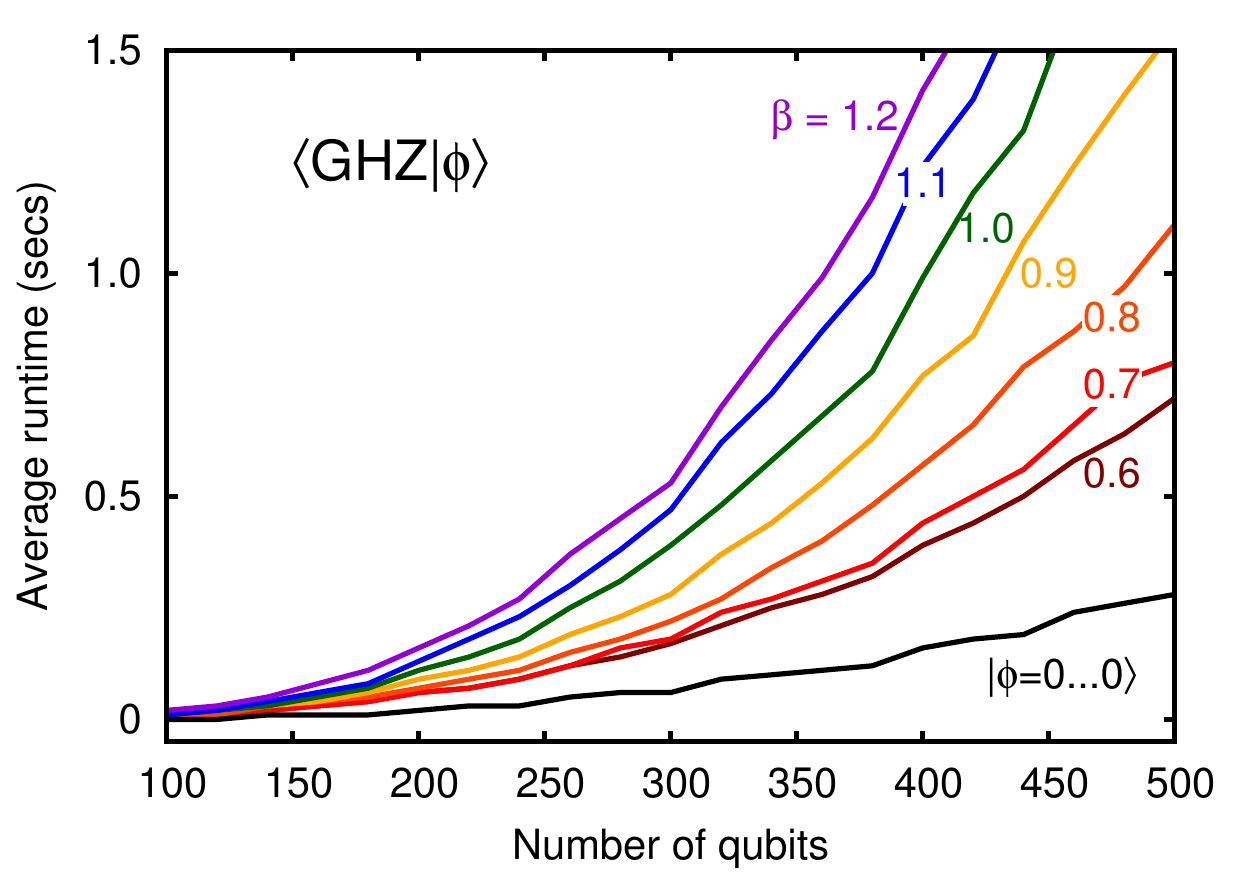} \\
		{\bf\small (a)} & {\bf\small (b)} \\
		&
		\vspace{-.25cm}
	\end{tabular}
	\fcaption{\label{fig:stabip_ghz0all} Average runtime for Algorithm \ref{alg:inprod}
	to compute the inner product between {\bf\small (a)} $\ket{0^{\otimes n}}$ and random 
	stabilizer state $\ket{\phi}$ and {\bf\small (b)} the $n$-qubit GHZ state and random 
	stabilizer state $\ket{\phi}$.}
\end{figure}

\begin{figure}[!t]
	\centering
	\begin{tabular}{cc}
		\hspace{-.7cm} 
		\includegraphics[scale=.6]{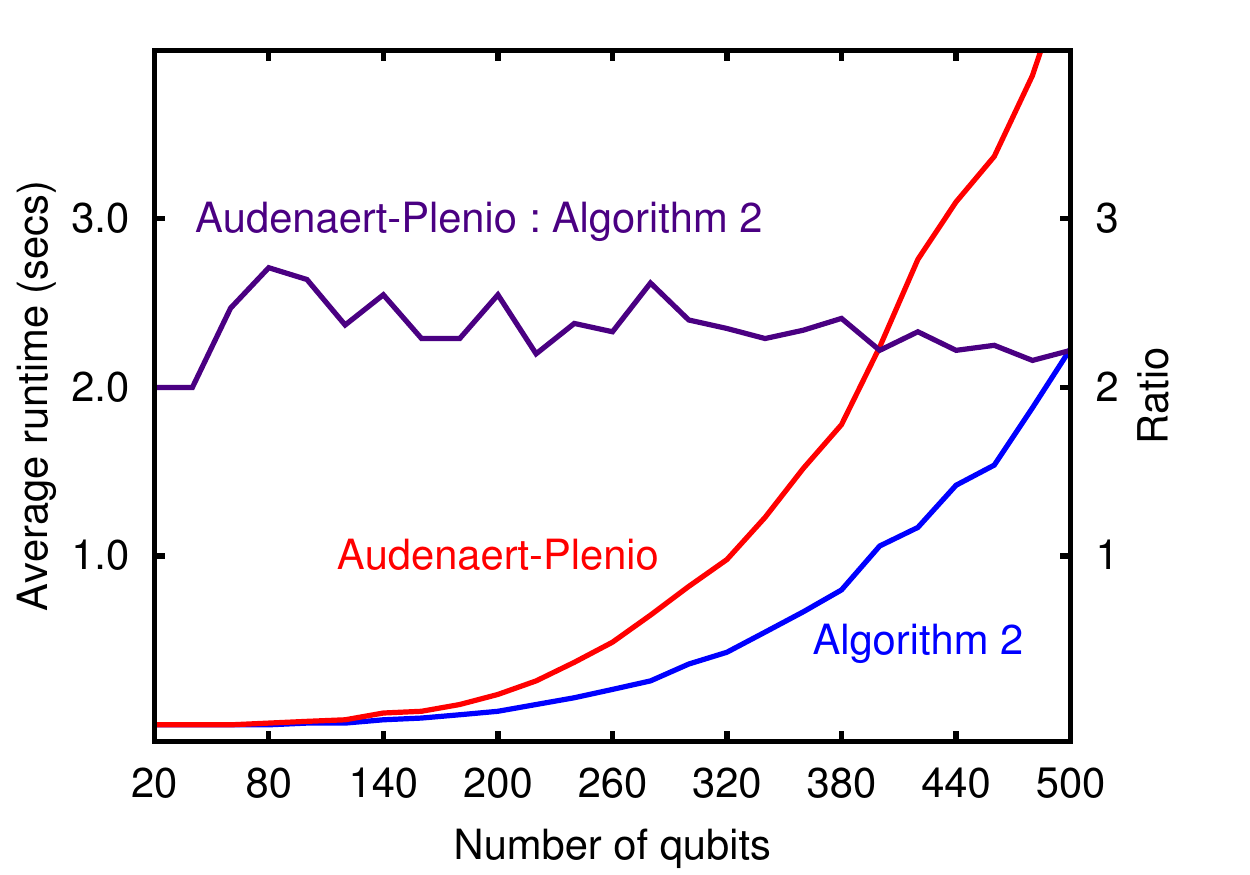} &
		\hspace{-7mm} 
		\includegraphics[scale=.6]{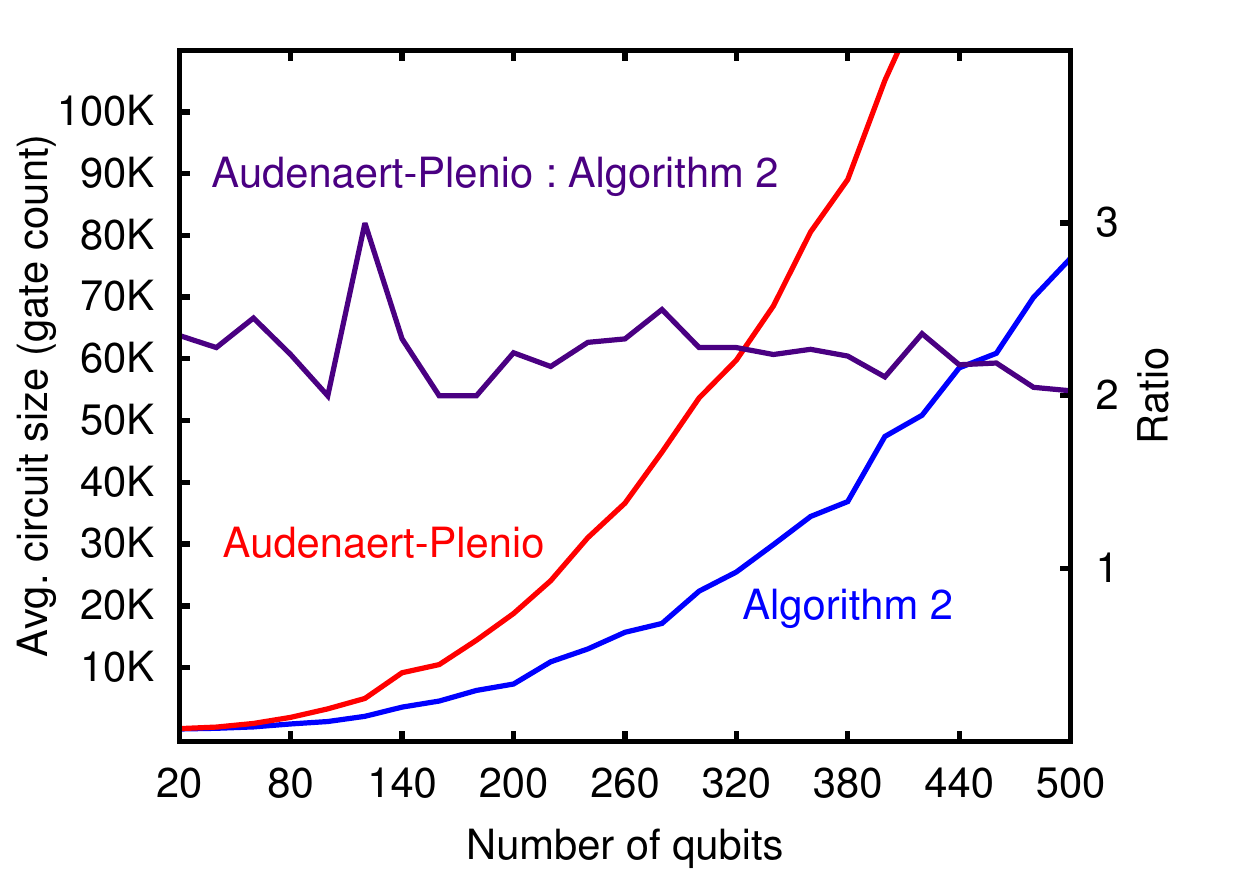} \\
		 {\bf\small (a)} & {\bf\small (b)} \\
		 &
		 \vspace{-.25cm}
	\end{tabular}
	\fcaption{\label{fig:stabip_comp} {\bf\small (a)} Runtime and {\bf\small (b)} circuit-size
	comparisons between Algorithm~\ref{alg:inprod_circ} 
	and the circuit synthesis portion of the Audenaert-Plenio inner-product algorithm. 
	On average, Algorithm~\ref{alg:inprod_circ} 
	runs roughly twice as fast and produces canonical circuits
	that contain less than half as many gates. Furthermore, the Audenaert-Plenio circuits 
	are not canonical.
	}
	\vspace{-.25cm}
\end{figure}

Our circuit-synthesis (Section~\ref{sec:synthesis}) and 
inner-product (Section~\ref{sec:ipalgo}) algorithms hold 
potential to be used in several practical applications 
including quantum error correction, quantum-circuit simulation,
and the computation of geometric measures of entanglement. 
Therefore, we implemented these algorithms in C++ and 
empirically validated their performance.
Recall that the runtime of Algorithm~\ref{alg:inprod_circ} 
is dominated by the two nested for-loops (lines 20-35). 
The number of times these loops execute depends on the amount of 
entanglement in the input stabilizer state. In turn, the number of entangled 
qubits depends on the the number of CNOT gates in the 
circuit $\cC$ used to generate the stabilizer state $\cC\ket{0^{\otimes n}}$
(Theorem \ref{th:stabst}). By a simple heuristic argument \cite{AaronGottes},
one generates highly entangled stabilizer states as long as the 
number of CNOT gates in $\cC$ is proportional to $n\log_2 n$. Therefore, 
we generated random $n$-qubit stabilizer circuits for $n\in\{20, 40, \ldots, 500\}$ as follows: 
fix a parameter $\beta > 0$; then choose $\beta \lceil n \log_2 n\rceil$ 
unitary gates (CNOT, Phase or Hadamard) each with probability $1/3$.
Then, each random $\cC$ is applied to the $\ket{00\ldots0}$ basis state to generate
random stabilizer matrices (states). The use of randomly generated benchmarks is 
justified for our experiments because (\emph{i}) our algorithms are not explicitly 
sensitive to circuit topology and (\emph{ii}) random stabilizer circuits are 
considered representative \cite{Knill}.
Both our inner-product and exterior-product algorithms exhibited
similar asymptotic runtime behavior since the bottleneck in
both algorithms is the execution of Algorithm~\ref{alg:inprod_circ}.
Therefore, for conciseness, we present experimental 
results only for our inner-product algorithm.
For each $n$, we applied Algorithm \ref{alg:inprod} to pairs of random
stabilizer matrices and measured the number of seconds needed to
compute the inner product. The entire procedure was repeated for
increasing degrees of entanglement by ranging $\beta$
from $0.6$ to $1.2$ in increments of $0.1$. Our results
are shown in Figure~\ref{fig:stabip}-a.
The runtime of Algorithm~\ref{alg:inprod} 
appears to grow quadratically in $n$ when $\beta = 0.6$. However, 
when we double the number of unitary gates
($\beta = 1.2$), the runtime exhibits cubic growth.
Therefore, Figure~\ref{fig:stabip}-a shows that 
the performance of Algorithm~\ref{alg:inprod} is highly 
dependent on the degree of entanglement in the input
stabilizer states. Figure~\ref{fig:stabip}-b shows the
average size of the basis-normalization circuit returned
by the calls to Algorithm~\ref{alg:inprod_circ}. As expected
(Proposition~\ref{prop:normform_circsize}), the size of the
circuit grows quadratically in $n$.
Figure~\ref{fig:stabip_ghz0all} shows the average runtime for 
Algorithm \ref{alg:inprod} to compute the inner 
product between: (\emph{i}) the all-zeros basis state        
and random $n$-qubit stabilizer
states, and (\emph{ii}) the $n$-qubit GHZ
state and random stabilizer states. GHZ states are maximally entangled
states of the form $\ket{GHZ}=\frac{\ket{0^{\otimes n}} + \ket{1^{\otimes n}}}{\sqrt{2}}$
that have been realized experimentally using several quantum technologies
and are often encountered in practical applications such as error-correcting 
codes and fault-tolerant architectures. Figure \ref{fig:stabip_ghz0all}
shows that, for such practical instances, Algorithm \ref{alg:inprod} 
computes the inner product in roughly $O(n^2)$ time (e.g., $\braket{GHZ}{0}$). 
However, without apriori information about the input stabilizer matrices
(e.g., a measure of the amount of the entanglement), 
one can only say that the performance of Algorithm \ref{alg:inprod} will 
be somewhere between quadratic and cubic in $n$.
We compared Algorithm~\ref{alg:inprod_circ} to the circuit synthesis 
approach developed by Audenaert and Plenio (AP) \cite{Audenaert} as part
of their inner-product algorithm. The benchmark consists of randomly
generated $n$-qubit stabilizer circuits with $n\log n$ unitary gates.
Figure \ref{fig:stabip_comp} shows
that the AP algorithm produces (non-canonical) circuits 
with more than twice as many gates as our canonical circuits
and takes roughly twice as long to produce them. 

\section{Conclusions} \label{sec:conclude}

The stabilizer formalism facilitates compact representation of 
stabilizer states and efficient simulation of stabilizer circuits~\cite{AaronGottes, Gottes, Gottes98}.
Stabilizer states arise in applications of quantum information processing,
and their efficient manipulation via geometric and linear-algebraic operations
may lead to additional insights in quantum entanglement, 
quantum error correction and quantum-circuit simulation. 
Furthermore, stabilizer states are closely related to valence-bond states~\cite{Verst04},
cluster/graph states~\cite{AnderBrieg, Dur, Hein}, and 
measurement-based/one-way quantum computation~\cite{Rauss}.
The emphasis of our work on the full set of (pure) $n$-qubit 
stabilizer states is justified by the desire to succinctly represent
as many quantum states as possible \cite{Aaron}. This is in contrast 
to identifying structured representations of more specialized
quantum states.  

In this work we characterized the nearest-neighbor structure 
of stabilizer states and quantified the distribution of angles 
between pairs of stabilizer states. We showed that, for any 
$n$-qubit stabilizer state $\ket{\psi}$, 
almost all stabilizer states are either orthogonal or nearly orthogonal 
to $\ket{\psi}$ as $n\rightarrow\infty$.
Although the geometric structure of stabilizer states is fairly uniform~\cite{DiVince, Klapp, Montanaro},
we showed that local search is not guaranteed to find 
the closest stabilizer state to an arbitrary quantum state.
Furthermore, we defined a family of unbiased states that cannot be approximated 
by polynomial-sized superpositions of stabilizer states, and proved that the maximal radius 
of any $2^n$-dimensional ball centered at a point on the unit sphere that does 
not contain any $n$-qubit stabilizer states cannot exceed $\sqrt{2}$,
but approaches $\sqrt{2}$ as $n\rightarrow\infty$.
We studied algorithms for: ($i$)~computing the inner product between 
stabilizer states, ($ii$)~orthogonalizing a set of stabilizer states, 
and ($iii$)~computing stabilizer bivectors.
A crucial step of our inner-product algorithm is the synthesis
of a circuit that transforms a stabilizer state into 
a computational-basis state. Our algorithm synthesizes such circuits
using a $5$-block canonical template structure using
$O(n^2)$ stabilizer gates. 
Our technique produces circuits
with half as many gates as the approach from \cite{Audenaert} using a block 
sequence that is shorter than the approach from \cite{AaronGottes}.
We analyzed the performance of our inner-product algorithm and 
showed that, although its runtime is $O(n^3)$, it can take
quadratic time in practice. 

\noindent
{\bf Open problems}. As reviewed in Section~\ref{sec:inprod_stab},
the work in~\cite{AaronGottes, Audenaert, Smith} describes how 
to represent and manipulate mixed states using the 
stabilizer formalism. To this end, extensions of our results 
to mixed states could be of interest.
Several approaches have been proposed for generalizing the
stabilizer representation to represent arbitrary quantum states and simulate
non-stabilizer circuits~\cite{AaronGottes, Vanden}.
An open question is whether the geometric properties of stabilizer states
shown in this work can be exploited to improve simulation of generic
quantum circuits while ensuring the scalability of simulation when 
stabilizer gates dominate. Another attractive direction for future work is 
generalizing presented results and algorithms to the case of 
$d$-dimensional qudit states~\cite{DeBeaudrap}. Further challenges include:

\begin{itemize}
	\vspace{-2mm}
	\item Characterization of possible volumes of high-dimensional 
	parallelotopes formed by more than two stabilizer vectors 
	(generalizing Proposition~\ref{prop:bivec_norm}).
	\vspace{2mm}
	\item Characterization of minimally-dependent sets of stabilizer states
    and algorithms for efficient detection of such sets (extending our
    discussion in Section~\ref{sec:lindepstab}).
    \vspace{-2mm}
	\item Efficiency improvement for computing 
	inner and exterior products of stabilizer states
	as well as the orthogonalization procedure from Algorithm~\ref{alg:stabortho}.
	\vspace{-2mm}
	\item Characterization of the complexity analysis of finding a 
	closest stabilizer state to a given non-stabilizer state 
	(for appropriate representations of non-stabilizer states). 
	\vspace{-2mm}
	\item Describing the Voronoi-diagram structure of
	stabilizer states (extending Theorem~\ref{th:num_near}).
	\vspace{-2mm}
\end{itemize}

\noindent
{\bf Acknowledgements}. The authors would like to thank 
the Microsoft Research QuArC group for questions and suggestions,
Dmitri Maslov for inspiring discussions, and anonymous reviewers
for suggested improvements.
This work was sponsored in part by the Air Force Research Laboratory
under agreement FA8750-11-2-0043. 

\nonumsection{References}
\noindent

\newpage
\appendix

\section*{The 1080 three-qubit stabilizer states}
\label{app:three_qbssts}

Shorthand notation represents a stabilizer state as $\alpha_0, \alpha_1, \alpha_2, \alpha_3$ where $\alpha_i$ are the normalized amplitudes of the basis states. Basis states are emphasized in bold. The $\angle$ column indicates the angle between that state and $\ket{000}$, which has $28$ $1$-neighbor states and $315$ orthogonal states ($\perp$). The tables are intended for on-screen viewing under magnification.

\begin{center}
	\scriptsize
	\scalebox{.5}[.4]{
\begin{tabular}{|c|c||c|c||c|c||c|c||c|c||c|c||c|c||c|c|}\hline
\sc State & \sc Gen'tors  & \sc State & \sc Gen'tors  & \sc State & \sc Gen'tors  & \sc State & \sc Gen'tors  & \sc State & \sc Gen'tors  & \sc State & \sc Gen'tors  & \sc State & \sc Gen'tors  & \sc State & \sc Gen'tors \\ \hline\hline
\hspace{-2mm}$1 1 1 1 1 1 1 1$\hspace{-2mm} & \hspace{-2mm}IIX IXI XII\hspace{-2mm} & \hspace{-2mm}$1 1 1 1 $-$ $-$ $-$ $-$$\hspace{-2mm} & \hspace{-2mm}IIX IXI -XII\hspace{-2mm} & \hspace{-2mm}$1 0 0 i $-$ 0 0 \bar{i} $\hspace{-2mm} & \hspace{-2mm}IXY IYX -XII\hspace{-2mm} & \hspace{-2mm}$0 \bar{i}  0 0 1 0 0 0$\hspace{-2mm} & \hspace{-2mm}IZI YIX -XIY\hspace{-2mm} & \hspace{-2mm}$1 1 $-$ $-$ $-$ $-$ 1 1$\hspace{-2mm} & \hspace{-2mm}IIX -IXI -XII\hspace{-2mm} & \hspace{-2mm}$1 \bar{i}  i $-$ $-$ i \bar{i}  1$\hspace{-2mm} & \hspace{-2mm}IYZ -IXX -XII\hspace{-2mm} & \hspace{-2mm}$1 0 $-$ 0 0 i 0 i$\hspace{-2mm} & \hspace{-2mm}XYX -IXZ -YYY\hspace{-2mm} & \hspace{-2mm}$1 $-$ $-$ 1 $-$ 1 1 $-$$\hspace{-2mm} & \hspace{-2mm}-IIX -IXI -XII\hspace{-2mm}\hspace{-2mm}\\ 
\hspace{-2mm}$1 1 1 1 i i i i$\hspace{-2mm} & \hspace{-2mm}IIX IXI YII\hspace{-2mm} & \hspace{-2mm}$1 1 1 1 \bar{i}  \bar{i}  \bar{i}  \bar{i} $\hspace{-2mm} & \hspace{-2mm}IIX IXI -YII\hspace{-2mm} & \hspace{-2mm}$1 0 0 i \bar{i}  0 0 1$\hspace{-2mm} & \hspace{-2mm}IXY IYX -YII\hspace{-2mm} & \hspace{-2mm}$1 i 0 0 $-$ i 0 0$\hspace{-2mm} & \hspace{-2mm}IZI YIX -XIZ\hspace{-2mm} & \hspace{-2mm}$1 1 $-$ $-$ \bar{i}  \bar{i}  i i$\hspace{-2mm} & \hspace{-2mm}IIX -IXI -YII\hspace{-2mm} & \hspace{-2mm}$1 \bar{i}  i $-$ \bar{i}  $-$ 1 i$\hspace{-2mm} & \hspace{-2mm}IYZ -IXX -YII\hspace{-2mm} & \hspace{-2mm}$1 i \bar{i}  $-$ 1 \bar{i}  i $-$$\hspace{-2mm} & \hspace{-2mm}XYY -IXX -YIX\hspace{-2mm} & \hspace{-2mm}$1 $-$ $-$ 1 \bar{i}  i i \bar{i} $\hspace{-2mm} & \hspace{-2mm}-IIX -IXI -YII\hspace{-2mm}\hspace{-2mm}\\
\hline\end{tabular}
}
\end{center}

\begin{center}
	\scriptsize
	\scalebox{.6}[.5]{\begin{tabular}{|c|c||c|c||c|c||c|c||c|c||c|c||c|c||c|c|}\hline
\sc State & $\angle$ & \sc State & $\angle$ & \sc State & $\angle$ & \sc State & $\angle$ & \sc State & $\angle$ & \sc State & $\angle$ & \sc State & $\angle$ & \sc State & $\angle$\\ \hline\hline
\hspace{-1mm}$1 1 1 1 1 1 1 1$\hspace{-1mm} & \hspace{-1mm}$\pi/2.59$\hspace{-1mm} & \hspace{-1mm}$1 1 1 1 $-$ $-$ $-$ $-$$\hspace{-1mm} & \hspace{-1mm}$\pi/2.59$\hspace{-1mm} & \hspace{-1mm}$1 0 0 i $-$ 0 0 \bar{i} $\hspace{-1mm} & \hspace{-1mm}$\pi/3$\hspace{-1mm} & \hspace{-1mm}$0 \bar{i}  0 0 1 0 0 0$\hspace{-1mm} & \hspace{-1mm} $\perp$ \hspace{-1mm} & \hspace{-1mm}$1 1 $-$ $-$ $-$ $-$ 1 1$\hspace{-1mm} & \hspace{-1mm}$\pi/2.59$\hspace{-1mm} & \hspace{-1mm}$1 \bar{i}  i $-$ $-$ i \bar{i}  1$\hspace{-1mm} & \hspace{-1mm}$\pi/2.59$\hspace{-1mm} & \hspace{-1mm}$1 0 $-$ 0 0 i 0 i$\hspace{-1mm} & \hspace{-1mm}$\pi/3$\hspace{-1mm} & \hspace{-1mm}$1 $-$ $-$ 1 $-$ 1 1 $-$$\hspace{-1mm} & \hspace{-1mm}$\pi/2.59$\hspace{-1mm}\hspace{-1mm}\\ 
\hspace{-1mm}$1 1 1 1 i i i i$\hspace{-1mm} & \hspace{-1mm}$\pi/2.59$\hspace{-1mm} & \hspace{-1mm}$1 1 1 1 \bar{i}  \bar{i}  \bar{i}  \bar{i} $\hspace{-1mm} & \hspace{-1mm}$\pi/2.59$\hspace{-1mm} & \hspace{-1mm}$1 0 0 i \bar{i}  0 0 1$\hspace{-1mm} & \hspace{-1mm}$\pi/3$\hspace{-1mm} & \hspace{-1mm}$1 i 0 0 $-$ i 0 0$\hspace{-1mm} & \hspace{-1mm}$\pi/3$\hspace{-1mm} & \hspace{-1mm}$1 1 $-$ $-$ \bar{i}  \bar{i}  i i$\hspace{-1mm} & \hspace{-1mm}$\pi/2.59$\hspace{-1mm} & \hspace{-1mm}$1 \bar{i}  i $-$ \bar{i}  $-$ 1 i$\hspace{-1mm} & \hspace{-1mm}$\pi/2.59$\hspace{-1mm} & \hspace{-1mm}$1 i \bar{i}  $-$ 1 \bar{i}  i $-$$\hspace{-1mm} & \hspace{-1mm}$\pi/2.59$\hspace{-1mm} & \hspace{-1mm}$1 $-$ $-$ 1 \bar{i}  i i \bar{i} $\hspace{-1mm} & \hspace{-1mm}$\pi/2.59$\hspace{-1mm}\hspace{-1mm}\\
\hline\end{tabular}
}
\end{center}

\begin{thebibliography}{000}

\bibitem{Aaron}
S. Aaronson, ``Multilinear formulas and skepticism of quantum computing,'' 
Proc. STOC '04, pp. 118-127 (2004).

\bibitem{AaronGottes}
S. Aaronson and D. Gottesman, ``Improved simulation of stabilizer circuits,''
Phys. Rev. A, vol.~70, no.~052328 (2004). 


\bibitem{AnderBrieg}
S. Anders and H. J. Briegel, ``Fast simulation of stabilizer circuits using a graph-state representation,''
Phys. Rev. A, vol.~73, no.~022334 (2006).

\bibitem{Audenaert}
K. M. R. Audenaert and M. B. Plenio, ``Entanglement on mixed stabiliser states: normal forms and reduction procedures,'' New J. Phys., vol.~7, no.~170 (2005).



\bibitem{Calder} A. Calderbank, E. Rains, P. Shor, and N. Sloane, ``Quantum error correction via codes over GF(4),'' IEEE Trans. Inf. Theory, vol.~44, pp. 1369--1387 (1998).

\bibitem{Cleve} R. Cleve and D. Gottesman, ``Efficient computations of encodings for quantum error correction,'' Phys. Rev. A, vol.~56, pp. 1201--1204 (1997).



\bibitem{Dahlsten}
O. Dahlsten and M. B. Plenio, ``Exact entanglement probability distribution of bi-partite randomised stabilizer states,'' Quant. Inf. Comp., vol.~6, no.~527 (2006).


\bibitem{Datta} A. Datta and G. Vidal, ``Role of entanglement and correlations in mixed-state quantum computation,'' Phys. Rev. A, vol.~75, no.~042310 (2007).


\bibitem{DeBeaudrap} N. de Beaudrap, ``A linearized stabilizer formalism for systems of finite 
dimension, '' Quant. Inf. Comp., vol.~13, pp. 73--115 (2013).


\bibitem{DehaeDemoor} J. Dehaene and B. De Moor, ``Clifford group, stabilizer states, and linear and quadratic operations over GF(2),'' Phys. Rev. A, vol.~68, no.~042318 (2003).

\bibitem{DiVince}
D. P. DiVincenzo, D. W. Leung and B. M. Terhal, ``Quantum data hiding,''
IEEE Trans. Inf Theory, vol.~48. no.~3, pp. 580--599 (2002).

\bibitem{Djor} I. Djordjevic, ``Quantum information processing and quantum error correction: 
an engineering approach, '' Academic press (2012). 

\bibitem{Dur} W. Dur, H. Aschauer and H.J. Briegel, ``Multiparticle entanglement purification 
for graph states,'' Phys. Rev. Lett., vol.~91, no.~107903 (2003). 

\bibitem{Emary}
C. Emary, ``A bipartite class of entanglement monotones for $N$-qubit pure states,''
J. Phys. A: Math. Gen., vol.~37, no.~8293 (2004).

\bibitem{Fattal} D. Fattal, T. S. Cubitt, Y. Yamamoto, S. Bravyi and I. L. Chuang, 
``Entanglement in the stabilizer formalism,'' arxiv:0406168 (2004).

\bibitem{Gottes}
D. Gottesman, ``Stabilizer codes and quantum error correction,'' Caltech Ph.D. thesis (1997).

\bibitem{Gottes98} D. Gottesman, ``The Heisenberg representation of quantum computers,''
arXiv:9807006v1 (1998).


\bibitem{Guehne} O. Guehne, G. Toth, P. Hyllus and H.J. Briegel,
``Bell inequalities for graph states,'' 	Phys. Rev. Lett. 95, 120405 (2005). 

\bibitem{Harrow} A. W. Harrow and R. A. Low, ``Random quantum circuits are approximate 2-designs,''
Comm. Math. Phys., vol.~291, no.~1, pp. 257--302 (2009).

\bibitem{Hayashi} M. Hayashi, et al., ``Entanglement of multiparty stabilizer, symmetric, and antisymmetric states,'' Phys. Rev. A, vol.~77, no.~012104 (2008).
 
\bibitem{Hein} M. Hein, J. Eisert, and H.J. Briegel,``Multi-party entanglement in graph states,'' 
Phys. Rev. A, vol.~69, no.~062311 (2004).

\bibitem{Heydari}
H. Heydari, ``Quantum entanglement measure based on wedge product,'' Quant. Inf. Comp., 
vol.~6, pp. 166-172 (2006).

\bibitem{Jozsa}
R. Jozsa, ``Embedding classical into quantum computation,'' arXiv:0812.4511 (2008).

\bibitem{Klapp}
A. Klappenecker and M. Roetteler, ``Mutually unbiased bases are complex projective 2-designs,'' 
arXiv:0502031 (2005).

\bibitem{Knill} E. Knill, et al., ``Randomized benchmarking of quantum gates,''
Phys. Rev. A, vol.~77, no.~1 (2007).





\bibitem{Montanaro}
A. Montanaro, ``On the distinguishability of random quantum states,'' 
Comm. in Math. Phys., vol.~273, no.~3, pp. 619--636 (2007). 

\bibitem{Moore}
C. Moore and M. Nilsson, ``Parallel quantum computation and quantum codes,''
SIAM Journal on Comp., vol.~31, no.~3, pp. 799--815 (2001).

\bibitem{NielChu}
M. A. Nielsen and I. L. Chuang, ``Quantum Computation and Quantum Information,''
(Cambridge  University Press, 2000).



\bibitem{PatelMarkov}
K. N. Patel, I. L. Markov and J. P. Hayes, ``Optimal synthesis of linear reversible circuits,'' 
Quant. Inf. Comp., vol.~8, no.~3 (2008).


\bibitem{Rauss} 
R. Raussendorf, D. E. Browne and H. J. Briegel, ``Measurement-based quantum computation on cluster states,'' Phys. Rev. A, vol.~68, no.~022312 (2003).






\bibitem{Smith} G. Smith and D. Leung, ``Typical entanglement of stabilizer states,''
Phys. Rev. A, vol.~74, no.~062314 (2006).

\bibitem{Vanden} M. Van den Nest, ``Classical simulation of quantum computation, the Gottesman-Knill theorem, and slightly beyond,'' Quant. Inf. Comp., vol.~10, pp. 0258--0271 (2010).

\bibitem{Vanden04} M. Van den Nest, J. Dehaene and B. De Moor, ``On local unitary versus local Clifford equivalence of stabilizer states,'' Phys. Rev. A, vol.~71, no.~062323 (2005).

\bibitem{Verst} F. Verstraete, J. I. Cirac and J. I. Latorre, 
``Quantum circuits for strongly correlated quantum systems,'' Phys. Rev. A, vol.~79, no.~032316 (2009).

\bibitem{Verst04} F. Verstraete and J. I. Cirac, ``Valence-bond states for quantum computation,''
Phys. Rev. A, vol.~70, no.~060302 (2004).


\bibitem{Wei} T.-C. Wei and P. M. Goldbart, ''Geometric measure of entanglement and applications to bipartite and multipartite quantum states,'' Phys. Rev. A, vol.~68, no.~042307 (2003). 



\bibitem{Wunder}
H. Wunderlich and M. B. Plenio, ``Quantitative verification of fidelities and entanglement from incomplete measurement data'', J. Mod. Opt., vol.~56, pp. 2100--2105 (2009). 


\end{thebibliography}
\end{document}